\documentclass[12pt,twoside]{article}   
\usepackage[super,sort,comma]{natbib}
\usepackage{gensymb}
\usepackage{booktabs} 
\usepackage{siunitx}
\usepackage[version=4]{mhchem}
\usepackage{widetable}
\usepackage[width=1.2\textwidth]{caption}

\usepackage{fancyhdr}		

\usepackage{soul}

\newcommand{\captionv}[3]{\begin{center}\parbox{#1cm}{\caption[#2]{{\sf #3}}}
        \end{center}}

\usepackage[section]{placeins}   %

\usepackage{graphicx}
\usepackage{subcaption}
\makeatletter \renewcommand\@biblabel[1]{$^{#1}$} \makeatother
\usepackage[mathlines]{lineno}

\usepackage{hyperref}
\hypersetup{ colorlinks,
    citecolor=blue,
    filecolor=blue,
    linkcolor=blue,
    urlcolor=blue
}

\usepackage{todonotes}
\usepackage{xcolor}

\definecolor{gray}{rgb}{0.6,0.6,0.6}
\definecolor{red}{rgb}{0.85,0,0}
\definecolor{green}{rgb}{0,0.85,0}
\definecolor{blue}{rgb}{0,0,0.85}
\definecolor{beige}{rgb}{0.92,0.87,0.78}
\usepackage[all]{hypcap}    

\begin{document}

\noindent
\sf {\Large {\bfseries Fast and accurate dose predictions for novel radiotherapy treatments in heterogeneous phantoms using conditional 3D-UNet generative adversarial networks} }\vspace*{10pt}\\ 
\noindent
\sf {Florian~Mentzel$^{\ast}$, Kevin~Kröninger}\\
\textit{\small {Department of Physics, TU Dortmund University, Germany}}\vspace*{5pt}\\
\sf {Michael~Lerch}\\
\textit{\small {Centre for Medical Radiation Physics, University of Wollongong, Australia}}\vspace*{5pt}\\
\sf {Olaf~Nackenhorst} \\
\textit{\small {Department of Physics, TU Dortmund University, Germany}} \vspace*{5pt}\\
\sf {Jason~Paino, Anatoly~Rosenfeld} \\
\textit{\small {Centre for Medical Radiation Physics, University of Wollongong, Australia}}\vspace*{5pt}\\
\sf {Ayu~Saraswati, Ah~Chung~Tsoi} \\
\textit{\small {School of Computing and Information Technology, University of Wollongong, Australia}} \vspace*{5pt}\\
\sf {Jens~Weingarten} \\
\textit{\small {Department of Physics, TU Dortmund University, Germany}} \vspace*{5pt}\\
\sf {Markus~Hagenbuchner} \\
\textit{\small {School of Computing and Information Technology, University of Wollongong, Australia}} \vspace*{5pt}\\
\sf {Susanna~Guatelli}\\
\textit{\small {Centre for Medical Radiation Physics, University of Wollongong, Australia}}\vspace*{5pt}\\

\noindent
Version typeset \today\\

\pagenumbering{roman}
\setcounter{page}{1}
\pagestyle{plain}
email: florian.mentzel@tu-dortmund.de \\

\newpage
\begin{abstract}

\noindent {\bf Purpose:} Novel radiotherapy techniques like synchrotron X-ray microbeam radiation therapy (MRT), require fast dose distribution predictions that are accurate at the sub-mm level, especially close to tissue/bone/air interfaces. Monte Carlo physics simulations are recognised to be one of the most accurate tools to predict the dose delivered in a target tissue but can be very time consuming and therefore prohibitive for treatment planning. Faster dose prediction algorithms are usually developed for clinically deployed treatments only. In this work, we explore a new approach for fast and accurate dose estimations suitable for novel treatments using digital phantoms used in pre-clinical development and modern machine learning techniques.
We develop a generative adversarial network (GAN) model, which is able to emulate the equivalent Geant4 Monte Carlo simulation with adequate accuracy, and use it to predict the radiation dose delivered by a broad synchrotron beam to various phantoms. 
\\\\ 
{\bf Methods:} 
The energy depositions used for the training of the GAN are obtained using full Geant4 Monte Carlo simulations of a synchrotron radiation broad beam passing through the phantoms. The energy deposition is scored and predicted in voxel matrices of size 140x18x18 with a voxel edge length of 1$\,$mm.\\
The GAN model consists of two competing 3D convolutional neural networks, which are conditioned on the photon beam and phantom properties. The generator network has a U-Net structure and is designed to predict the energy depositions of the photon beam inside three phantoms of variable geometry with increasing complexity. 
The critic network is a relatively simple convolutional network, which is trained to distinguish energy depositions predicted by the generator from the ones obtained with the full Monte Carlo simulation.
Because for future applications in treatment planning the model is required to predict the energy depositions inside phantoms or patients, which have not been used for training the model,
special attention is placed on the interpolation capabilities of the network.  \\\\
{\bf Results:} The energy deposition predictions inside all phantom geometries under investigation show deviations of less than 3\% of the maximum deposited energy from the simulation for roughly 99\% of the voxels in the field of the beam.
Inside the most realistic phantom, a simple pediatric head, the model predictions deviate by less than 1\% of the maximal energy deposition from the simulations in more than 96\% of the in-field voxels.
For all three phantoms, the model generalizes the energy deposition predictions well to phantom geometries, which have not been used for training the model but are interpolations of the training data in multiple dimensions.
The computing time for a single prediction is reduced from several hundred hours using Geant4 simulation to less than a second using the GAN model.
\\\\
{\bf Conclusions:} 
The proposed GAN model predicts dose distributions inside unknown phantoms with only small deviations from the full MC simulation with computations times of less than a second. It demonstrates good interpolation ability to unseen but similar phantom geometries and is flexible enough to be trained on data with different radiation scenarios without the need for optimisation of the model parameter. This proof-of-concept encourages to apply and further develop the model for the use in MRT treatment planning, which requires fast and accurate predictions with sub-mm resolutions.

\end{abstract}
%
%
%
%

\newpage

\setlength{\baselineskip}{0.7cm}      

\pagenumbering{arabic}
\setcounter{page}{1}
\pagestyle{fancy}

\section{Introduction}
\label{S:1}

A variety of highly conformal treatment techniques such as intensity-modulated radiation therapy~\cite{Brahme1982} (IMRT) and volumetric modulated arc therapy~\cite{Otto2008} (VMAT) have been developed in recent decades~\cite{Bucci2005}. These techniques have helped to reduce the healthy tissue dose during the irradiation of tumour cells and have led to strong improvements in the treatment of many types of cancer\cite{Jani2006, Deman2012}. However, for certain types of cancer, such as glioblastoma, the overall clinical outcome has seen little improvement \cite{VonNeubeck2015}. One potential treatment technique that has shown promising results in pre-clinical studies to treat these aggressive types of cancer, is microbeam radiation therapy (MRT), which is based on polarized high-intensity, spatially-fractioned synchrotron X-ray beams\cite{Brauer-Krisch2015, Engels2020, Grotzer2015}. The  treatment planning of such beams requires dose calculations with a spatial resolution of a few micrometers and are currently performed using time-consuming Monte Carlo (MC) simulations~\cite{Engels2020}. Accurate predictions of the dose distributions inside a phantom can take days using such MC simulations~\cite{Donzelli2018}. 
\\
Several faster approximations for MRT dose distribution computations have been investigated in recent years. The fastest reported computation method comprises the integration of MRT in the Varian Eclipse treatment planning software~\cite{Poole2017}. The approach calculates energy depositions based on the radiological depth and yields short computation times, but cannot provide accurate dose predictions in heterogeneous phantoms (overestimation of approximately 20\% of the dose in lung tissue inside water) and at high resolution (only macroscopic treatment of the dose calculation). Another approach uses a point kernel algorithm, which results in an execution time of approximately five minutes for the prediction of the dose inside a phantom, but is found to lead to large deviations compared to the full MC simulations, which are especially present at interfaces between tissues~\cite{Debus2017}. The agreement to full MC simulations is improved using a hybrid model, which combines a simplified MC simulation with an analytical dose kernel approach\cite{Donzelli2018}. This method is reported to take approximately 30 minutes to compute the dose distribution inside a phantom, which is still relatively long for a clinical application of a treatment planning system.
\\
The goal of our work is to develop an algorithm, which is fast enough for real-world treatment planning systems and accurate enough for high-resolution radiation therapy such as MRT. In the presented study, we take a first step into that direction by developing a general and flexible model for the fast and accurate prediction of dose distributions inside phantoms using modern machine learning techniques.\\
In recent years, machine learning has been used in an increasing number of publications for fast treatment planning of commercially-available therapies~\cite{Wang2020}. Most of these studies have in common, that neural networks are used to generate a full treatment plan based on a database of already delivered treatments \cite{Shiraishi2016}. For novel treatments, such as MRT, existing treatment plans are usually not available. 
More recent developments include the computation of dose distributions using computed tomography (CT) scans of a large number of patients as training data for neural networks \cite{Peng2019, Barragan-Montero2019, Kontaxis2020}. 
In the case of novel treatments like MRT, many different targets from slab phantoms to rats are usually subject to in-silico and experimental investigations (e. g. by Engels et al.\cite{Engels2020}). To allow development work on the machine learning model prior to the acquisition of dedicated CT data, we present a fully MC simulation-based approach using Geant4~\cite{Agostinelli2003} and digital phantoms which is possible for any kind of new therapy method, as long as a model for the MC simulation exists. The use of digital phantoms, rather than CT-based phantom geometries, in addition allows to generate accurate MC training data faster and to address eventual development problems more efficiently.\\
The machine learning algorithm presented in this paper is based on generative adversarial networks (GANs)~\cite{Goodfellow2014}. GANs have been successfully applied to a variety of complex data generation tasks~\cite{Lan2020} and have recently also been employed in the field of dose distribution caomputations for radiotherapy~\cite{Sarrut2019, Kearney2020}.
The basic concept of GANs is to use two competing neural networks: A generator network that learns to generate new data samples similar to given training samples, and a discriminator (or critic) network, which is trained to distinguish those generated samples from the ones of the training data.
Feedback from the discriminator enables the generator to learn to generate output which is indistinguishable from the training data, while the discriminator improves its level of distinction by being presented with more and more similar data from the generator.
A recent study~\cite{Kearney2020} found that, especially for the case of dose predictions in heterogeneous material resulting in high dose gradients, GANs are more accurate in their predictions compared to regression networks due to their dynamically changing objective function, which can capture more volume-level information than a localized loss function.
While the training of GANs is a complex process and often challenging and time-consuming, the prediction using only the generator is usually very fast, which makes them an ideal candidate for our purposes.
To our knowledge, there s no prior published study which includes every step of the development process: the creation of the required digital phantoms, the design of full Monte Carlo simulations which are adapted to a special irradiation modality and are experimentally validated and, finally, training of the studied ML model with detailed investigation of its performance.  
While there are many publications about dose predictions using deep learning models, only few publications that we know of aim at similar research goals. For example, Zang et al.~\cite{Zhang2021} published a very notable study on dose predictions with a GAN in the case of proton irradiation, they did not compare their predictions though on a voxel-by-voxel base, especially around tissue borders. This makes it difficult to inspect the usability for high-gradient applications. In the work of Kearney et al.~\cite{Kearney2020}, GANs were used for dose prediction and results were compared with Monte Carlos simulations as well, but no detailed comparison of depth-dose curves is shown as the focus remains on clinical parameters like dose-volume histograms. More detailed comparisons were conducted e. g. in the work of Kontaxis et al.~\cite{Kontaxis2020}, but as they use only prostate CT scans for all their data, it is difficult to understand how the training and test data really differ, even if taken from individual patients. Our approach using digital phantoms allows for a more straightforward understanding of differences between training and test. 
\\
As a proof-of-concept, we apply our GAN model to three simplified irradiation scenarios, which are incrementally increasing in their complexity: a water phantom with a rotated bone slab, a water phantom with a rotated bone slab of variable thickness and a pediatric head phantom. For simplicity, we use a broad beam instead of a microbeam for the irradiation of the phantoms and require a spatial resolution of $1\,$mm$^3$ in the dose prediction.
\\
This paper is structured as follows: Section~\ref{section:MaterialsAndMethods} describes the MC simulation using Geant4 and the developed machine learning model. 
In Section~\ref{section:Results} the dose predictions of the GAN obtained for the three scenarios are presented and compared to the full MC simulation. Finally, in Section~\ref{section:Discussion} the results are discussed and in Section~\ref{section:Conclusion} conclusions are made.
\section{Materials and methods\label{section:MaterialsAndMethods}}
The development of the digital phantoms and the dose computation model follows an iterative-incremental approach which is recognized to a successful software process, adopted also in the development of Geant4~\cite{Allison2016, geant4Whitepaper}. Following this software process, the machine learning dose prediction model is trained on the simpler configurations described in this section. This allows to generate accurate MC data more rapidly compared to CT-based simulations and thereby to address eventual development problems more easily and to develop a more robust software product. As MRT is at pre-clinical stage and dosimetric studies are performed in phantoms and in pets, two digital phantoms which are commonly used in pre-clinical research are presented in this work: a slab and an ellipsoid head phantom.

\subsection{Monte Carlo simulation}
\begin{figure} [!tb]
	\centering
	\begin{subfigure}[t]{0.45\textwidth}
	    \includegraphics[width=\linewidth]{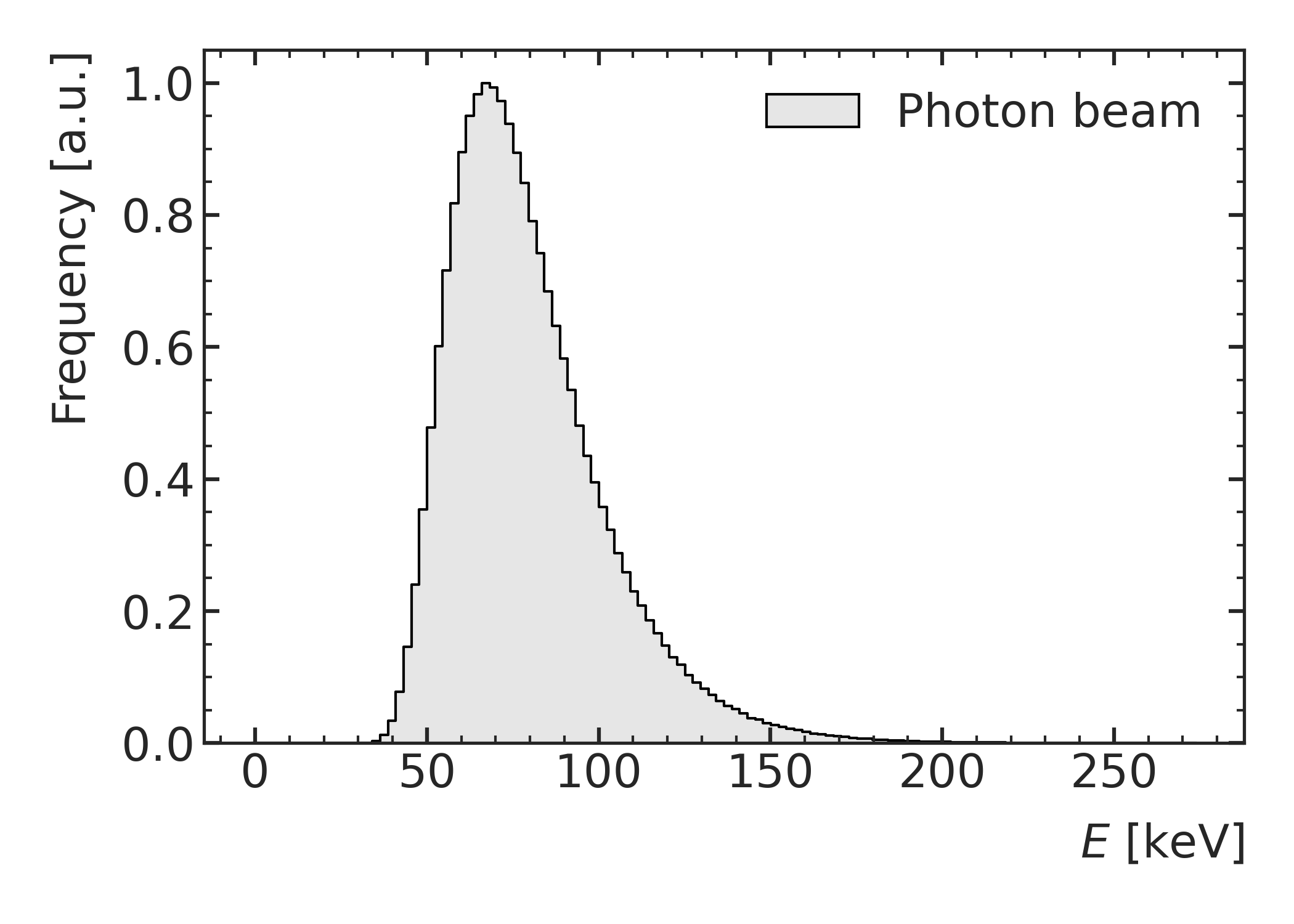}
	    \caption{}
		\label{fig:MaterialsAndMethods:Geant4:psf:a}
    \end{subfigure}
    \begin{subfigure}[t]{0.45\textwidth}
	    \includegraphics[width=\linewidth]{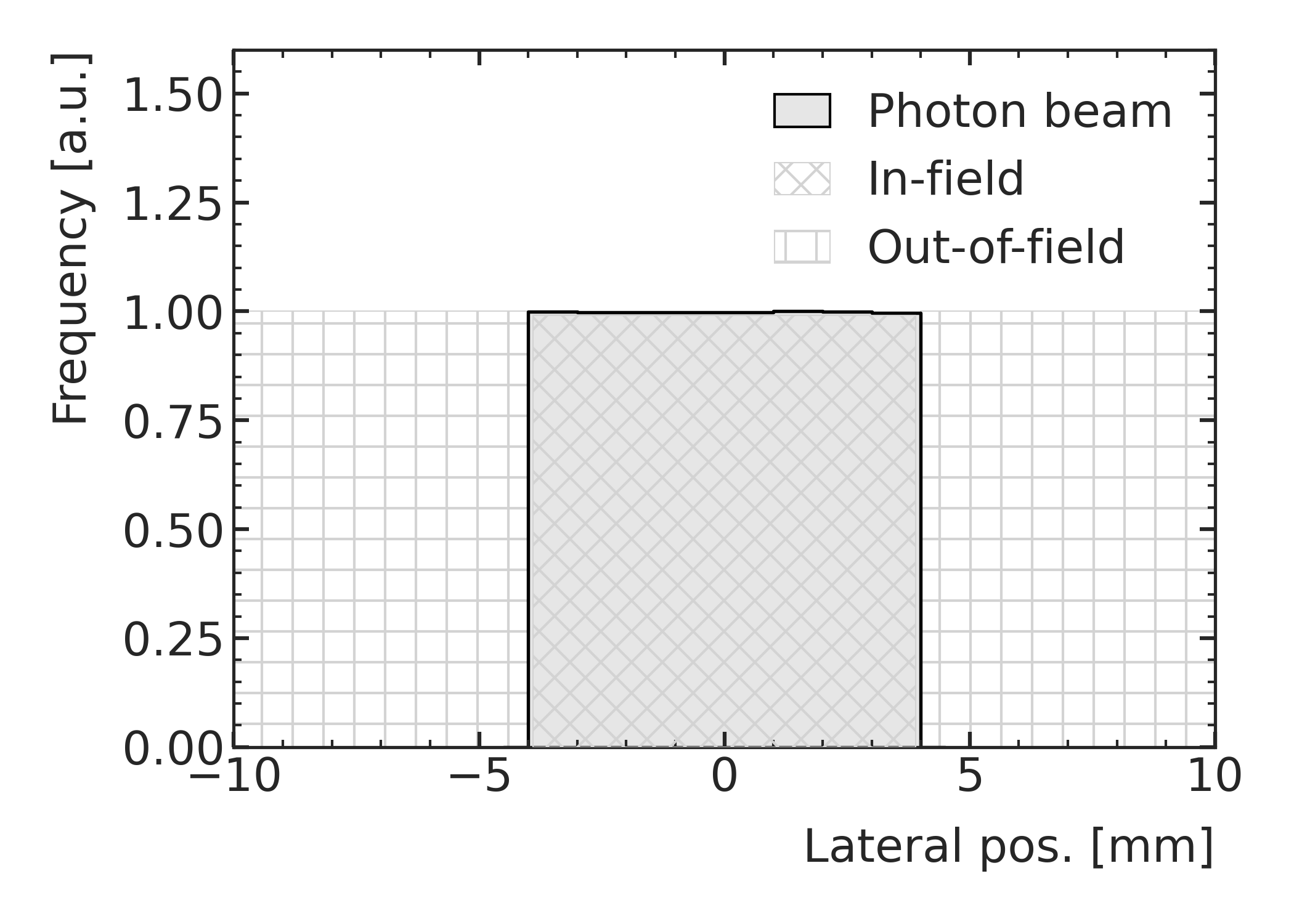}
	    \caption{}
		\label{fig:MaterialsAndMethods:Geant4:psf:b}
    \end{subfigure}

	\caption{(a) Energy spectrum of X-ray photons in keV as used as input to the simulation and as derived from Dipuglia et. al\cite{Dipuglia2019}. (b) Lateral beam profile after the tungsten mask indicating the in-field and out-of-field region of the beam.\bigskip}
	\label{fig:MaterialsAndMethods:Geant4:psf}
\end{figure}

\begin{figure}[!tb]
	\centering
	\begin{subfigure}[t]{0.31\textwidth}
		\includegraphics[width=\linewidth]{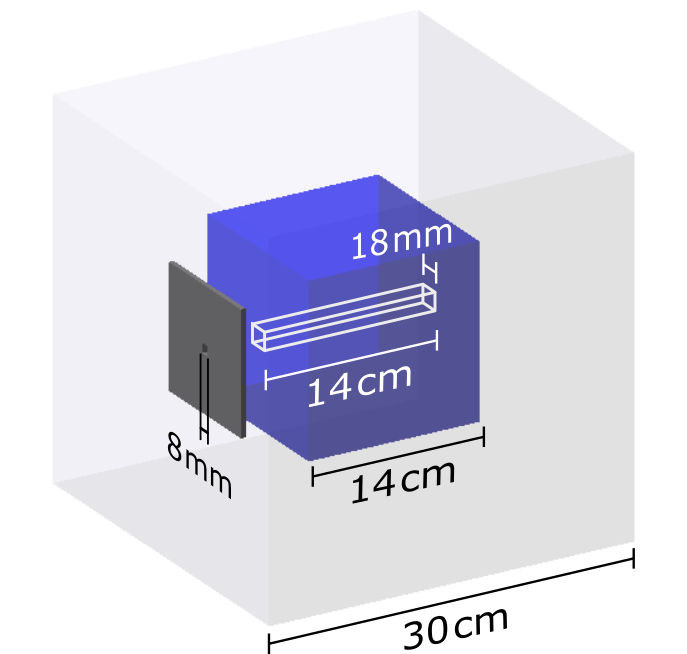}
		\caption{}
		\label{fig:MaterialsAndMethods:Geant4:wateronlybeamcenter:a}
	\end{subfigure}
	\hfill
	\begin{subfigure}[t]{0.32\textwidth}
		\includegraphics[width=\linewidth]{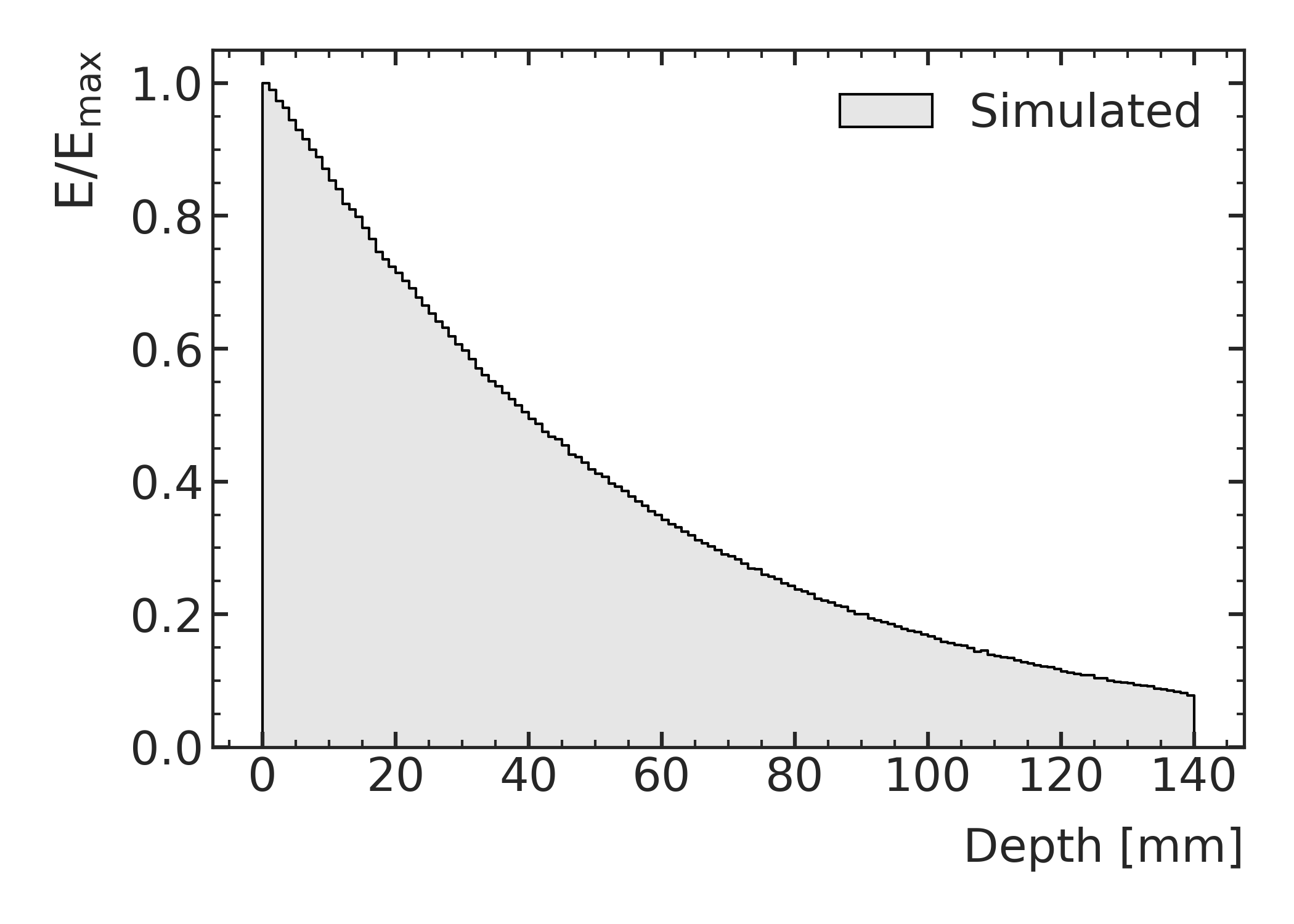}
		\caption{}
		\label{fig:MaterialsAndMethods:Geant4:wateronlybeamcenter:b}
	\end{subfigure}
	\hfill
	\begin{subfigure}[t]{0.32\textwidth}
		\includegraphics[width=\linewidth]{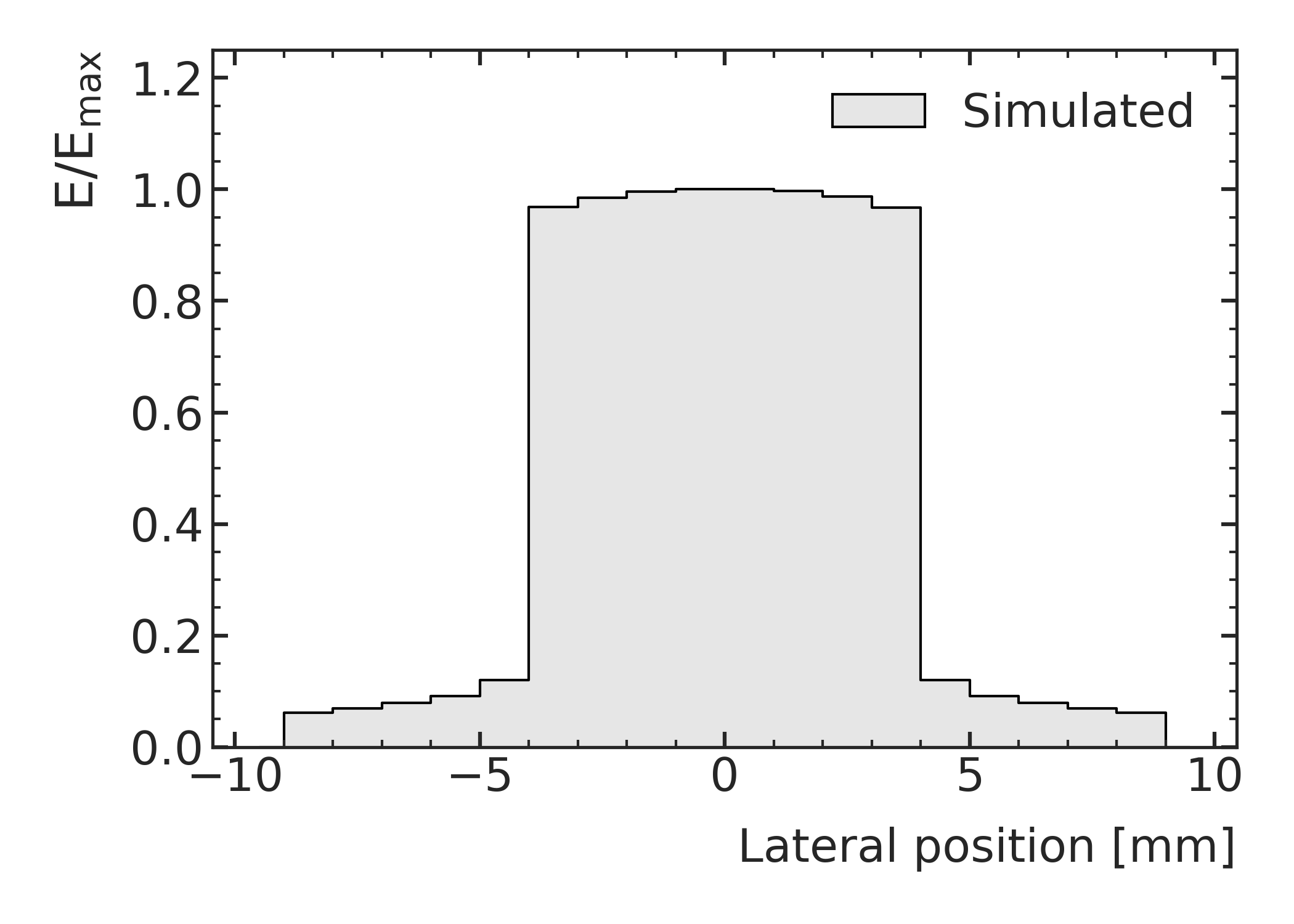}
		\caption{}
		\label{fig:MaterialsAndMethods:Geant4:wateronlybeamcenter:c}
	\end{subfigure}
	\caption{(a) Schematic of the simulation setup of a cubic water phantom including the world volume (light grey), the tungsten mask (dark grey), the water phantom (blue) and the scoring volume (outlined in white). (b) Normalized energy deposition inside the water phantom along the depth at the center of the beam. (c) Normalized lateral energy deposition inside the water phantom integrated along beam direction. The energy deposition is normalised to the maximum value $E_\mathrm{max}$ of the in-field region of the beam.\bigskip}
	\label{fig:MaterialsAndMethods:Geant4:wateronlybeamcenter}
\end{figure}

The photons of the broad beam and their interactions with the material of the various phantoms are simulated using Geant4 10.6p01~\cite{Agostinelli2003} with option 4 of the standard electromagnetic physics constructor, which is adapted to model the interactions of the polarized photons with the target (Livermore Polarized Physics)\cite{Geant4PhysicsManual}.
The beam is simulated using the description of the Imaging and Medical beam line (IMBL) at the Australian Synchrotron~\cite{Stevenson2017} and corresponds to a typical highly-polarized synchrotron beam~\cite{Spiga2009} as used for pre-clinical research for MRT~\cite{Engels2020, Dipuglia2019}, before passed through a mulit-slit collimator to produce the microbeams. The energy spectrum of the \st{of the} X-ray beam is shown in Figure~\ref{fig:MaterialsAndMethods:Geant4:psf:a}.
The initial beam traverses a tungsten mask in order to obtain a beam cross section of 8x8$\,$mm$^2$ before it hits the various phantoms. The lateral beam profile together with the in-field and out-of-field definition is shown in Figure~\ref{fig:MaterialsAndMethods:Geant4:psf:b}.
The phantoms are placed at a distance of 4$\,$cm behind the mask to reduce back-scattering effects, which is shown exemplary in Figure~\ref{fig:MaterialsAndMethods:Geant4:wateronlybeamcenter:a} for a cubic water phantom with a size of 14x14x14 cm$^3$.
The tungsten mask (dark grey) and the water phantom (blue) are surrounded by a total simulation volume (light grey) of size 30x30x30$\,$cm$^3$.
For all simulated phantoms, the energy depositions are calculated for each voxel of size 1x1x1$\,$mm$^3$ in the scoring volume of DxWxH=140x18x18$\,$mm$^3$, which starts at the surface of the phantom and is centered around the beam. The one-dimensional projection of the normalized energy depositions of the beam inside this water phantom along the beam (depth) and the profile perpendicular to the beam (lateral) are shown in Figure~\ref{fig:MaterialsAndMethods:Geant4:wateronlybeamcenter:b} and Figure~\ref{fig:MaterialsAndMethods:Geant4:wateronlybeamcenter:c}, respectively.\\
%
The Geant4-based data samples were simulated using the LiDO3 cluster at TU Dortmund University which took approximately two weeks using on average 500 Intel Xeon cores simultaneously.
All simulated data is divided into \textit{training data}, which is presented to the neural network for weight optimization, \textit{validation data}, which is used to identify the best model parameters and
\textit{test data}, which is exclusively used to evaluate the performances after the best machine learning model was chosen. 

\subsubsection{Water phantom with rotated bone slab of constant thickness}
\label{subsection:matMeth:rotBone}
\begin{figure}[!tb]
	\centering
	\begin{subfigure}[t]{0.31\textwidth}
		\includegraphics[width=\linewidth]{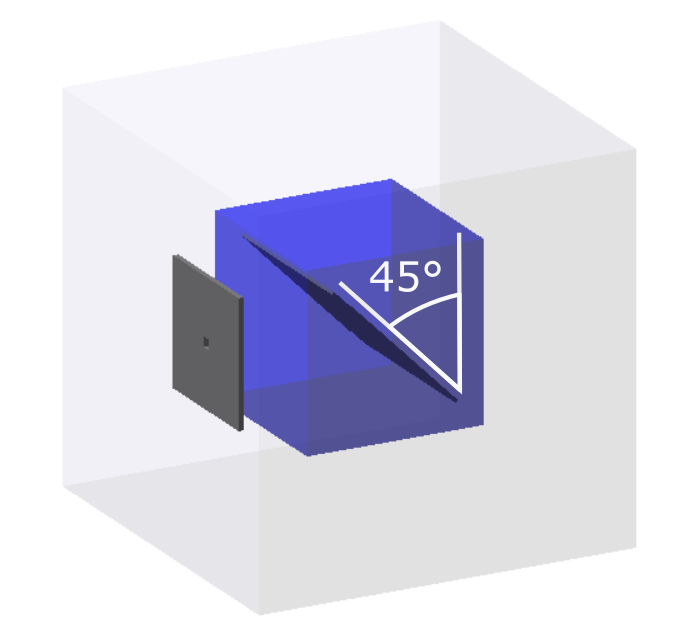}
		\caption{}
        \label{fig:MaterialsAndMethods:Geant4:bone_slab_sim:a}
    \end{subfigure}
	\hfill
	\begin{subfigure}[t]{0.32\textwidth}
		\includegraphics[width=\linewidth]{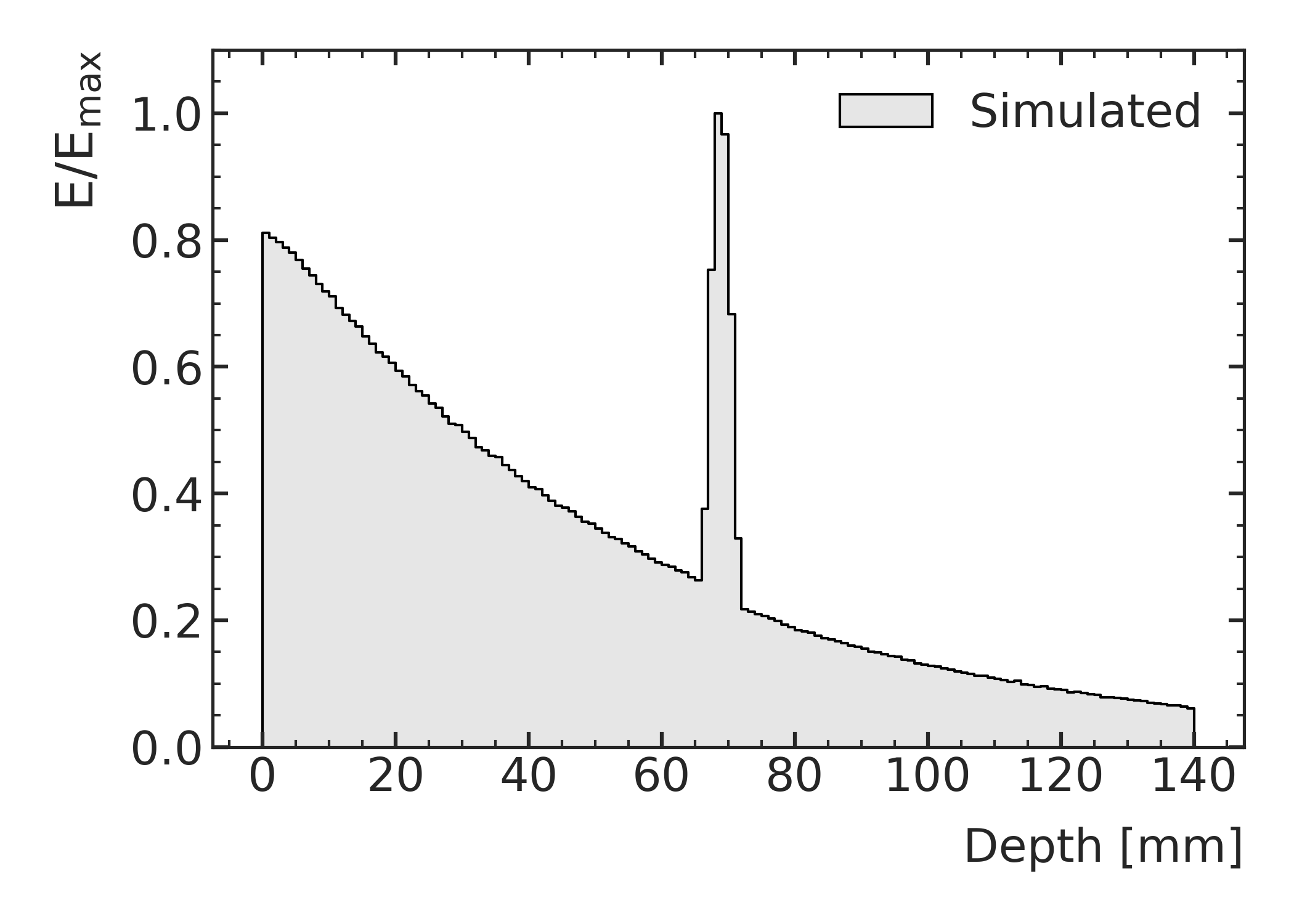}
		\caption{}
		\label{fig:MaterialsAndMethods:Geant4:bone_slab_sim:b}
	\end{subfigure}
	\hfill
	\begin{subfigure}[t]{0.32\textwidth}
		\includegraphics[width=\linewidth]{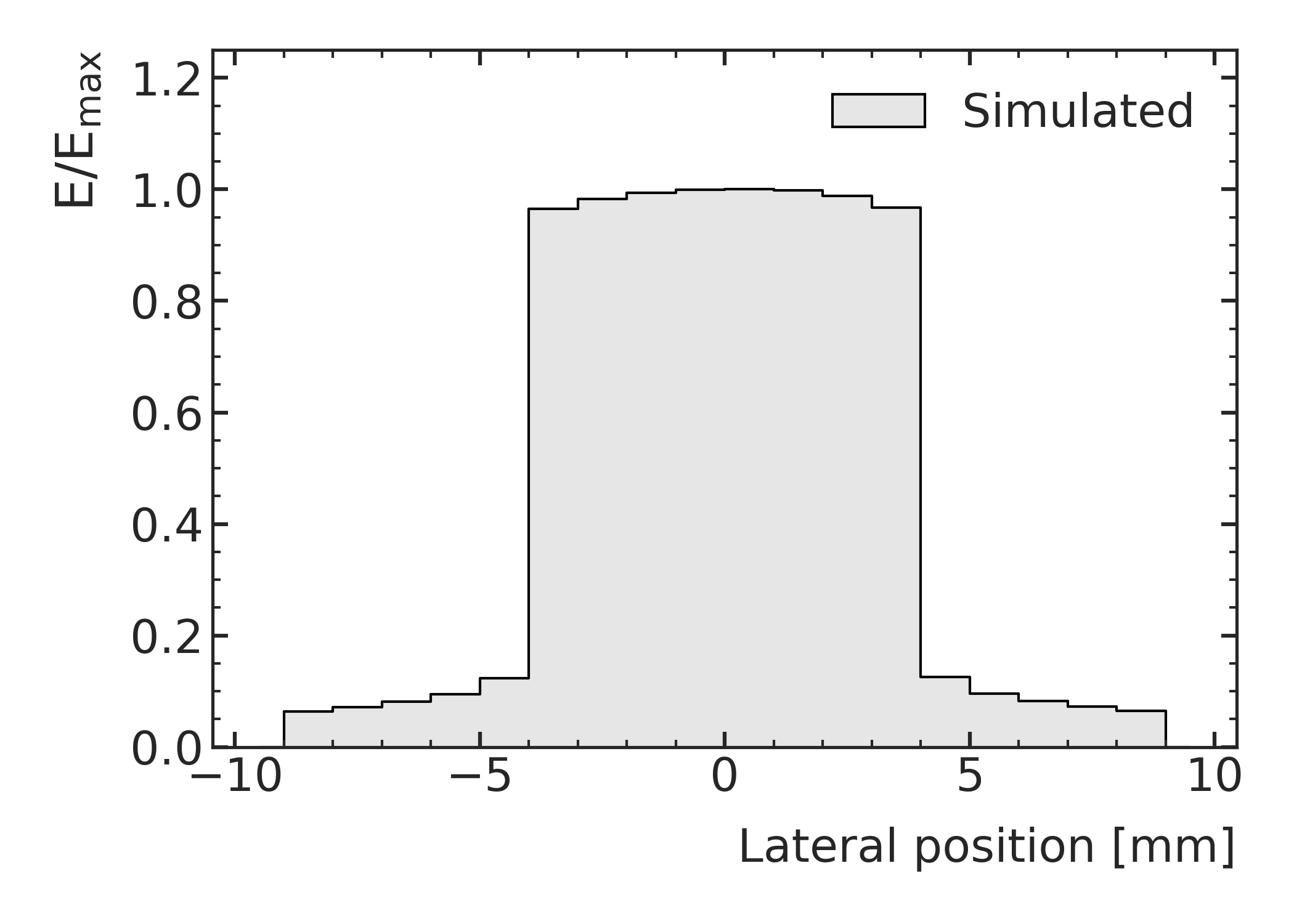}
		\caption{}
		\label{fig:MaterialsAndMethods:Geant4:bone_slab_sim:c}
	\end{subfigure}
	\caption{
	(a) Schematic of the simulation setup of a cubic water phantom with inserted bone slab including the world volume (light grey), the tungsten mask (dark grey), the water phantom (blue) and the bone slab (dark blue) rotated at 45$\degree$ (white). (b) Normalized energy deposition inside the water phantom with the rotated bone slab along the depth at the center of the beam. (c) Normalized lateral energy deposition inside the water phantom with the rotated bone slab integrated along beam direction. The energy deposition is normalised to the maximum value $E_\mathrm{max}$ of the in-field region of the beam.
    \bigskip}
	\label{fig:MaterialsAndMethods:Geant4:bone_slab_sim}
\end{figure}

\begin{figure}[!tb]
	\centering
	\includegraphics[width=.45\linewidth]{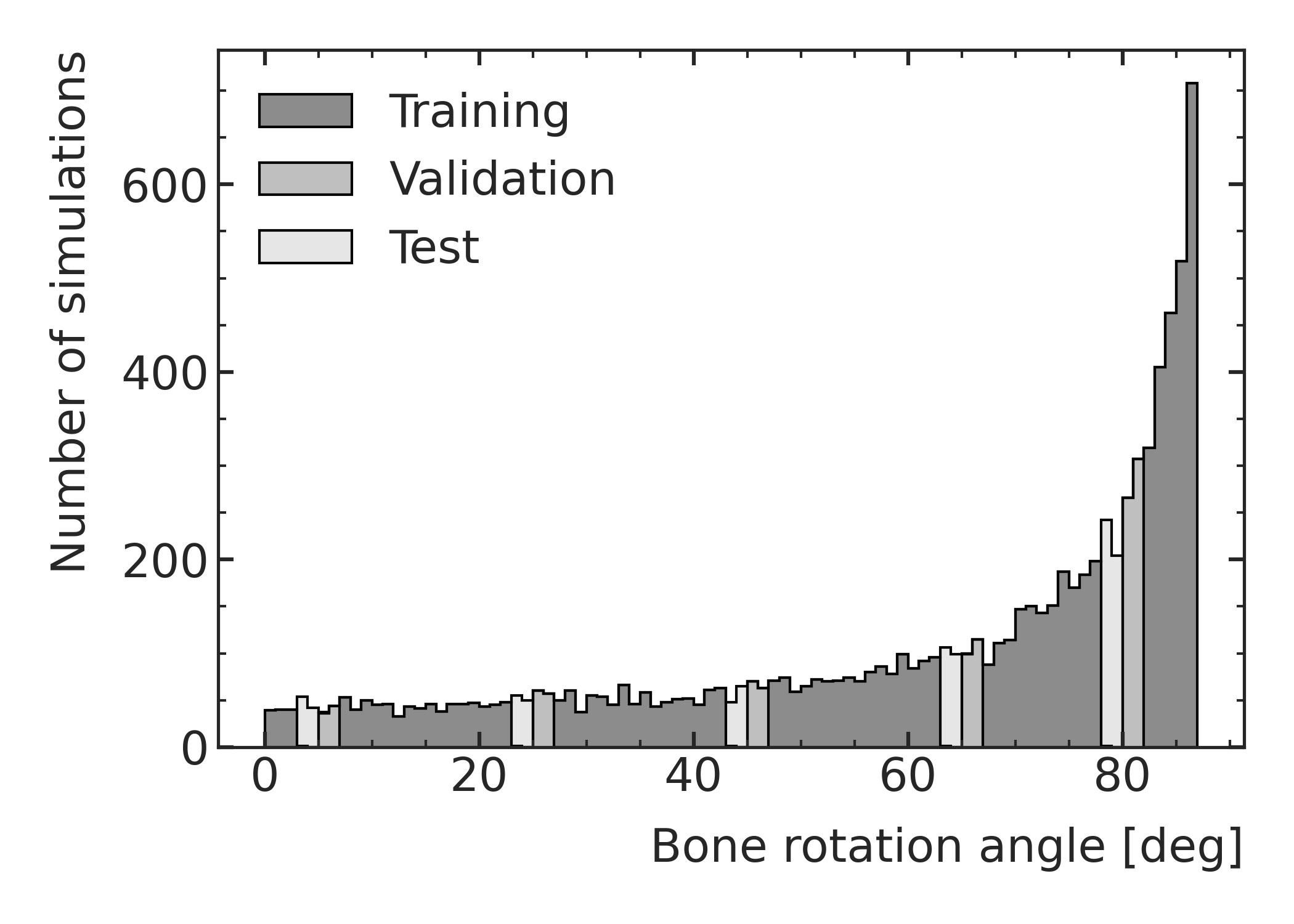}
	\caption{Number of simulated training (dark grey), validation (medium grey) and test samples (light grey) as a function of bone slab rotation angle.\bigskip}
	\label{fig:MaterialsAndMethods:Geant4:ang_dist_all_dist}
\end{figure}

The simplest phantom considered in this studies, consist of a water cube as described above with a centrally inserted compact bone slab (ICRU, $\rho = 1.85\,\mathrm{g/cm^3}$) with a constant thickness of $d=2.5\,$mm, which is rotated continuously with angles of \mbox{$\alpha \in [0,87]\degree$} around an axis perpendicular to the beam axis.
A schematic of the phantom including the tungsten mask is shown in Figure~\ref{fig:MaterialsAndMethods:Geant4:bone_slab_sim:a} with an exemplary rotation angle of \mbox{$\alpha=45\degree$}.
Figure~\ref{fig:MaterialsAndMethods:Geant4:bone_slab_sim:b} and \ref{fig:MaterialsAndMethods:Geant4:bone_slab_sim:c} shows the normalized energy depositions of the beam inside the phantom along the beam axis and along the rotation axis integrated over the beam axis, respectively.
Due to the rotation of the bone slab, the energy deposition peaks slightly before the center of the phantom ($70\;$mm in depths), where the bone slab is inserted.\\
The simulated data of this phantom serves as a feasibility study to develop a flexible GAN model, which is general enough to be applied to more complex phantom geometries and predicts at the same time accurately the simulated energy depositions inside the phantom. 
The traversed distance of the beam within the bone material and, as a result, the energy depositions change more drastically at larger rotation angles. To increase the sampling density of the simulation at those larger rotation angles, the rotation angles of the simulation are determined by drawing random numbers from a $1/\cos(\alpha)$ distribution.
The simulated data are split into training data with rotation angles \mbox{$\alpha \in \{[0,3)\cup[7,23)\cup[27,43)\cup[47,63)\cup[67,78)\cup[82,87]\}\degree$}, validation data with rotation angles of \mbox{$\alpha \in \{[5,7)\cup[25,27)\cup[45,47)\cup[65,67)\cup[80,82)\}\degree$} and test data with rotation angles of \mbox{$\alpha \in \{[3,5)\cup[23,25)\cup[43,45)\cup[63,65)\cup[78, 80)\}\degree$}. Here, the symbol '[' denotes an inclusive interval limit, ')' means an exclusive interval limit and '$\cup$' the union of the individual intervals.
The resulting number of simulated samples and the split in training, validation and test set is shown in Figure~\ref{fig:MaterialsAndMethods:Geant4:ang_dist_all_dist}.
The purpose of this split is to have unbiased simulation data for the performance evaluation during the training and hyperparameter optimization (validation data) as well as for the final assessment of the performance (test data). At the same time this allows for investigating how well the GAN model can interpolate in one dimension.

\subsubsection{Water phantom with rotated bone slab of variable thickness\label{subsection:matMeth:4thickness}}
\begin{figure}[!tb]
	\centering
	\begin{subfigure}[t]{0.45\textwidth}
		\includegraphics[width=\linewidth]{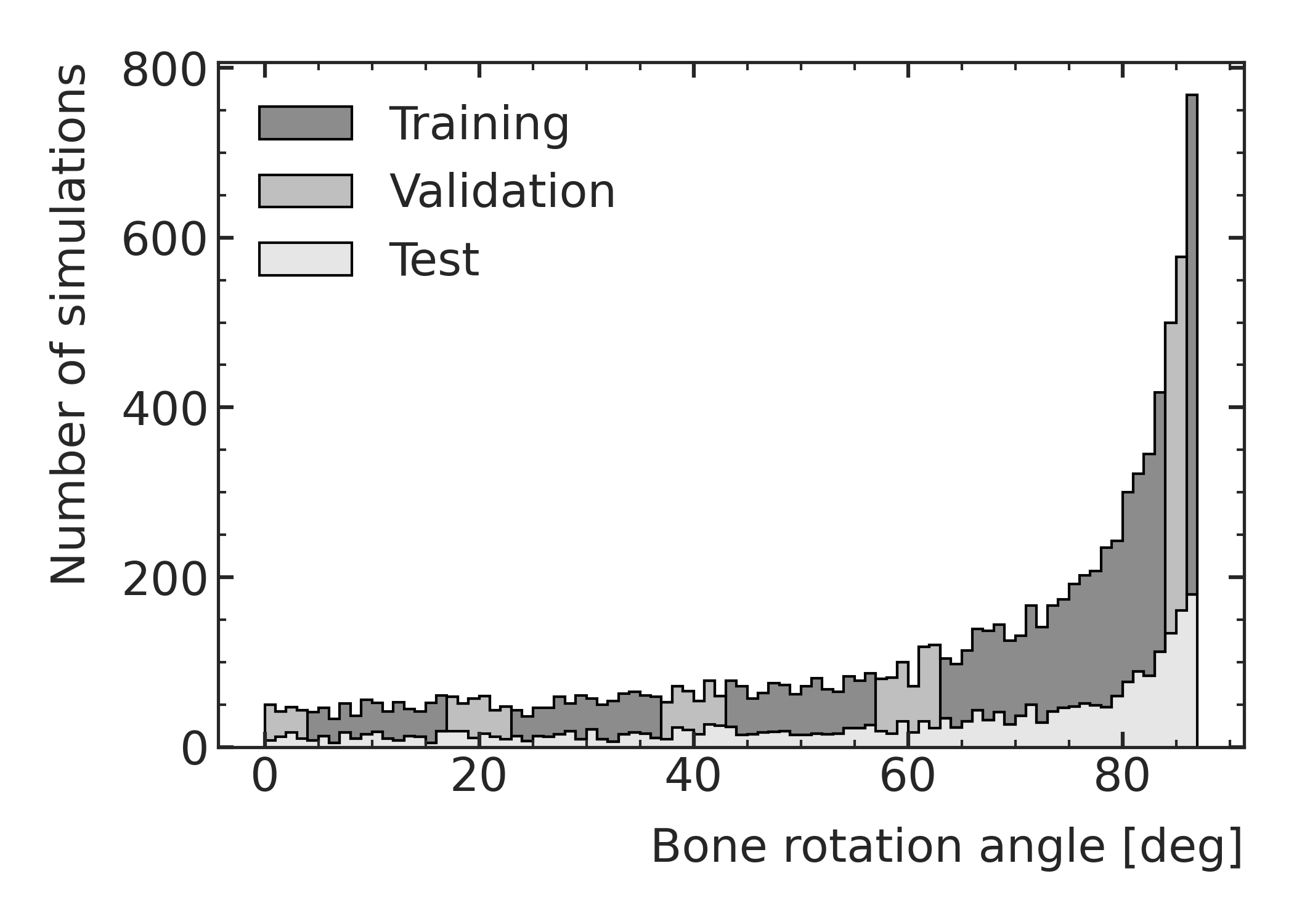}
		\caption{}
        \label{fig:MaterialsAndMethods:Geant4:datset_variable_thicknesses:a}
    \end{subfigure}
    \begin{subfigure}[t]{0.45\textwidth}
		\includegraphics[width=\linewidth]{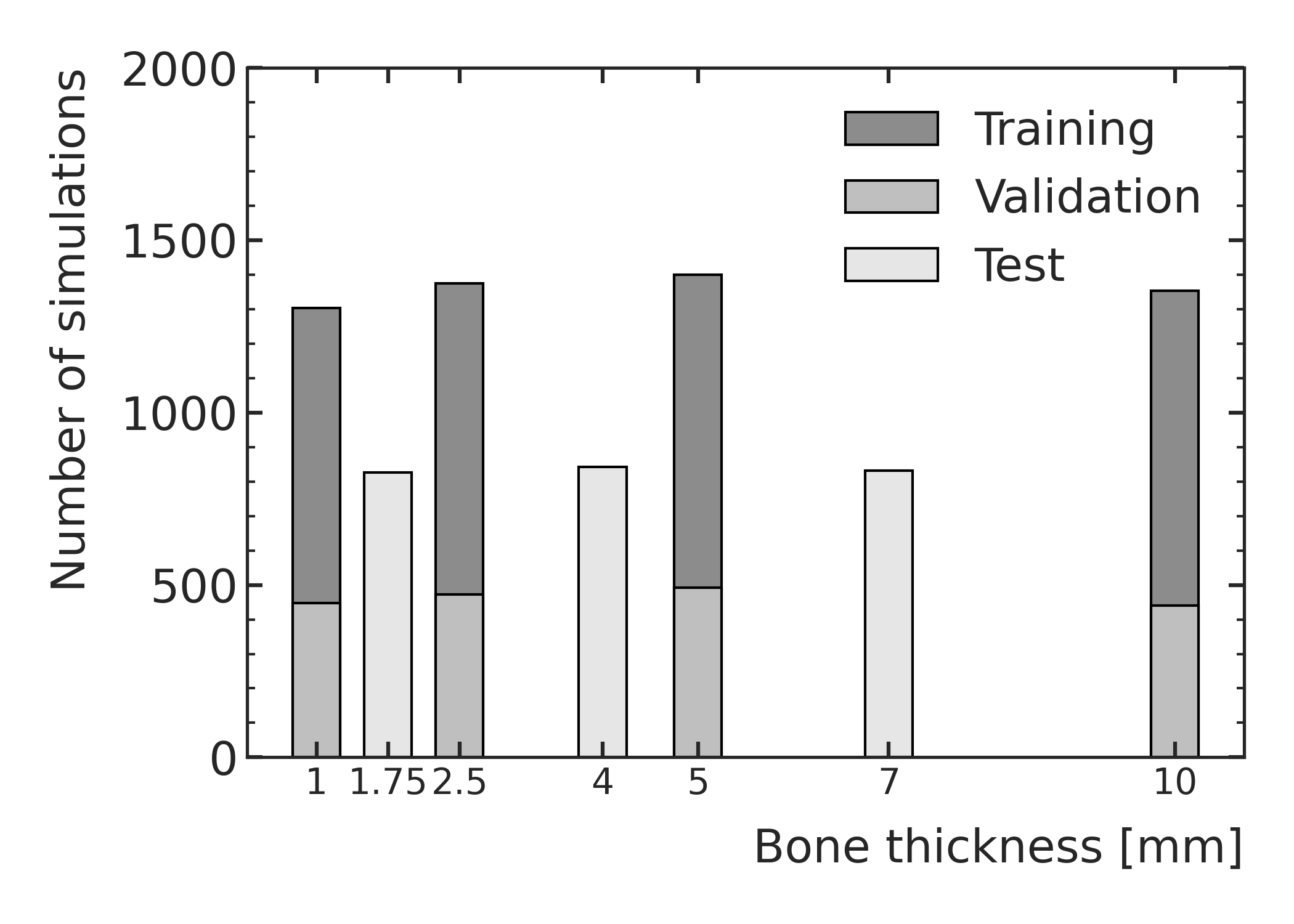}
		\caption{}
        \label{fig:MaterialsAndMethods:Geant4:datset_variable_thicknesses:b}
    \end{subfigure}
	
	\caption{Number of simulated training (dark grey), validation (medium grey) and test samples (light grey) as a function of (a) bone slab rotation angle and (b) bone slab thickness.\bigskip}
	\label{fig:MaterialsAndMethods:Geant4:datset_variable_thicknesses}
\end{figure}

The second phantom is identical to the first phantom with the exception that the thickness of the bone slab is varied in addition to varying rotation angle.
The purpose of the simulated data using this phantom is to demonstrate the interpolation capabilities of the GAN model in two dimensions. This is of particular importance, because for future applications in treatment planning, the model needs to be able to accurately predict energy depositions for unknown geometries, which have not been used for the training of the GAN. Multi-dimensional interpolation is in general not an easy task for a generative model, which is why we consider this as an important step before moving to more complex phantom geometries.
For this purpose, the simulated data is split into training data with bone slab thicknesses of
$d \in \{1, 2.5, 5, 10\}$mm and rotation angles of \mbox{$\alpha \in \{[4,17)\cup[23,37)\cup[43,57)\cup[63,84)\cup[86,87)\}\degree$}.
The simulated data of the same thicknesses but with rotation angles not used during the training and in the range of $[0,87]\degree$ are used as validation data.
In order to investigate the ability of the GAN to predict the energy depositions inside unknown phantom geometries, the test data is simulated with bone slab thicknesses of $d \in \{1.75, 4, 7\}$mm but continuously in the full rotation angle range of $[0,87]\degree$. The split of the simulated data is illustrated in  Figure~\ref{fig:MaterialsAndMethods:Geant4:datset_variable_thicknesses}.
\subsubsection{Simple paediatric head phantom with translation} \label{subsection:matMeth:simpleHead}
\begin{figure}[!tb]
 	\centering
 	\begin{subfigure}[t]{0.28\textwidth}
		\includegraphics[width=\linewidth]{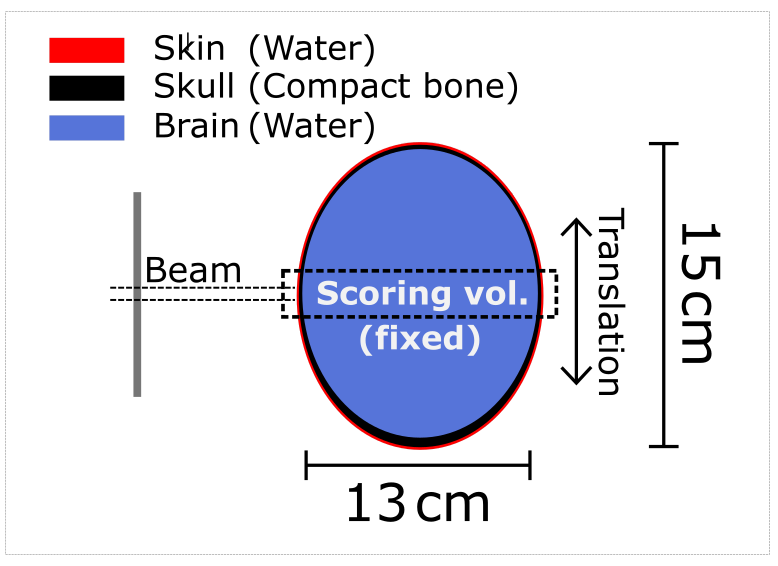}
		\caption{}
        \label{fig:MaterialsAndMethods:Geant4:translateheadsimwitharrows:a}
    \end{subfigure}
	\hfill
	\begin{subfigure}[t]{0.32\textwidth}
		\includegraphics[width=\linewidth]{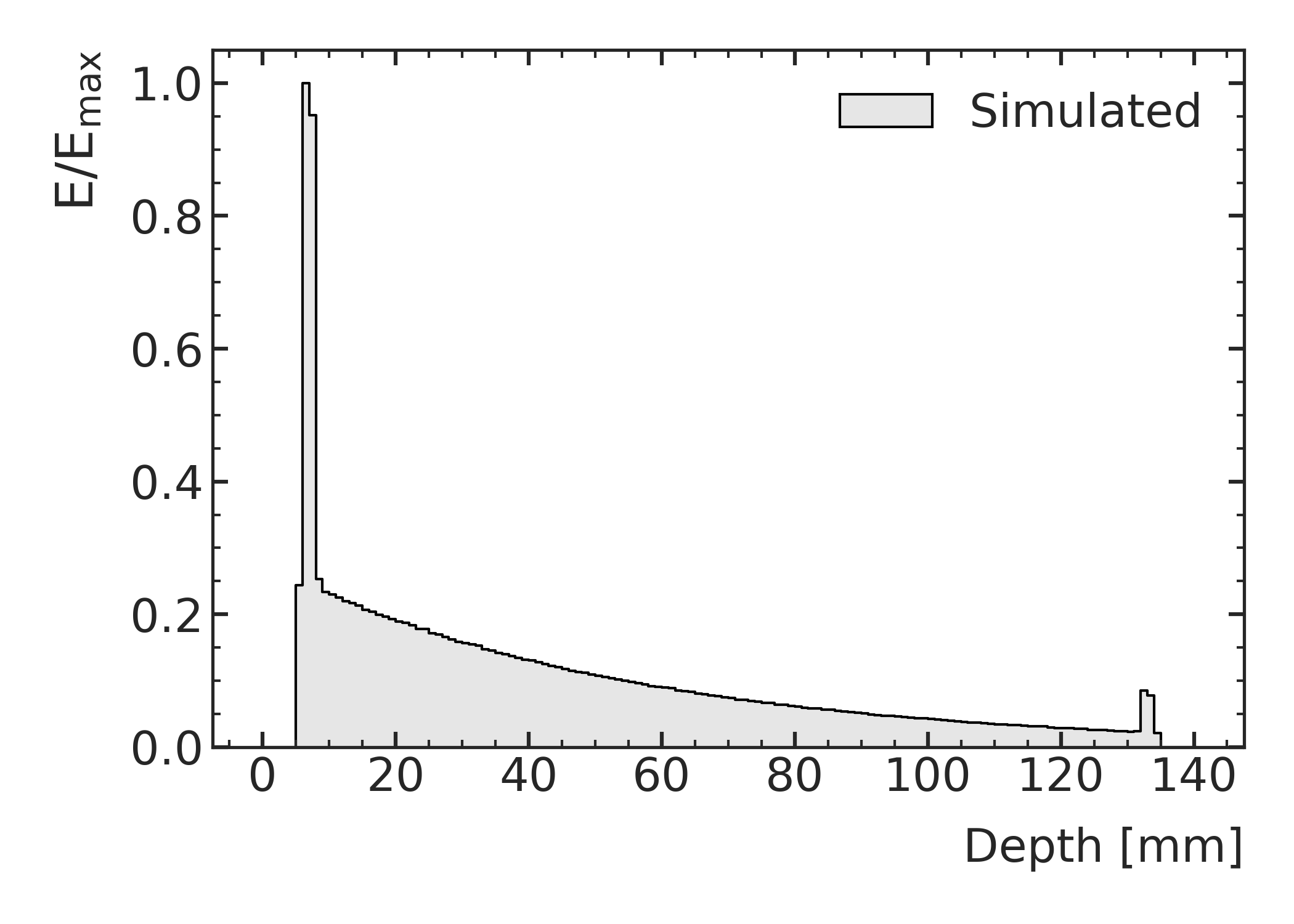}
		\caption{}
		\label{fig:MaterialsAndMethods:Geant4:translateheadsimwitharrows:b}
	\end{subfigure}
	\hfill
	\begin{subfigure}[t]{0.32\textwidth}
		\includegraphics[width=\linewidth]{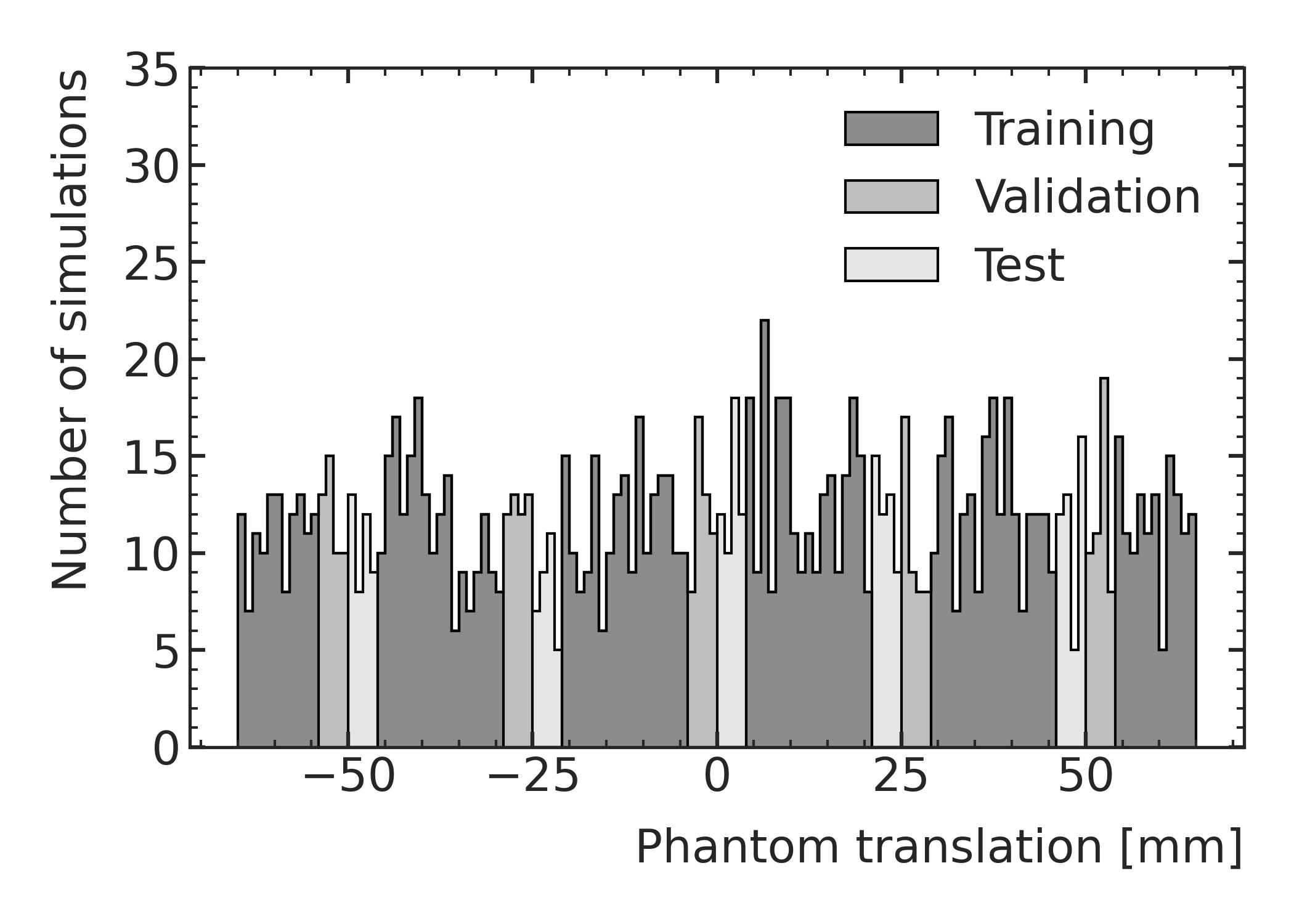}
		\caption{}
		\label{fig:MaterialsAndMethods:Geant4:translateheadsimwitharrows:c}
	\end{subfigure}
 	
 	\caption{(a) Schematic of the simulation setup of a simple paediatric head phantom showing the tungsten mask (dark grey), the incoming beam, the skin (red), the skull (black), the brain (blue) and the scoring volume and the fixed scoring volume (outlined by dotted line). (b) Normalized energy deposition inside the water phantom along the depth at the center of the beam with translation $t=0$\,mm. The energy deposition is normalised to the maximum value $E_\mathrm{max}$ of the in-field region of the beam. (c) Number of simulated training (dark grey), validation (medium grey) and test samples (light grey) as a function of phantom translation. Normalized lateral energy deposition inside the water phantom integrated along the beam direction. The energy deposition is normalised to the maximum value $E_\mathrm{max}$ of the in-field region of the beam.\bigskip}
 	\label{fig:MaterialsAndMethods:Geant4:translateheadsimwitharrows}
 \end{figure}

The third phantom is a simple model of a pedeatric head. The skull of the head is constructed using two rotational ellipsoids (spheroids). 
The first spheroid has a length of 15$\,$cm and a width/height of 13$\,$cm. The second spheroid with a length of 14.25$\,$cm and width/height of 12.6$\,$cm is located inside the first one and is slightly shifted by 1.25$\,$mm from the center to the front. The volume in between the two spheroids corresponds to the skull and consists of compact bone (ICRU, $\rho = 1.85\,\mathrm{g/cm^3}$), 
resulting in bone thicknesses of 2.5$\,$mm at the forehead, of 5$\,$mm at the back of the head and of 2$\,$mm at the sides.
The obtained skull is filled with water in order to mimic the brain tissue and surrounded by a thin layer (1$\,$mm) of water in order to model the skin. The head is positioned orthogonal to the beam, such that the radiation would be performed centrally from the side. A schematic of the phantom and the tungsten mask is shown in Figure~\ref{fig:MaterialsAndMethods:Geant4:translateheadsimwitharrows:a}.
The simulation is performed with translations $t$ of the head orthogonal to the beam in the range of $t\in [-65, 65]$\,mm along the longitudinal axis of the spheroid in order to simulate any possible position of the beam along the equatorial ellipse.
The number of simulated samples is uniformly distributed over the full translation range. Due to the translation of the head, the thickness, shape, angle and position of the radiated bone changes at the same time. 
An exemplary normalized energy deposition inside the phantom at the center of the head with $t=0$\,mm is shown in Figure~\ref{fig:MaterialsAndMethods:Geant4:translateheadsimwitharrows:b}. The two energy peaks result from the bones at the front and the back of the skull.\\
The purpose of this simulated data is to investigate the GANs ability to predict energy depositions in more complex geometries and to interpolate the predictions in dynamically changing geometries in multiple dimensions. Although the head phantom has a relatively simple geometry compared to realistic heads, this study is an important milestone and a proof-of-concept for future applications aiming for real-life treatment planning.
The simulated data is split in training data with \mbox{$t \in \{[-65,-54)\cup[-46,-29)\cup[-21,-4)\cup[4,21)\cup[29,46)\cup[54,65]\}$\,mm}, in validation data with \mbox{$t \in \{[-54,-50)\cup[-29,-25)\cup[-4,0)\cup[25,29)\cup[50,54)\}$\,mm} and in test data with \mbox{$t \in \{[-50,-46)\cup[-25,-21)\cup[0,4)\cup[21,25)\cup[46,50)\}$\,mm}) as visualized in Figure~\ref{fig:MaterialsAndMethods:Geant4:translateheadsimwitharrows:c}.

\subsection{The Generative Adversarial Network Model}
In the following we describe and motivate our model, which was using the simulated data of the first phantom. 
We use a conditional GAN~\cite{Mirza2014} with a Wasserstein loss function~\cite{Arjovsky2017} and a Gradient Penalty term\cite{Gulrajani2017} to regularize and stabilize the weight updates,
which are performed using the Adam optimizer~\cite{Kingma2015} with an initial learning rate of $\alpha=2\cdot10^{-5}$ and a batch size of 32.
As the output of Wasserstein GANs (WGANs) reflects not only a qualitative but also a quantitative assessment by using an approximation of the Earth-Mover's distance as loss, the discriminator is usually called critic instead.
One batch of simulation data is used to update the weights of the critic five consecutive times, before it is used to update the weights of the generator a single time.
This ensures that the critic is strong enough to provide valuable information to the training of the generator.
\subsubsection{The generator model}
\begin{figure}[!tb]
	\centering
	\includegraphics[width=\linewidth]{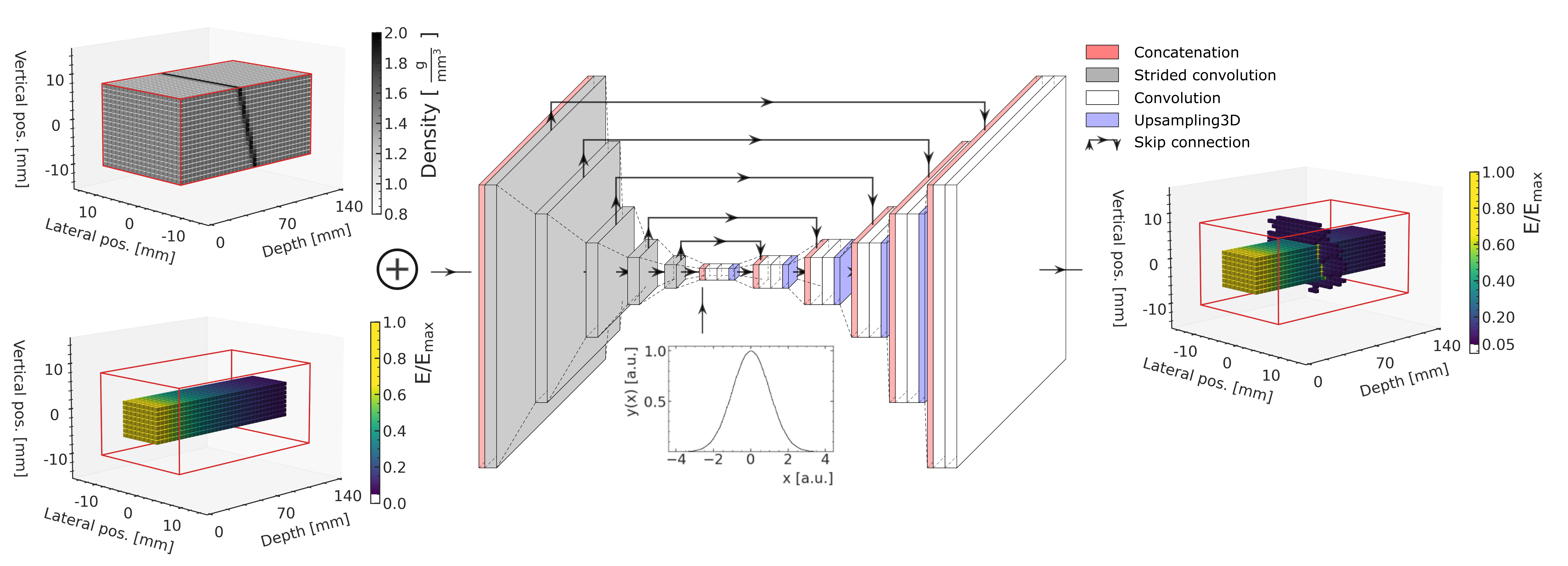}
	\caption{Schematic of the generator 3D U-Net structure with input and output data. Left: Conditional generator input consisting of a 3D representation of a material density matrix (top) and a 3D matrix of the energy deposition of the photon beam in a water phantom (bottom). Center:  Convolutional encoder and decoder connected by skip connections. The normal distribution below indicates the injection of random noise in the bottleneck layer. Right: Generator output consisting of a 3D representation of energy depositions inside the phantom.\bigskip}
	\label{fig:MaterialsAndMethods:ML:gen}
\end{figure}

A schematic of the generator model is shown in Figure \ref{fig:MaterialsAndMethods:ML:gen}.
The goal of the generator is to predict the energy depositions of the photon beam used for the radiation therapy inside a three-dimensional (3D) body, e.g. a phantom or a patient. 
In our case, the output of the generator is thus a 3D matrix of energy depositions of size 140x18x18 corresponding to the voxelized scoring volume of the simulation (right).
The prediction is dependent on the beam properties as well as the localisation and properties of the material inside the body, e.g. the localisation and density of bone, skin or water.
Hence, the generator network of the GAN is conditioned on this information~\cite{Mirza2014}. The first conditional information of the generator are the 3D energy depositions of a beam inside a pure water phantom (lower left), which describes the relevant properties of the beam (energy spectrum, beam profile, polarization, etc.), and at the same time gives the network information about how the beam interacts with a default material (water) in the same format (140x18x18) as the prediction is made. Introducing this conditional information considerably accelerates and stabilizes the training and yields more accurate predictions.
The second conditional information is the 3D material density matrix of the phantom model (upper left) of dimension 140x18x18 as used for the simulation. This information is of course not available for a real-life application but would be replaced by the 3D medical imaging information of the patient, such as the Hounsfield units of a CT scan.
As the conditional input of the generator consists of two high dimensional matrices (140x18x18x2), the dimensionality of this information is reduced by encoding the 3D images with multiple strided 3D convolutions into low-dimensional representations (3x3x3) of the data combining the geometrical properties of the phantom with the information obtained from the energy depositions of the beam in water (center). 
The compressed features are then concatenated with 100 values of random noise drawn from a Gaussian distribution in order to statistically predict the energy depositions inside the phantom by decoding the images again to dimension 140x18x18 using 3D up-samplings with subsequent convolutions. Each convolutional layers of the generator consists of 64 filters of variable size, is activated using the Swish function~\cite{Ramachandran2017}, stabilized using batch normalization~\cite{Ioffe2015} and regularized using dropout~\cite{Srivastava2014} with a rate of 15\%.
Because not only the low-dimensional features contain important information for the prediction but also the intermediately obtained features of the encoder, we introduce skip connections to pass information from the same level of the encoder to the same level of the decoder by concatenating the two layers and effectively doubling the number of the filters in the decoder. This obtained structure using a convolutional encoder-decoder with skip connection is usually referred to as U-Net structure, which was first introduced for fast 2D biomedical image segmentation~\cite{Ronneberger2015}.
A similar structure of a 3D U-Net~\cite{Cicek2016} has recently been applied for dose distribution estimation in IMRT~\cite{Kontaxis2020}. 
\subsubsection{The critic model}
\begin{figure}[tb!]
	\centering
	\includegraphics[width=\linewidth]{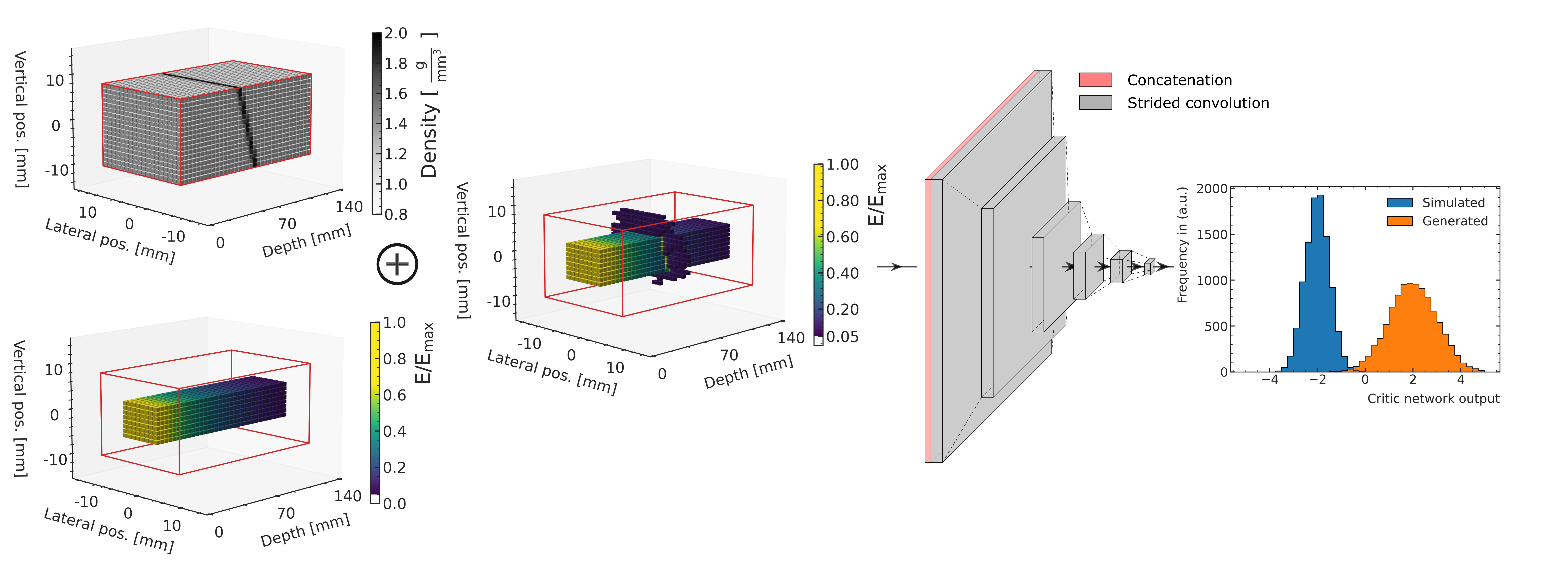}
	\caption{Schematic of the critic network with input and output data. Left: Conditional generator input consisting of a 3D representation of a material density matrix (top left) and a 3D matrix of the energy deposition of the photon beam in a water phantom (bottom left) as well as the simulated or generated 3D energy depositions inside the phantom (right). Center: The critic architecture with six subsequent convolutional layers. Right: The output is a Wasserstein loss rating the input energy depositions as simulated or generated.\bigskip}
	\label{fig:MaterialsAndMethods:ML:crit}
\end{figure}

The critic of the GAN is built as a relatively simple 3D convolutional network. A schematic of the critic model is shown in Figure~\ref{fig:MaterialsAndMethods:ML:crit}. 
In addition to the same conditional information as given to the generator, the critic receives either the simulated or generated energy depositions inside the phantom as input in order to classify these two types of data (left). The three input matrices of dimension 140x18x18 are concatenated and passed through six consecutive transposed 3D convolutions of decreasing filter size (center). Each convolutional layer of the critic consists of 128 filters, is activated using the Swish function and regularized using a dropout rate of 15\%.
The last layer of the critic consist of a single linear unit resulting in a continuous and quantitative rating distinguishing simulated from generated samples (right). 

\subsubsection{Performance measure}
In many machine learning applications, the development of the loss over the training process is used as an indication whether the optimization of the weights is converged. In the training of GANs, the loss is usually not a good measure to assess the convergence of the two networks to a stable equilibrium nor to judge about the performance of the generator.
Therefore, we define the delta index, which is inspired by the global gamma index~\cite{Low1998}, from which we can derive a performance measure for the prediction quality of the generator. The delta index is defined as
\begin{equation}
    \delta = \frac{D_\mathrm{gen} - D_\mathrm{sim}}{D_\mathrm{sim}^{max}}\,,
    \label{Eq:GammaIndex}
\end{equation}
where $D_\mathrm{gen}$ is the dose predicted by the generator network of the GAN, $D_\mathrm{sim}$ is the dose obtained by the Geant4 simulation, and $D^{max}_\mathrm{sim}$ is the maximum dose deposition in the phantom as calculated in the MC simulation. The delta index is calculated for each voxel of the in-field area of the beam inside the scoring volume. As global performance measure we use the 1\% (3\%) {\em passing rate}, which is the fraction of the voxels with a delta index smaller than 1\%  (3\%).
The model of the generator is chosen for the final performance evaluation, which maximizes the 1\% passing rate on the validation data during the training.

\section{Results \label{section:Results}}
In this section we compare the predictions of the GAN model for the three different phantoms with the MC simulation and use the passing rate in order to summarize the prediction accuracy.
The machine learning models are implemented using the Keras\cite{chollet2015keras} interface to Tensorflow 2.2\cite{tensorflow2015-whitepaper} on a Nvidia GeForce GTX 1080i graphical processing unit (GPU). Training took several days for each model.

\subsection{Water phantom with rotated bone slab of constant thickness}

\begin{figure}[!tb]
	\centering
	\begin{subfigure}[t]{0.48\textwidth}
		\includegraphics[width=\linewidth]{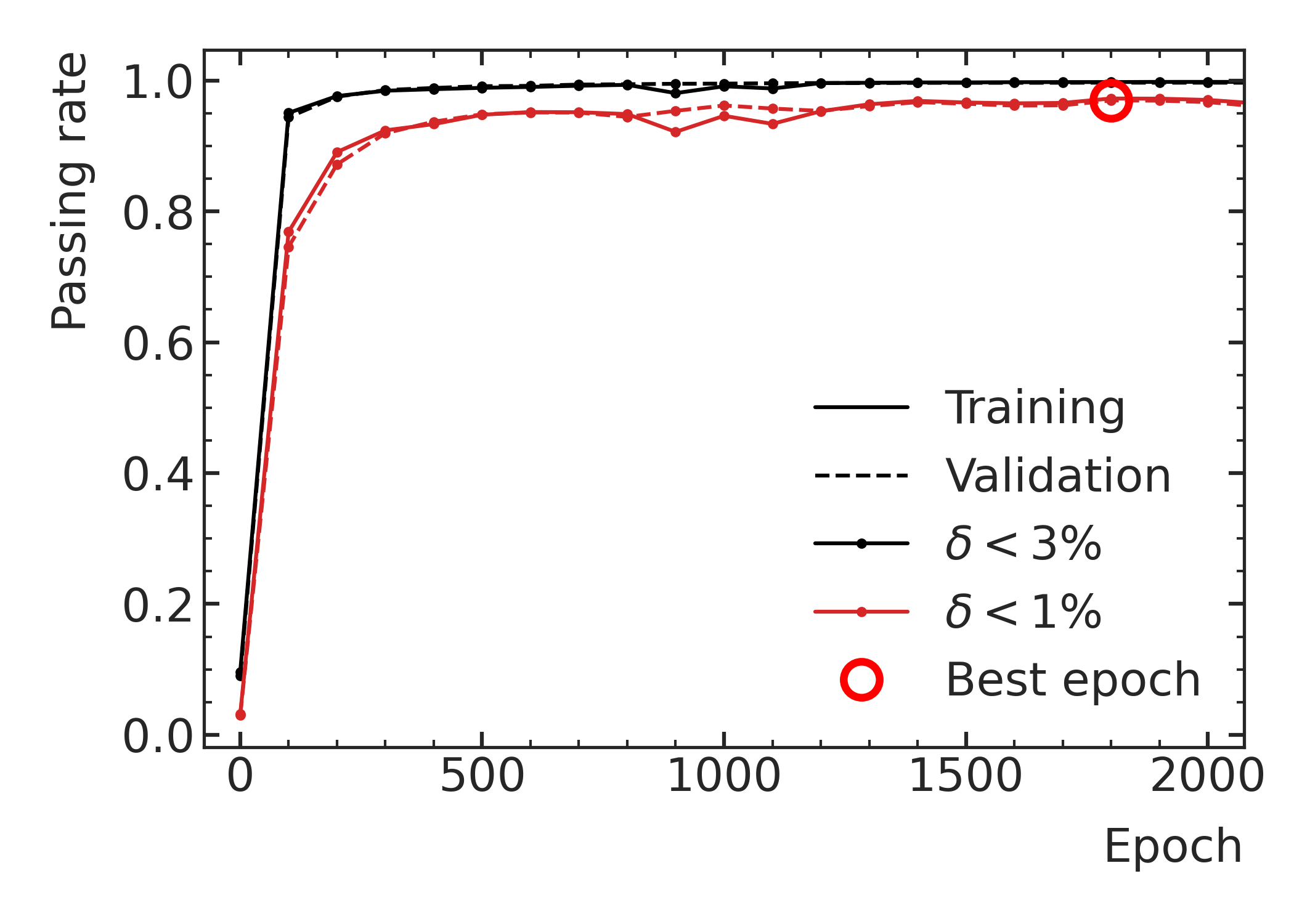}
		\caption{}
		\label{fig:all_angles_progress:a}
	\end{subfigure}
	\begin{subfigure}[t]{0.48\textwidth}
		\includegraphics[width=\linewidth]{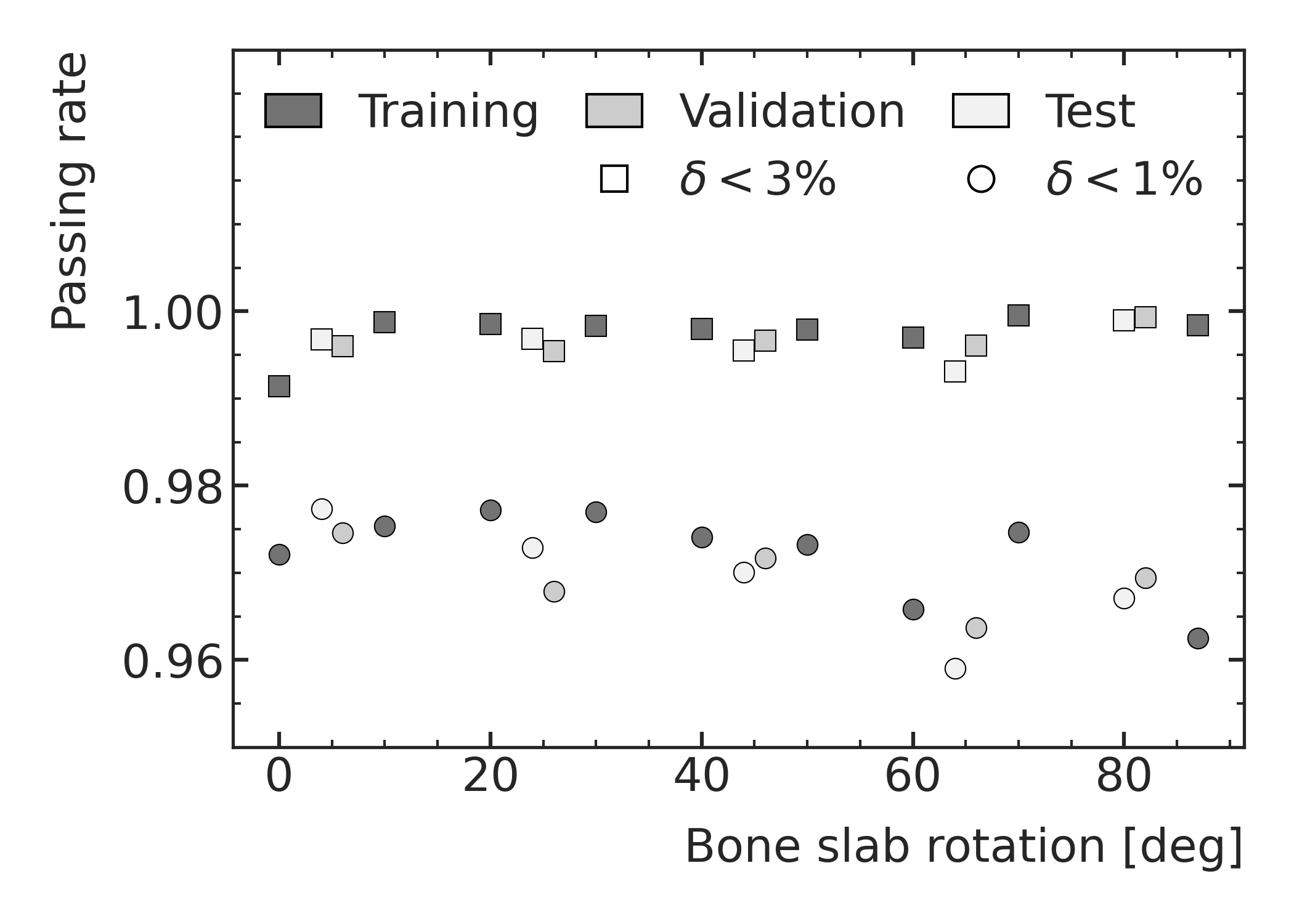}
		\caption{}
		\label{fig:all_angles_progress:b}
	\end{subfigure}
	\caption{(a) Global 1\% (red) and 3\% (black) passing rates for the training (solid) and validation (dashed) data as a function of training epochs. The best model on the validation data is highlighted with a red circle. (b) Passing rates as a function of the bone slab rotation angle for training (dark grey), validation (medium grey) and test (light grey) data.\bigskip}
	\label{fig:all_angles_progress}
\end{figure}

\begin{figure}[!tb]
	\centering
	\begin{subfigure}[t]{.32\textwidth}
		\includegraphics[width=\linewidth]{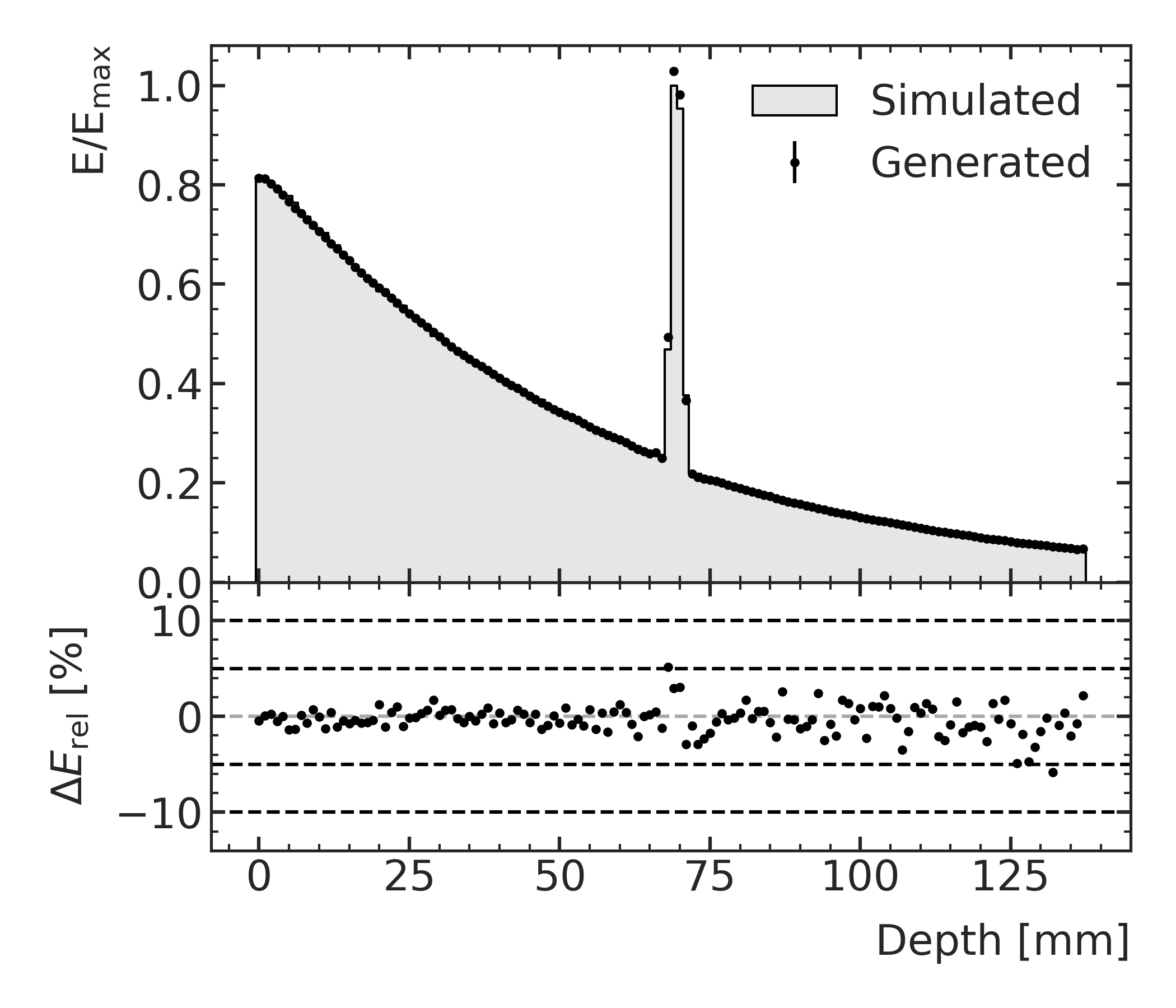}
		\caption{}
		\label{fig:all_ang_pred_results:a}
	\end{subfigure}
	\begin{subfigure}[t]{.32\textwidth}
		\includegraphics[width=\linewidth]{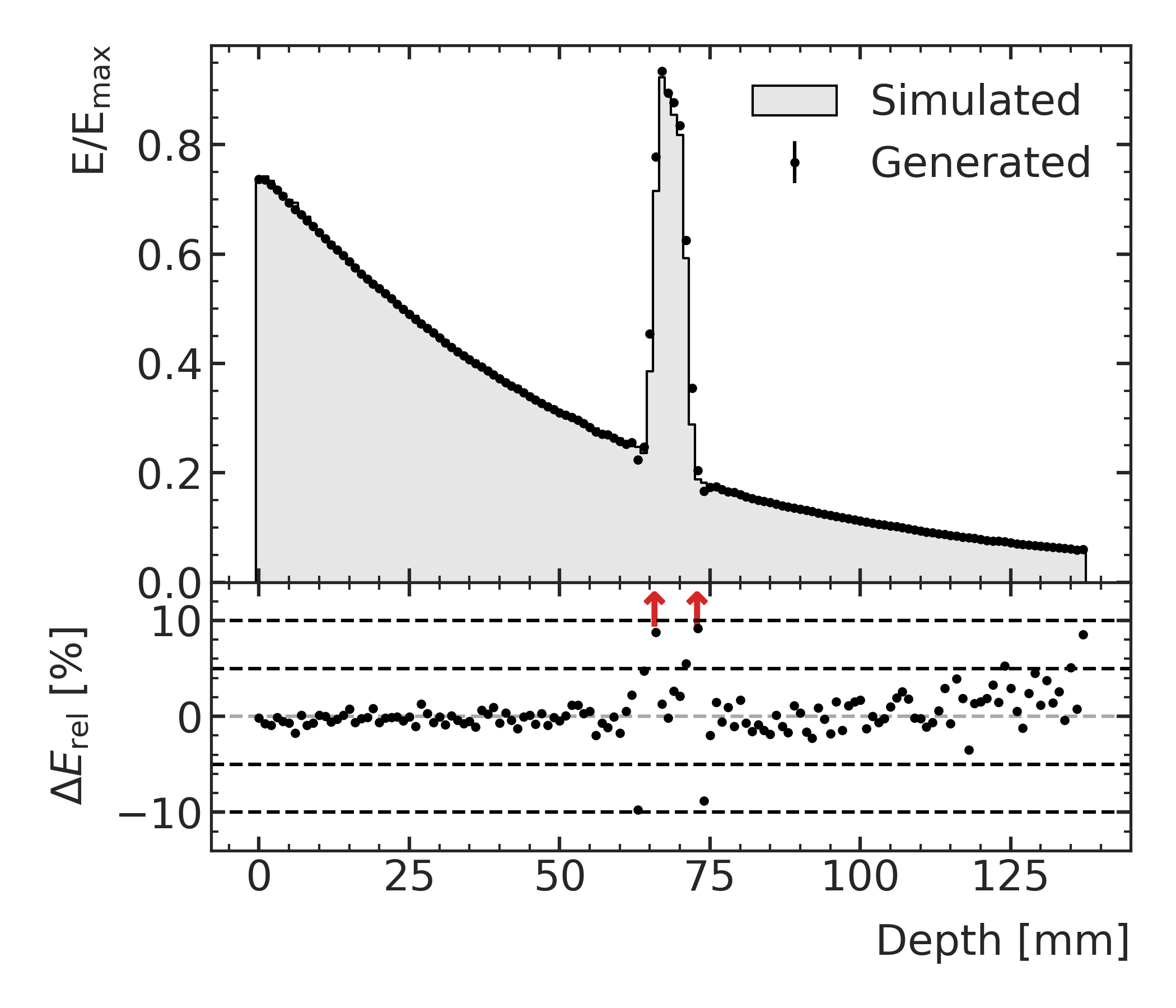}
		\caption{}
		\label{fig:all_ang_pred_results:b}
	\end{subfigure}
	\begin{subfigure}[t]{.32\textwidth}
		\includegraphics[width=\linewidth]{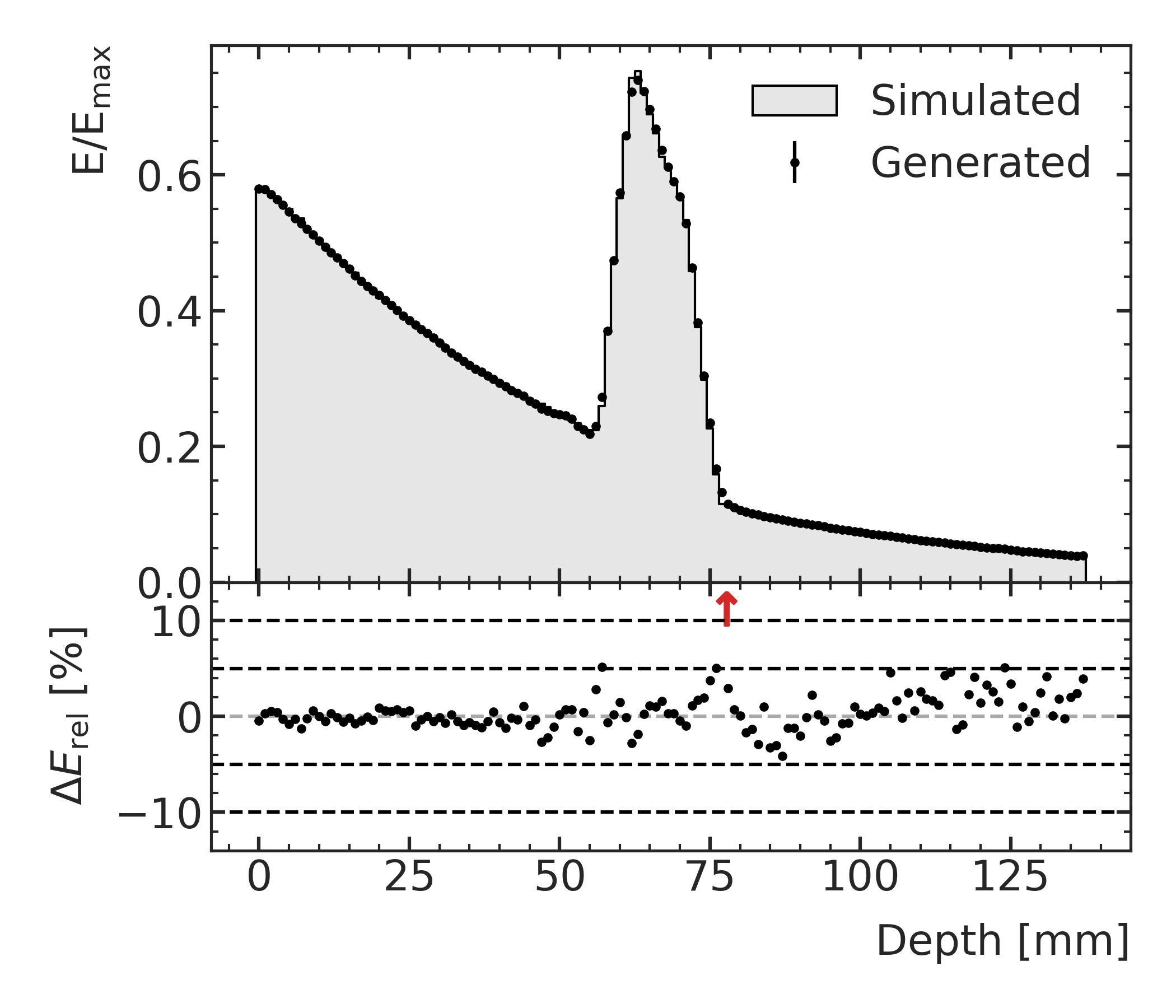}
		\caption{}
		\label{fig:all_ang_pred_results:c}
	\end{subfigure}
	\begin{subfigure}[t]{.32\textwidth}
		\includegraphics[width=\linewidth]{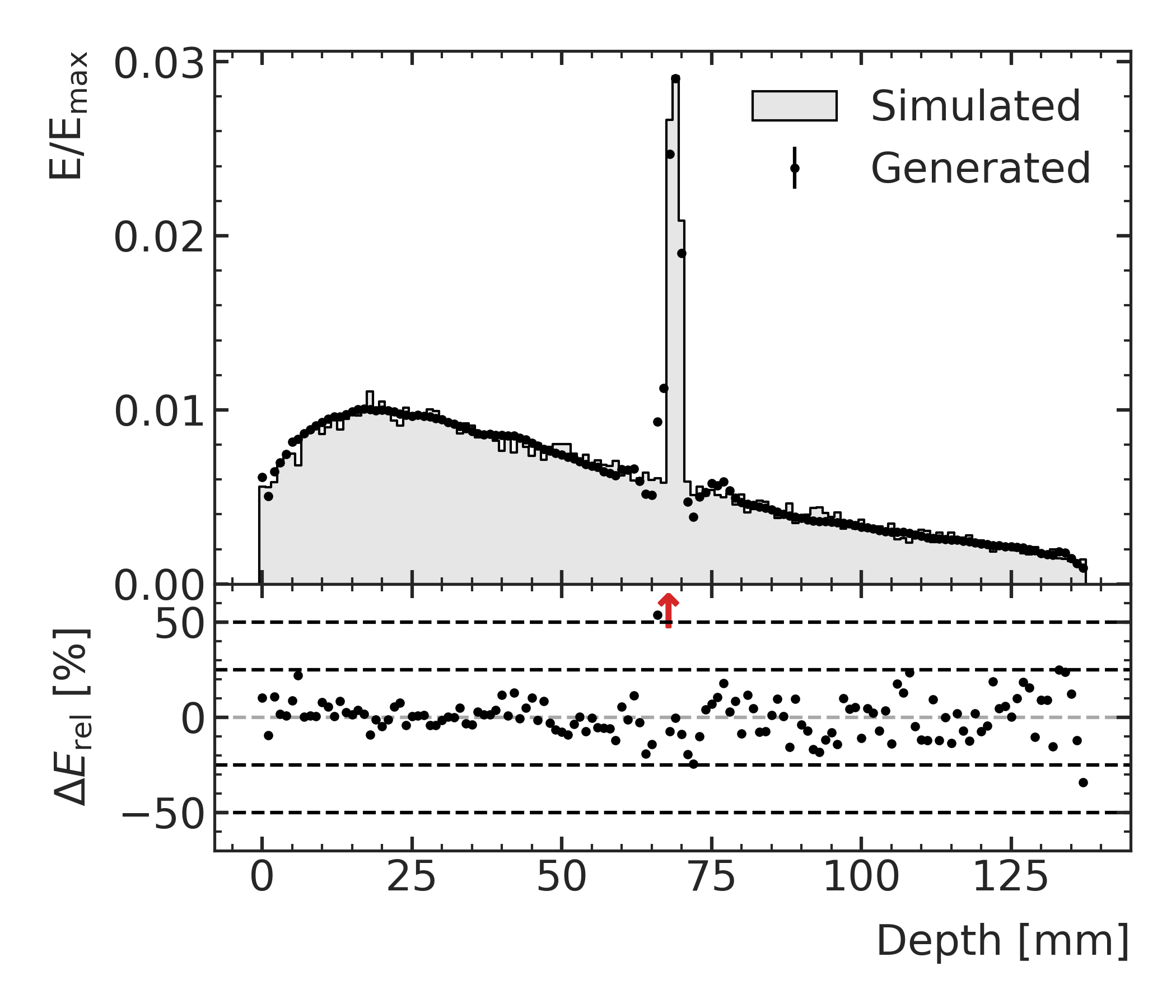}
		\caption{}
		\label{fig:all_ang_pred_results:d}
	\end{subfigure}
	\begin{subfigure}[t]{.32\textwidth}
		\includegraphics[width=\linewidth]{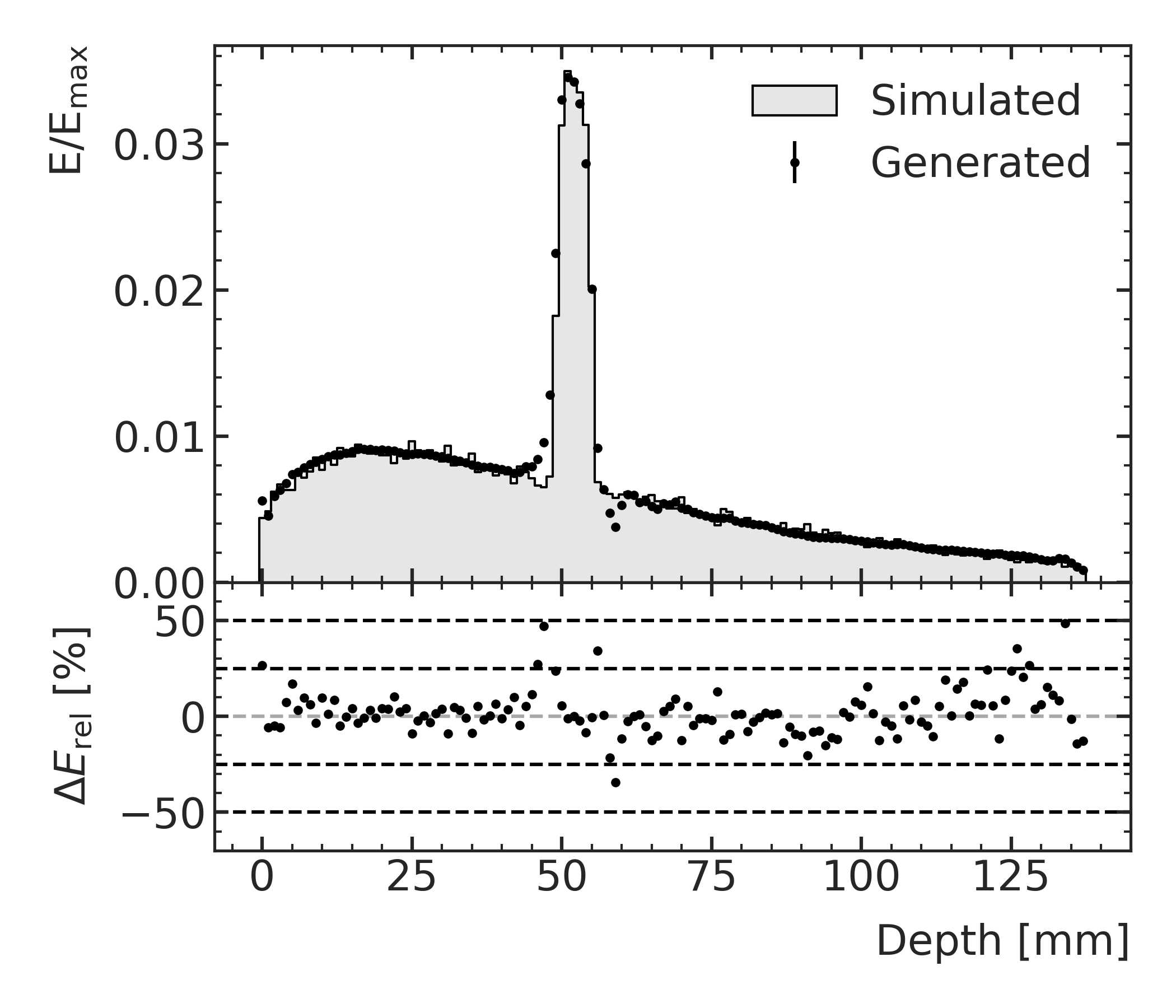}
		\caption{}
		\label{fig:all_ang_pred_results:e}
	\end{subfigure}
	\begin{subfigure}[t]{.32\textwidth}
		\includegraphics[width=\linewidth]{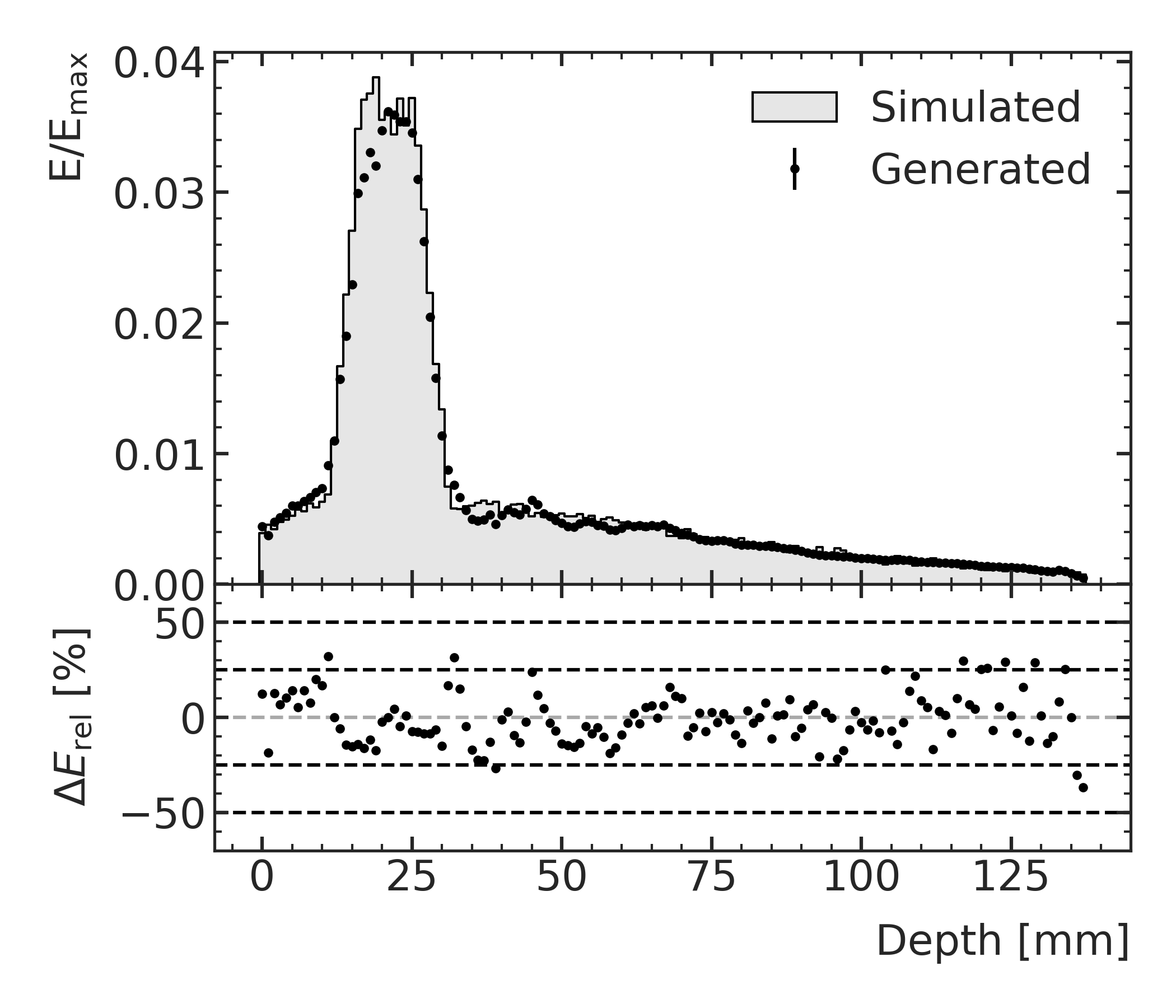}
		\caption{}
		\label{fig:all_ang_pred_results:f}
	\end{subfigure}
	\caption{Comparisons of normalized simulated (grey) and generated (black) energy depositions of the test data as a depth profile in-field (a-c) and out-of-field (d-f) of the beam with bone slab rotation angles of $\alpha = [4, 64, 80]^\circ$ (a+d, b+e, c+f). The lower part of the plots shows the relative energy deposition deviation $\Delta E_\mathrm{rel} = E_\mathrm{gen}/E_\mathrm{sim}-1$ in percent. Outliers are indicated with a red arrow.\bigskip} 
	\label{fig:all_ang_pred_results}
\end{figure}

\begin{figure}[!tb]
	\centering
	\begin{subfigure}[t]{.48\textwidth}
		\includegraphics[width=\linewidth]{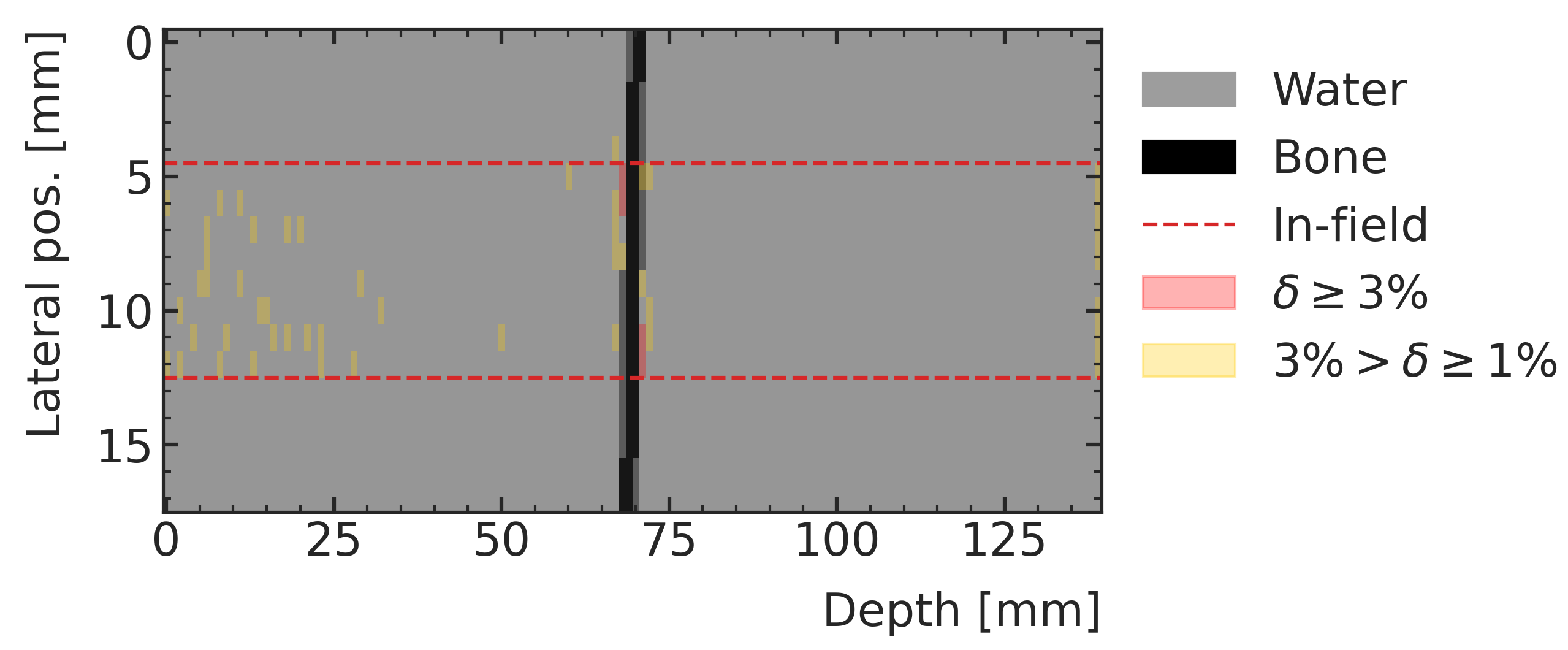}
		\caption{}
		\label{fig:2d_0_degree:a}
	\end{subfigure}
	\begin{subfigure}[t]{.48\textwidth}
		\includegraphics[width=\linewidth]{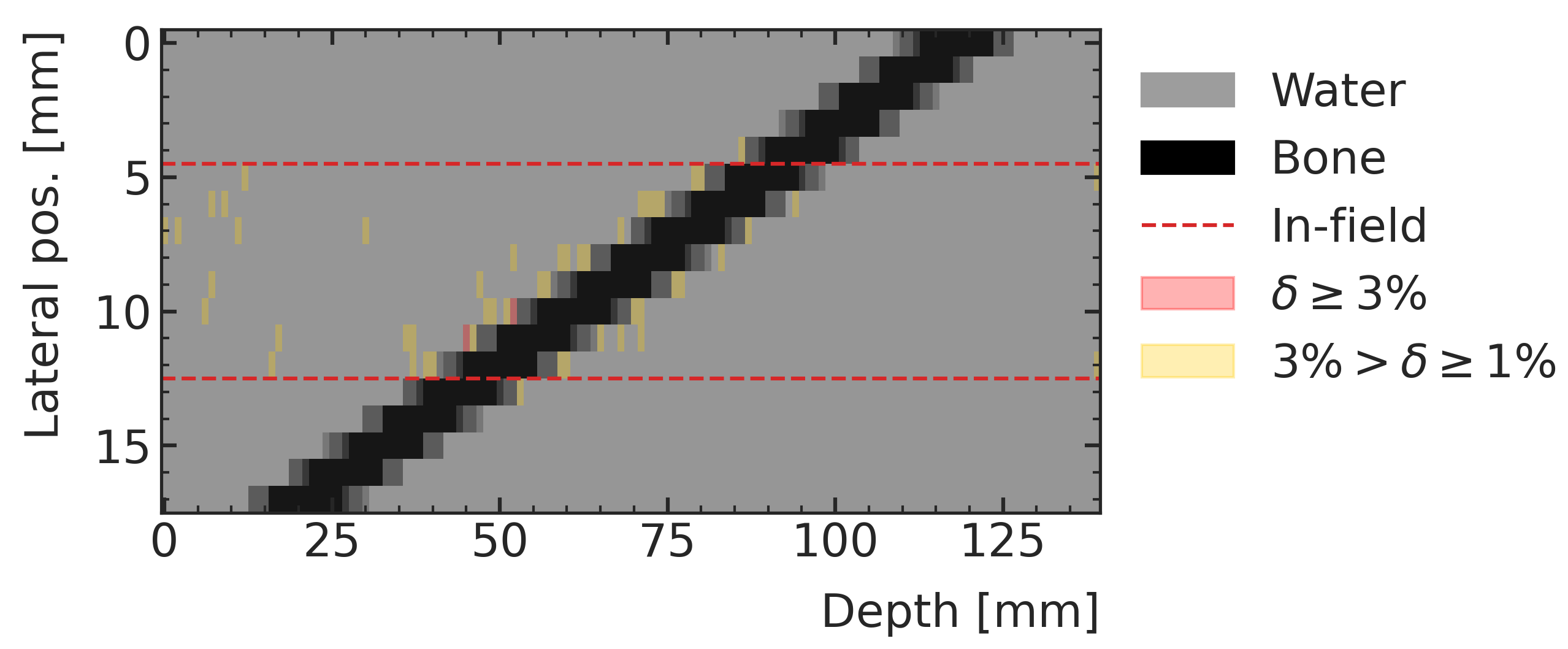}
		\caption{}
		\label{fig:2d_0_degree:b}
	\end{subfigure}
	\caption{2D slices of the phantom with a rotation angles of (a) $\alpha=4^\circ$ and (b) $\alpha=79^\circ$ as part of the test data. 
		Voxels with deviations in their dose predictions of larger than 3\% (1\%) are shown in red (yellow). Whenever the deviation is smaller than 1\%, the gray-scale color of the material density matrix is shown. The in-field part of the beam is located between the red dashed lines.
		\bigskip}
	\label{fig:2d_0_degree}
\end{figure}

The development of the 1\% (3\%) passing rate averaged over the bone slab rotation angles during the training process using the data of the first phantom is shown in Figure~ \ref{fig:all_angles_progress:a} for the training and validation data, where one epoch means, that the full training data has been used for updating the weights of the GAN model a single time. 
Relatively high passing rates are already obtained after a few hundred epochs, the highest 1\% passing rate of $(96.9\pm0.4)\%$ is achieved on the validation data after 2,800 training epochs. The model from this epoch is chosen as the best model for evaluating the performance of the GAN. 
Table~\ref{tab:4rotation_results} summarizes the passing rates of the training, validation and test data averaged over bone slab rotation angles of this model.
\begin{table}[!htb]
	\begin{center}
		\captionv{10}{}{Delta index passing rates with standard deviation for training, validation and test data for the water phantom with the rotated bone slab. 
			\label{tab:4rotation_results}
			\vspace*{1ex}
		}
		\begin{widetable}{\columnwidth}{r
				S[table-format=2.1(2)]
				S[table-format=2.1(2)]
				S[table-format=2.1(2)]
			}
			\toprule
			{}& \multicolumn{3}{c}{Passing rate [\%]}\\
			{} &    {Training} &  {Validation} &          {Test}\\
			\midrule
			$\delta < 1\%$    &   97.2 \pm 0.5 &  96.9 \pm 0.4 &    96.9 \pm 0.7 \\
			$\delta < 3\%$    &   99.8 \pm 0.2 &  99.7 \pm 0.2 &    99.6 \pm 0.2 \\
			\bottomrule
		\end{widetable}
	\end{center}
\end{table}
The passing rates are all at a very high level and agree within their standard deviations, which indicates that the model is not over-trained and is capable of generalizing well to the unseen bone slab rotation angles of the validation and test data.
The passing rates of the selected model dependent on the bone slab rotation angle are shown in Figure \ref{fig:all_angles_progress:a} for the training, validation and test data. Over the full rotation angle range more than 99\% of the voxel deviate by less than 3\% of the maximal simulated dose. Although slightly smaller passing rates are obtained for the validation and test data compared to the training data the performance is relatively stable over the full rotation angle range and the GAN shows good ability to interpolate to geometries of unknown rotation angles.
Figure~\ref{fig:all_ang_pred_results} shows comparisons of normalized simulated and generated energy depositions of the test data as a depth profile in-field (a-c) and out-of-field (d-f) of the beam with bone slab rotation angles of $\alpha = [4, 64, 80]^\circ$ (a+d, b+e, c+f).
In the in-field region of the beam, the relative energy deposition deviations $\Delta E_\mathrm{rel} = E_\mathrm{gen}/E_\mathrm{sim}-1$ are mostly well within 5\% for all rotation angles except for the boundary regions of the bone, where the deviations in individual voxels are larger, and the end-region of the phantom, where the absolute energy depositions are relatively small. Considering, that the energy depositions change most drastically in the transition regions of water and bone in particular for larger rotation angles
and the model could not learn from data around the chosen angles, the agreement over the full depths of the phantom is remarkable and demonstrates a good interpolation capability of the model.\\
Out-of-field, the relative deviations of the generated energies from the simulated energies are larger but mostly within 25\% except for the transition regions. However, the deposited energy is more than one order of magnitude smaller in the out-of-field regions, which means deviations of this magnitude will result in relatively small penalties during the training. Considering this, the energy depositions in the out-of-field region are qualitatively and quantitatively well predicted.\\
In order to get an idea of the accuracy of the prediction in two dimensions, Figure~\ref{fig:2d_0_degree} shows the relative dose deviations of voxels as 2D slices of the phantom with a rotation angles of $\alpha=4^\circ$ (a) and $\alpha=79^\circ$ (b), which are part of the test data.
Mostly voxel in the transition region of water and bone deviate by more than 1\% and only few voxel exactly at the edge of the bone deviate by more than 3\% in their predictions from the simulation.

\subsection{Water phantom with rotated bone slab of variable thickness}
\begin{figure}[!tb]
	\centering
	\begin{subfigure}[t]{.32\textwidth}
		\includegraphics[width=\linewidth]{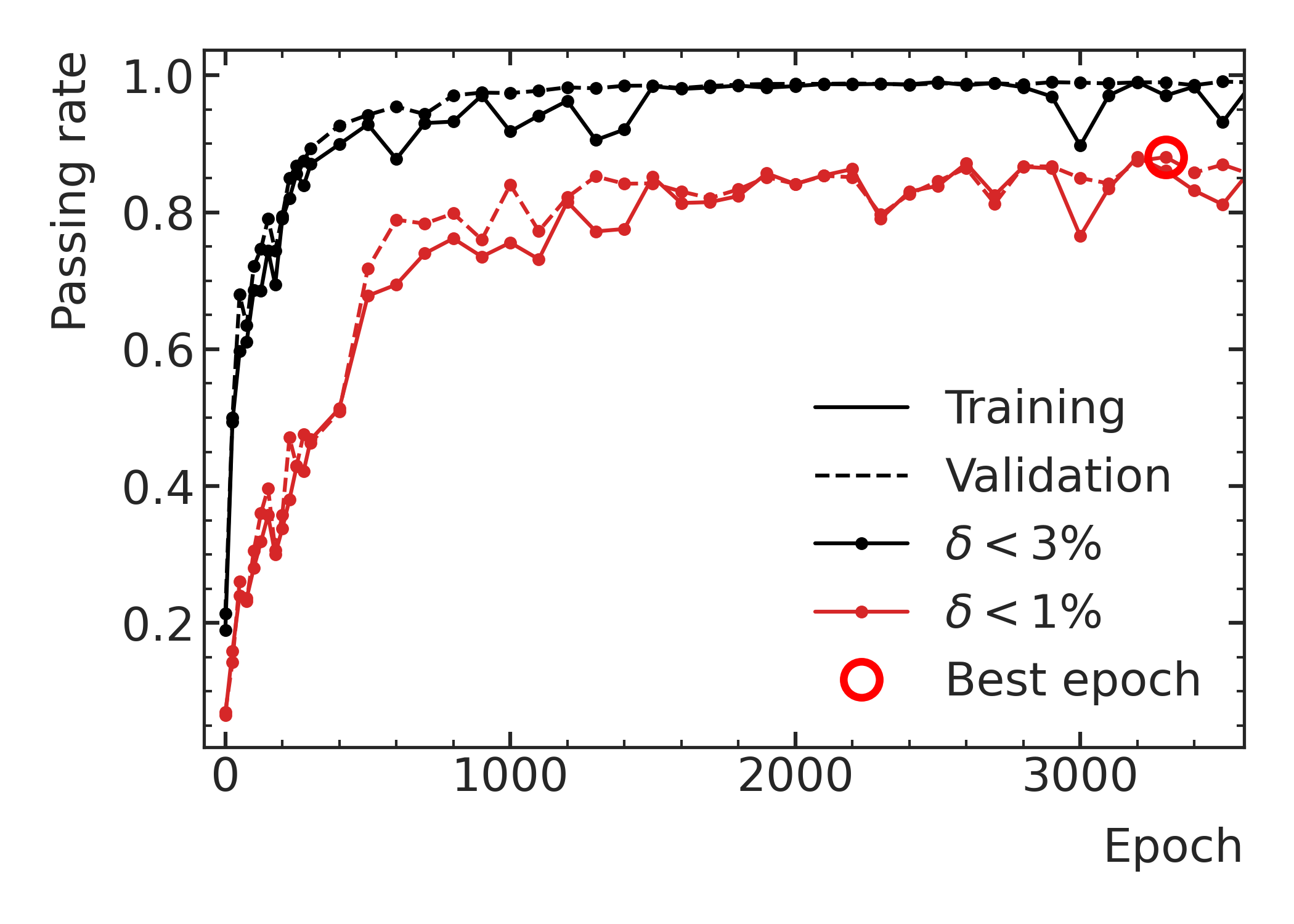}
		\caption{}
		\label{fig:4thickness_val_test_progress:a}
	\end{subfigure}
	\begin{subfigure}[t]{.32\textwidth}
		\includegraphics[width=\linewidth]{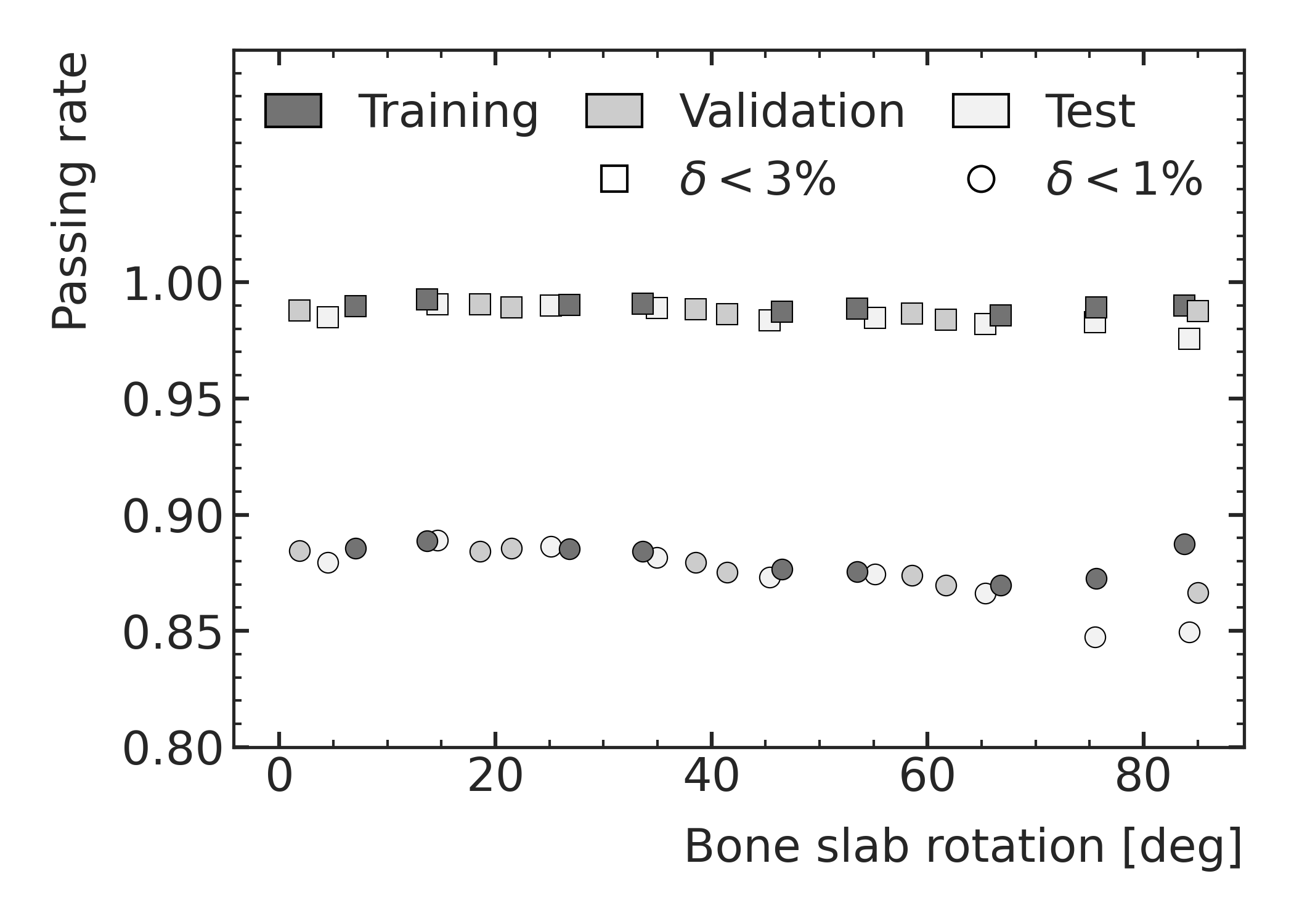}
		\caption{}
		\label{fig:4thickness_val_test_progress:b}
	\end{subfigure}
	\begin{subfigure}[t]{.32\textwidth}
		\includegraphics[width=\linewidth]{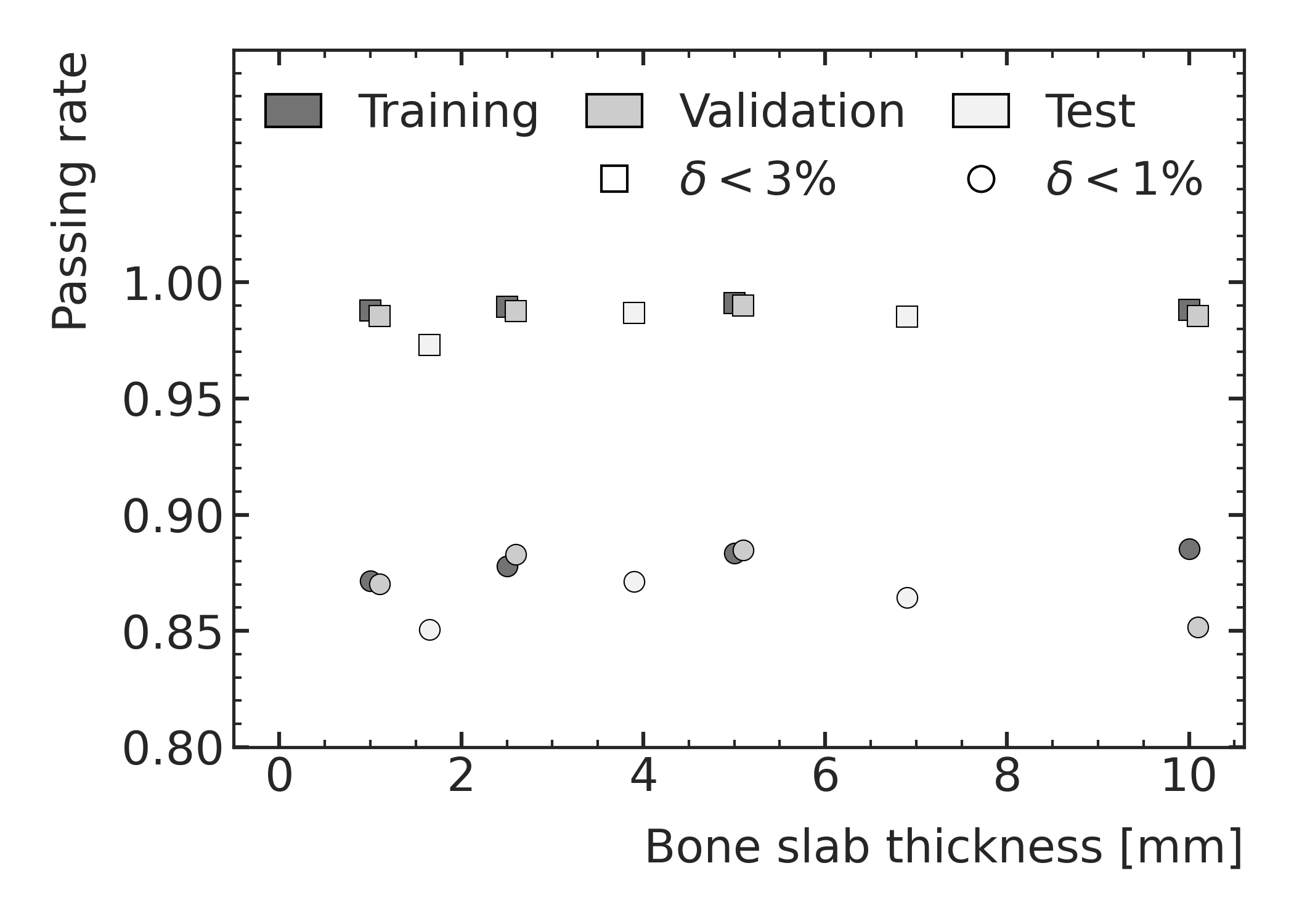}
		\caption{}
		\label{fig:4thickness_val_test_progress:c}
	\end{subfigure}
	\caption{(a) Global 1\% (red) and 3\% (black) passing rates for the training (solid) and validation (dashed) data as a function of training epochs. The best model on the validation data is highlighted with a red circle. (b) Passing rates as a function of the bone slab rotation angle	for training (dark grey), validation (medium grey) and test (light grey) data.
		(C) Passing rates as a function of the bone slab thickness for training (dark grey), validation (medium grey) and test (light grey) data.
		\bigskip}
\label{fig:4thickness_val_test_progress}
\end{figure}

\begin{figure}[!tb]
\centering
\begin{subfigure}[t]{.32\textwidth}
\includegraphics[width=\linewidth]{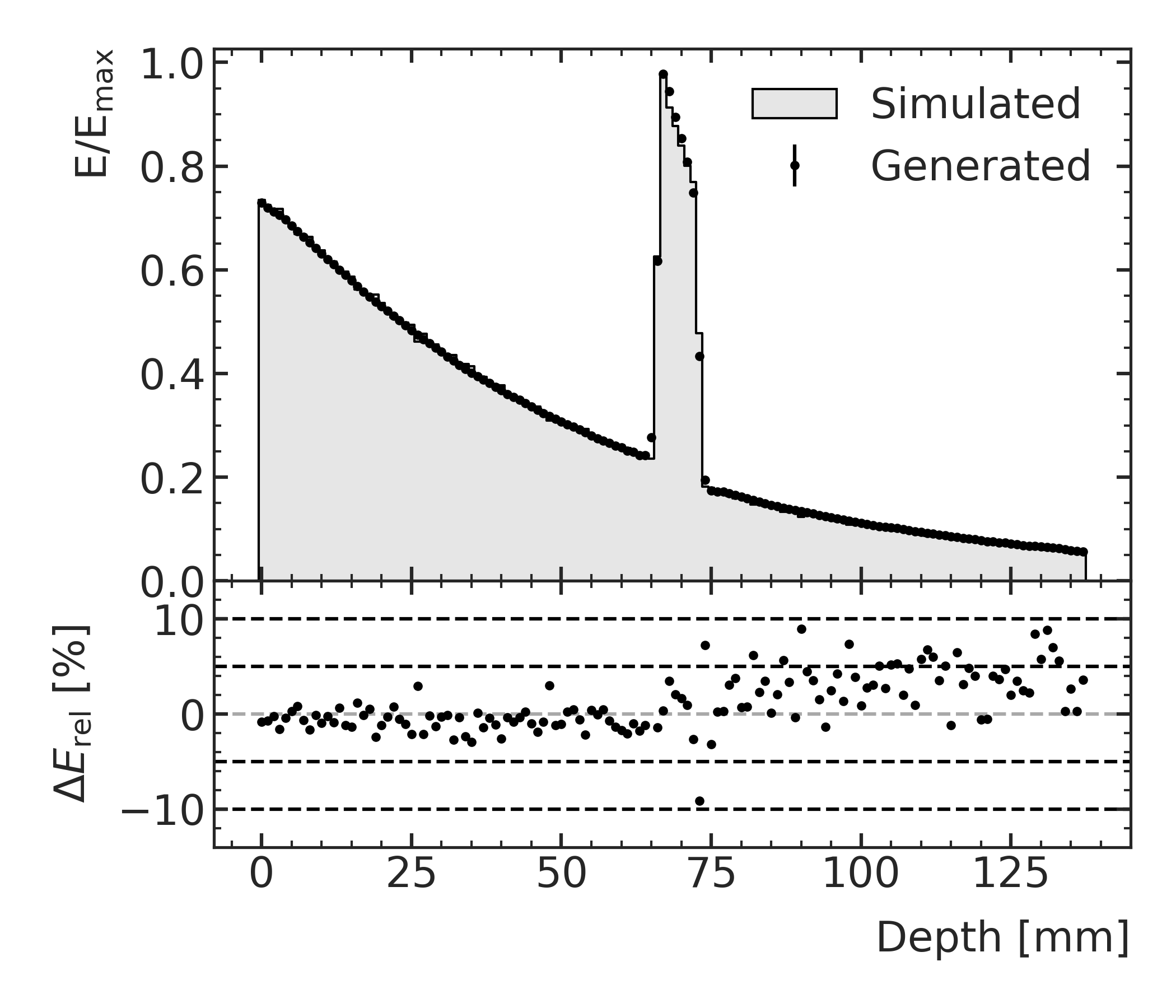}
\caption{}
\label{fig:4thicknesses_in-and-out-of-field:a}
\end{subfigure}
\begin{subfigure}[t]{.32\textwidth}
\includegraphics[width=\linewidth]{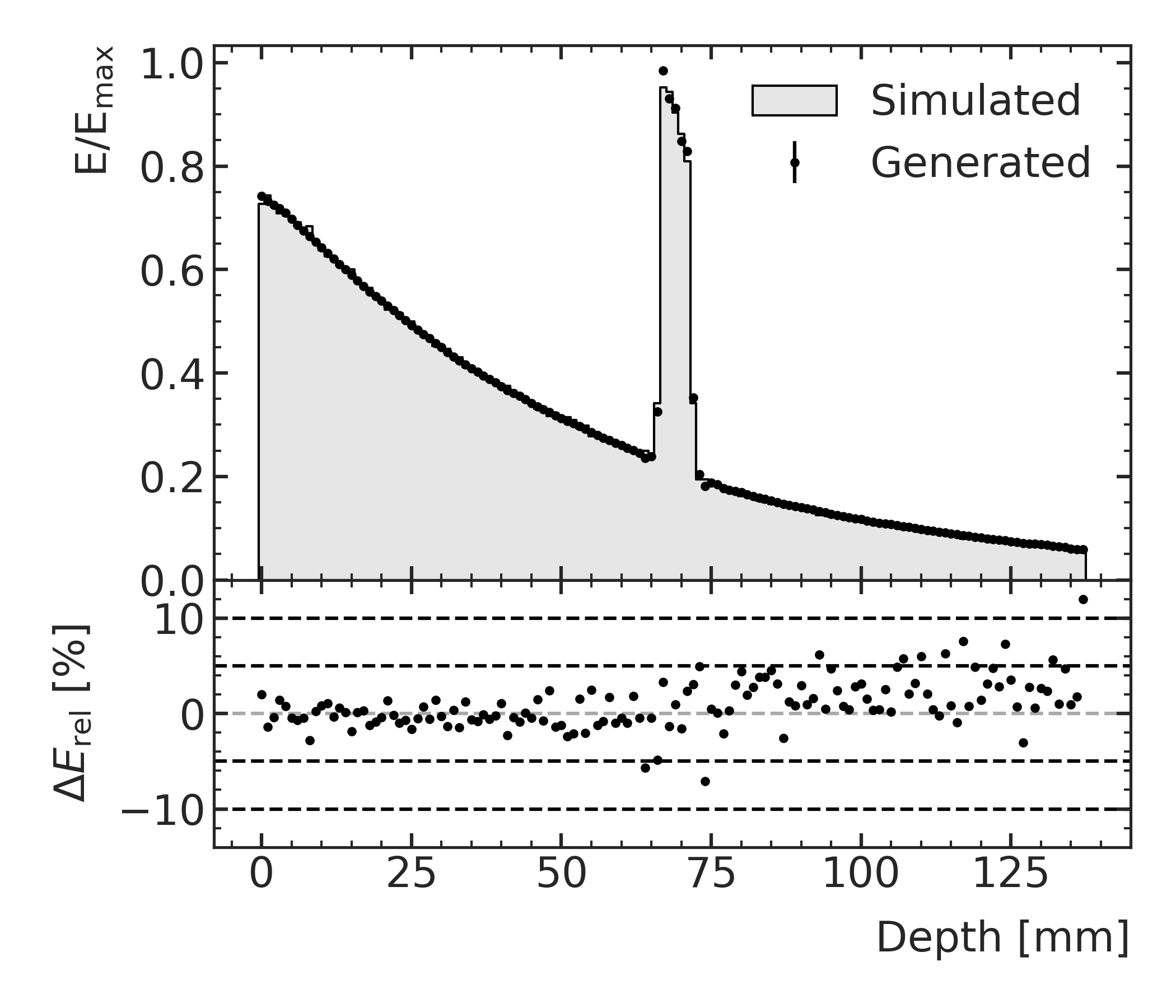}
\caption{}
\label{fig:4thicknesses_in-and-out-of-field:b}
\end{subfigure}
\begin{subfigure}[t]{.32\textwidth}
\includegraphics[width=\linewidth]{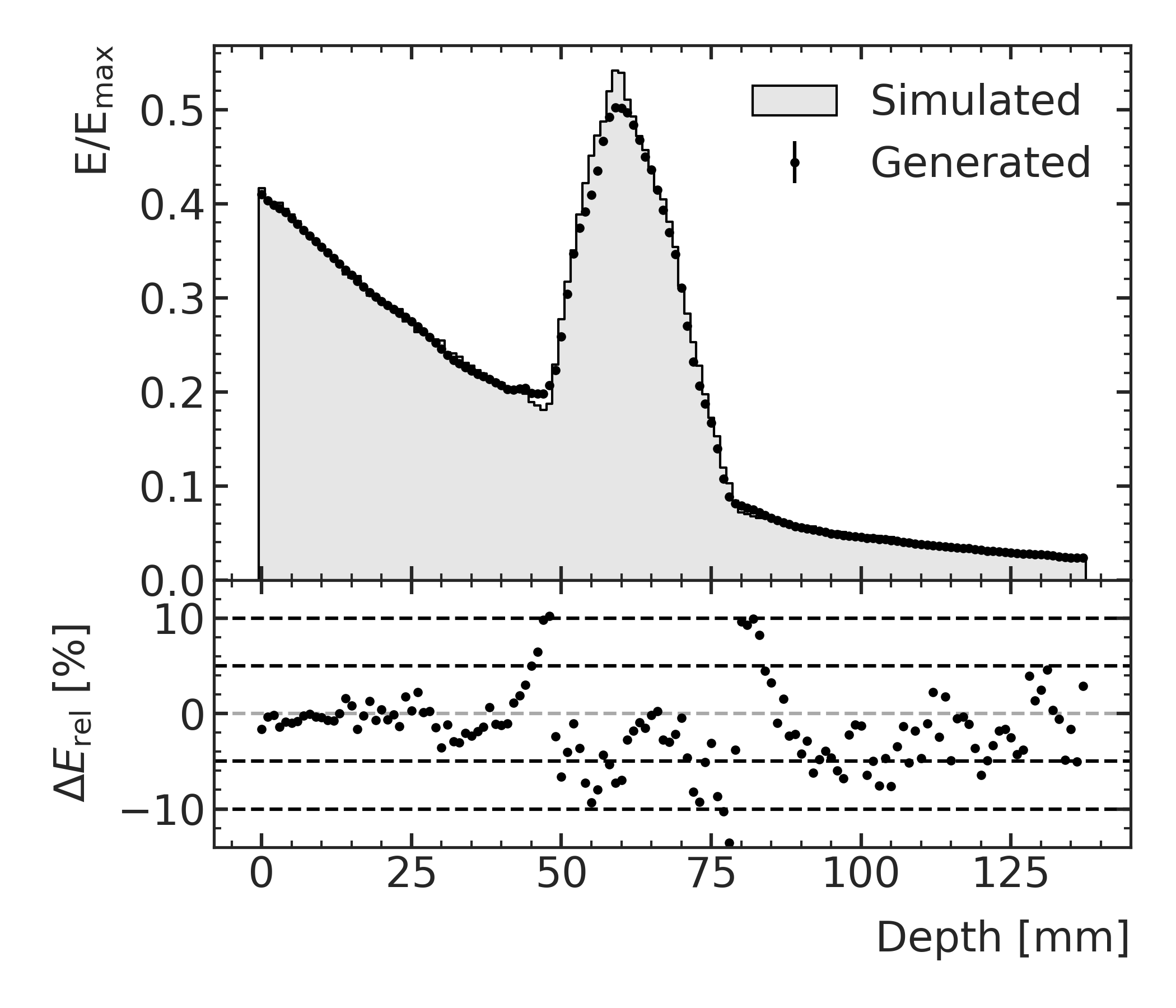}
\caption{}
\label{fig:4thicknesses_in-and-out-of-field:c}
\end{subfigure}
\begin{subfigure}[t]{.32\textwidth}
\includegraphics[width=\linewidth]{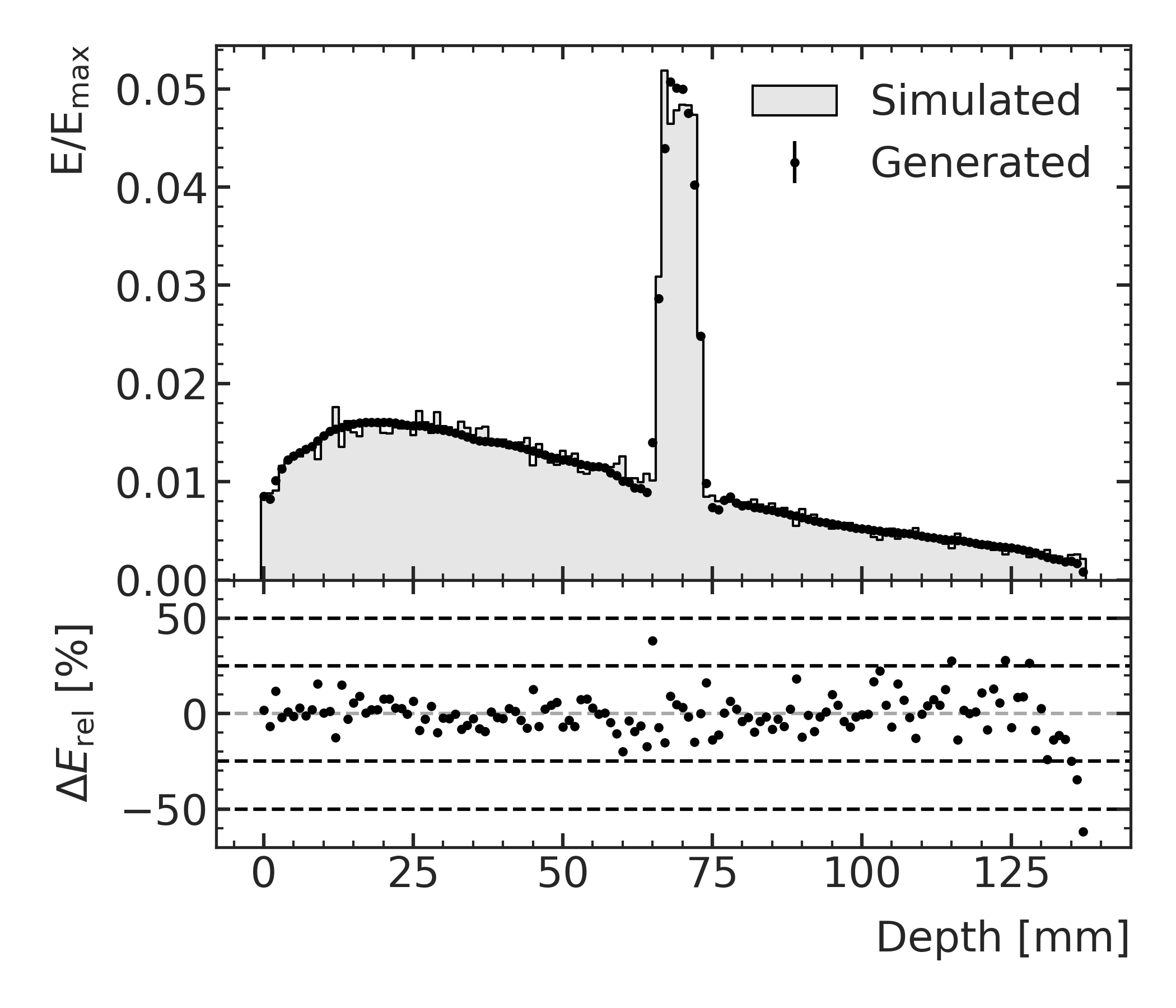}
\caption{}
\label{fig:4thicknesses_in-and-out-of-field:d}
\end{subfigure}
\begin{subfigure}[t]{.32\textwidth}
\includegraphics[width=\linewidth]{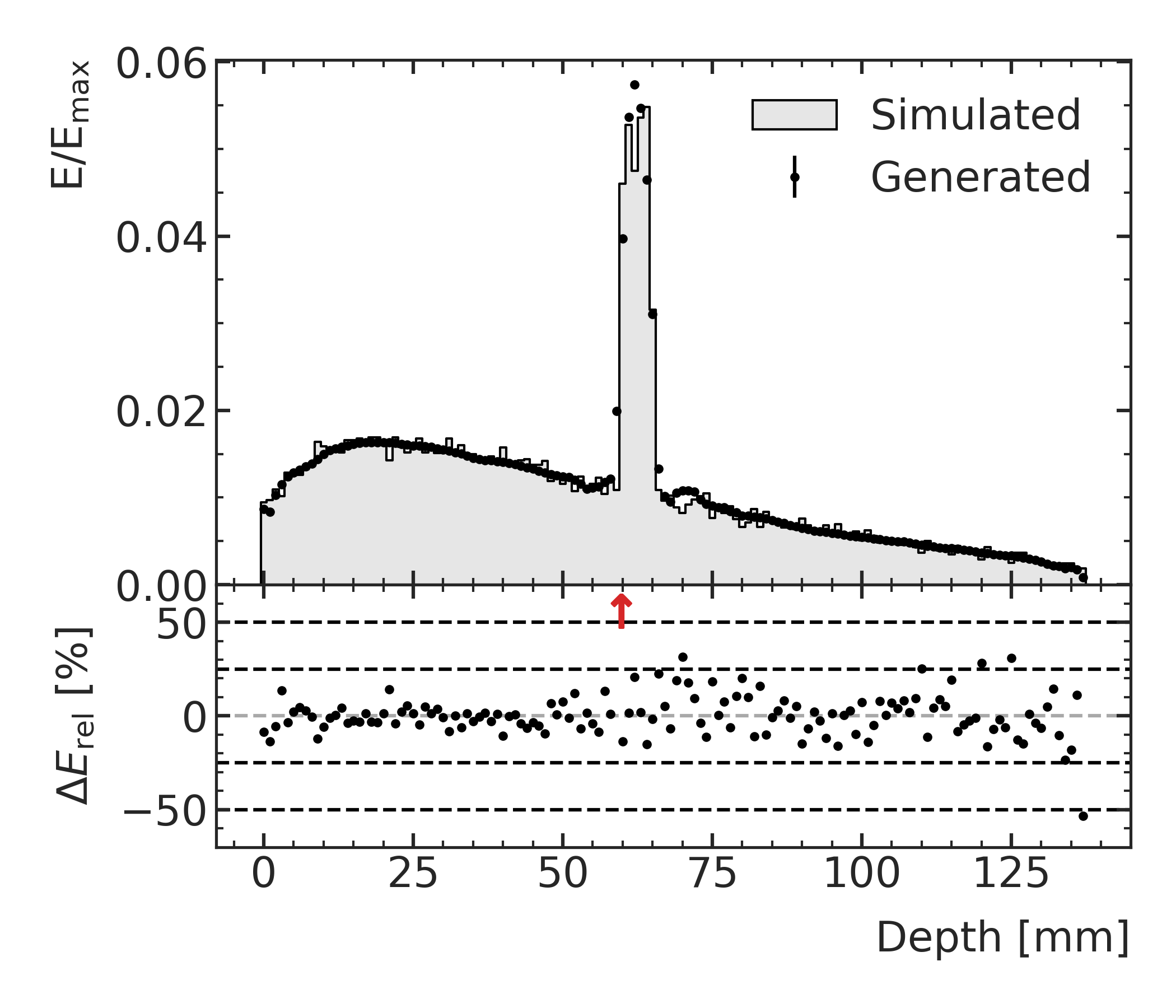}
\caption{}
\label{fig:4thicknesses_in-and-out-of-field:e}
\end{subfigure}
\begin{subfigure}[t]{.32\textwidth}
\includegraphics[width=\linewidth]{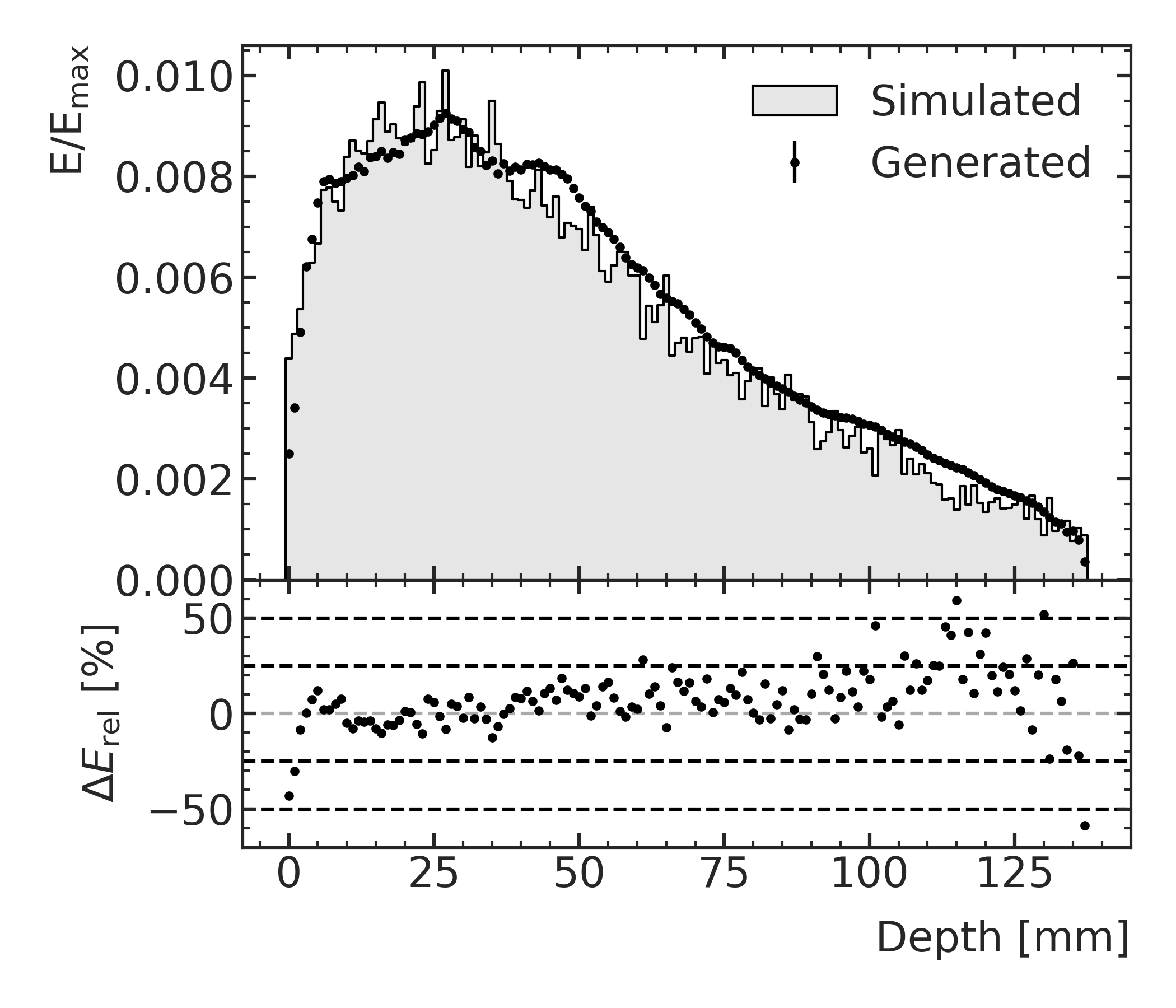}
\caption{}
\label{fig:4thicknesses_in-and-out-of-field:f}
\end{subfigure}
\caption{Comparisons of normalized simulated and generated energy depositions inside the phantom along the beam using the test data for the in-field (a-c) and out-of-field (d-f) region of the beam with bone slab rotation angles of $\alpha = [0, 40, 85]^\circ$ (a+d, b+e, c+f) and thicknesses of  $d = [7, 4, 1.75]$\,mm (a+d, b+e, c+f). The lower part of the plots shows the relative energy deposition deviation $\Delta E_\mathrm{rel}$ in percent. Outliers are indicated with a red arrow.\bigskip} 
\label{fig:4thicknesses_in-and-out-of-field}
\end{figure}

The development of the passing rates averaged over bone slab rotation angles and thicknesses during the training process using the data of the second phantom are shown in Figure~\ref{fig:4thickness_val_test_progress:a} for the training and validation data.
The training takes significantly longer until a plateau is reached and is less stable. The highest 1\% passing rate on the validation data is achieved at epoch 3,300, which is chosen as the best model. Table~\ref{tab:4thicknesses_results} summarizes the passing rates of the training, validation and test data averaged over bone slab rotation angles and thicknesses of this model.
\begin{table}[!htb]
\begin{center}
	\captionv{10}{}{Delta index passing rates with standard deviation for training, validation and test data for the rotated bone slab phantom with different thicknesses. 
		\label{tab:4thicknesses_results}
		\vspace*{2ex}
	}
	
	\begin{widetable}{\columnwidth}{r
			S[table-format=2.1(2)]
			S[table-format=2.1(2)]
			S[table-format=2.1(2)]
		}
		\toprule
		{}& \multicolumn{3}{c}{Passing rate [\%]}\\
		{} &    {Training} &  {Validation} &          {Test}\\
		\midrule
		$\delta < 1\%$    &   88.1 \pm 0.7 &  87.7 \pm 0.2 &    87.2 \pm 1.4 \\
		$\delta < 3\%$    &   98.9 \pm 0.2 &  98.8 \pm 0.2 &    98.5 \pm 0.4 \\
		\bottomrule
	\end{widetable}
\end{center}
\end{table}
The passing rates agree within the statistical uncertainties, hence there is no indication of over-fitting and the model seems to generalize well to the unseen bone slab rotation angles and thicknesses of the validation and test data.
The passing rates in dependence of the bone slab rotation angle but averaged over the bone thickness is shown in \ref{fig:4thickness_val_test_progress:b} for the training, validation and test data.
While the 3\% passing rate is still constantly above 98\% for all rotation angles, the 1\% passing rate is considerably lower at values between 85\% and 90\% with a small decrease at higher rotation angles. The model generalizes well to the rotation angles and thicknesses of the validation and test data.
In Figure \ref{fig:4thickness_val_test_progress:c} the passing rates are shown in dependence of the bone slab thickness but averaged over the rotation angle. The performance is stable for all bone thicknesses of the training and validation data set. However, the obtained passing rates are slightly lower for the interpolated bone thicknesses of the test data, in particular for thin bones with large rotation angles. These geometries result in dose distributions which are particularly challenging to predict.\\
Figure \ref{fig:4thicknesses_in-and-out-of-field} shows exemplary normalized simulated (grey) and generated (black) energy depositions inside the phantom along the beam using the test data for the in-field (a-c) and out-of-field (d-f) region of the beam with bone slab rotation angles of $\alpha = [0, 40, 85]^\circ$ (a+d, b+e, c+f) and thicknesses of  $d = [7, 4, 1.75]$\,mm (a+d, b+e, c+f).
While the energies are generally well predicted in front of the inserted bone, the relative deviations $\Delta E_\mathrm{rel}$ are larger at the bone slab transition region and behind the bone slab with a small trend to overpredict the deposited energies behind the bone slab in the in-field region. In particular the thinnest bone slab of $1.75\,$mm thickness with the largest rotation angle of $85^\circ$ results in larger deviations at the bone slab of up to 10\% but represents also the most challenging case of the interpolation.
In general, the model shows a good ability to interpolate to unknown geometries.

\subsection{Simple paediatric head phantom with translation}
\begin{figure}[!tb]
	\centering
	\begin{subfigure}[t]{0.48\textwidth}
		\includegraphics[width=\linewidth]{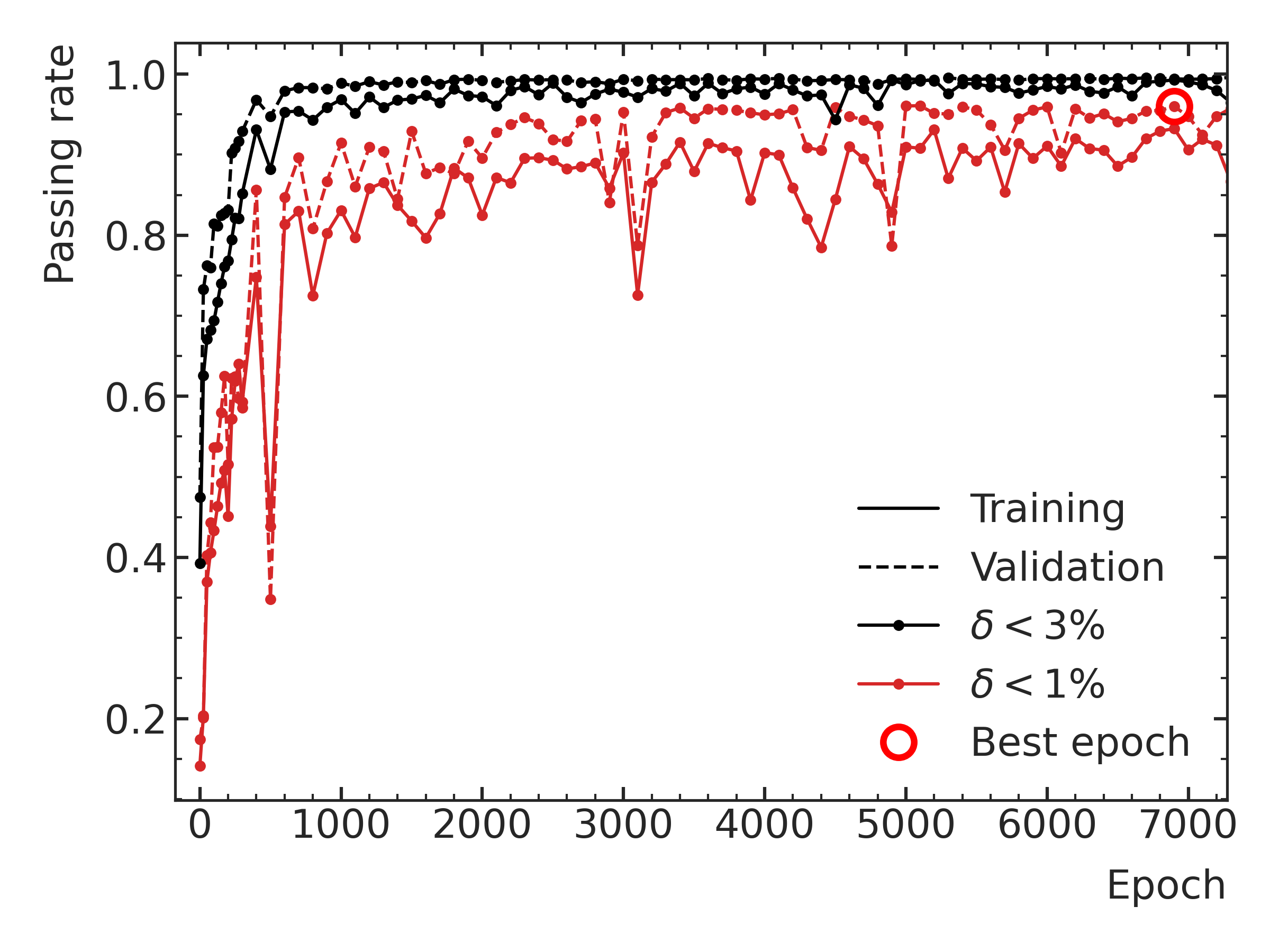}
		\caption{}
		\label{fig:simple_head:performance_train_val_test:a}
	\end{subfigure}
	\begin{subfigure}[t]{0.48\textwidth}
		\includegraphics[width=\linewidth]{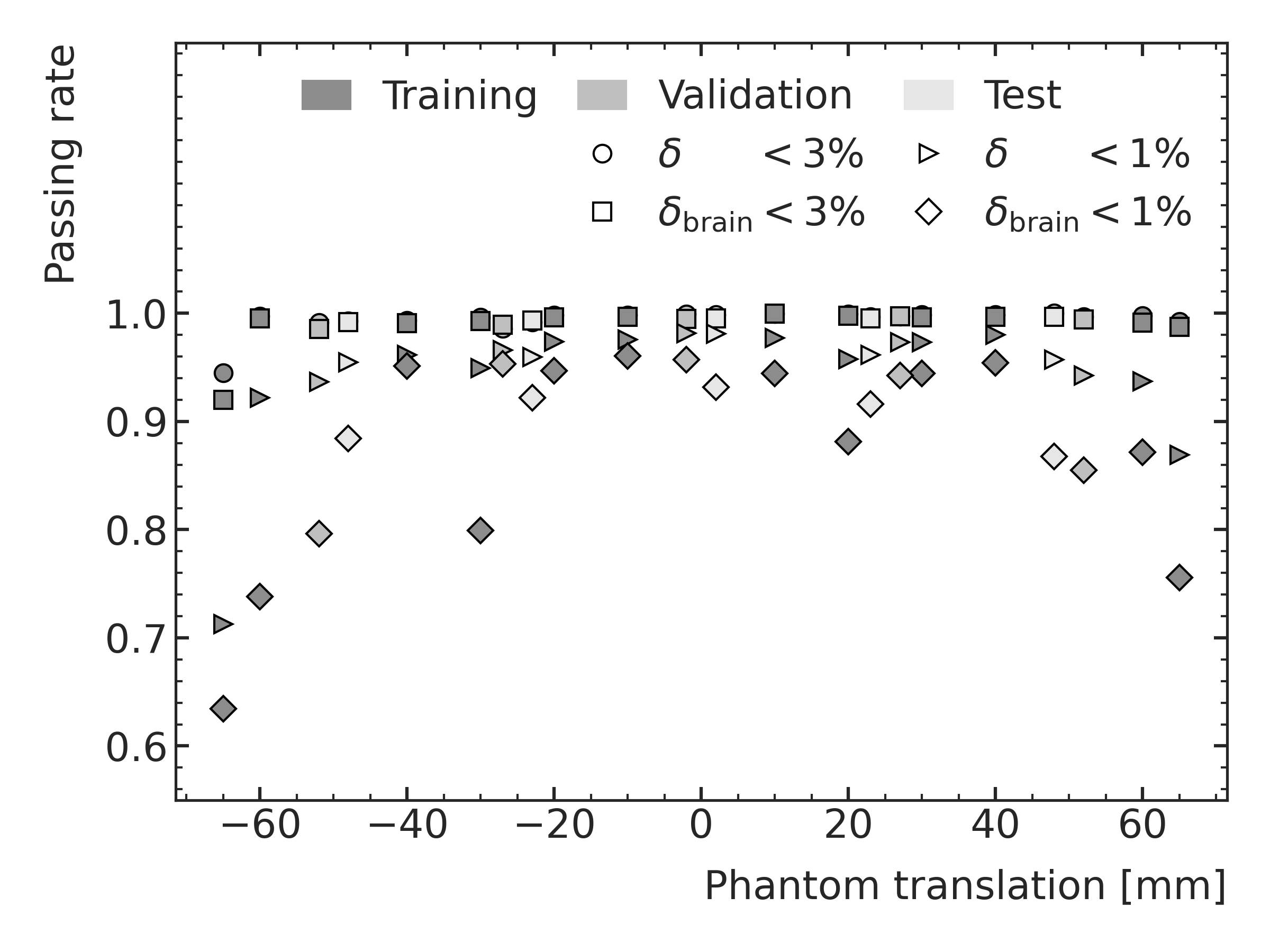}
		\caption{}
		\label{fig:simple_head:performance_train_val_test:b}
	\end{subfigure}
	\caption{(a) Global 1\% (red) and 3\% (black) passing rates for the training (solid) and validation (dashed) data as a function of training epochs. The best model on the validation data is highlighted with a red circle. (b) Passing rates as a function of the phantom translation for training (dark grey), validation (medium grey) and test (light grey) data.\bigskip}
	\label{fig:simple_head:performance_train_val_test}
\end{figure}

\begin{figure}[!tb]
\centering
\begin{subfigure}[t]{.49\textwidth}
	\includegraphics[width=\linewidth]{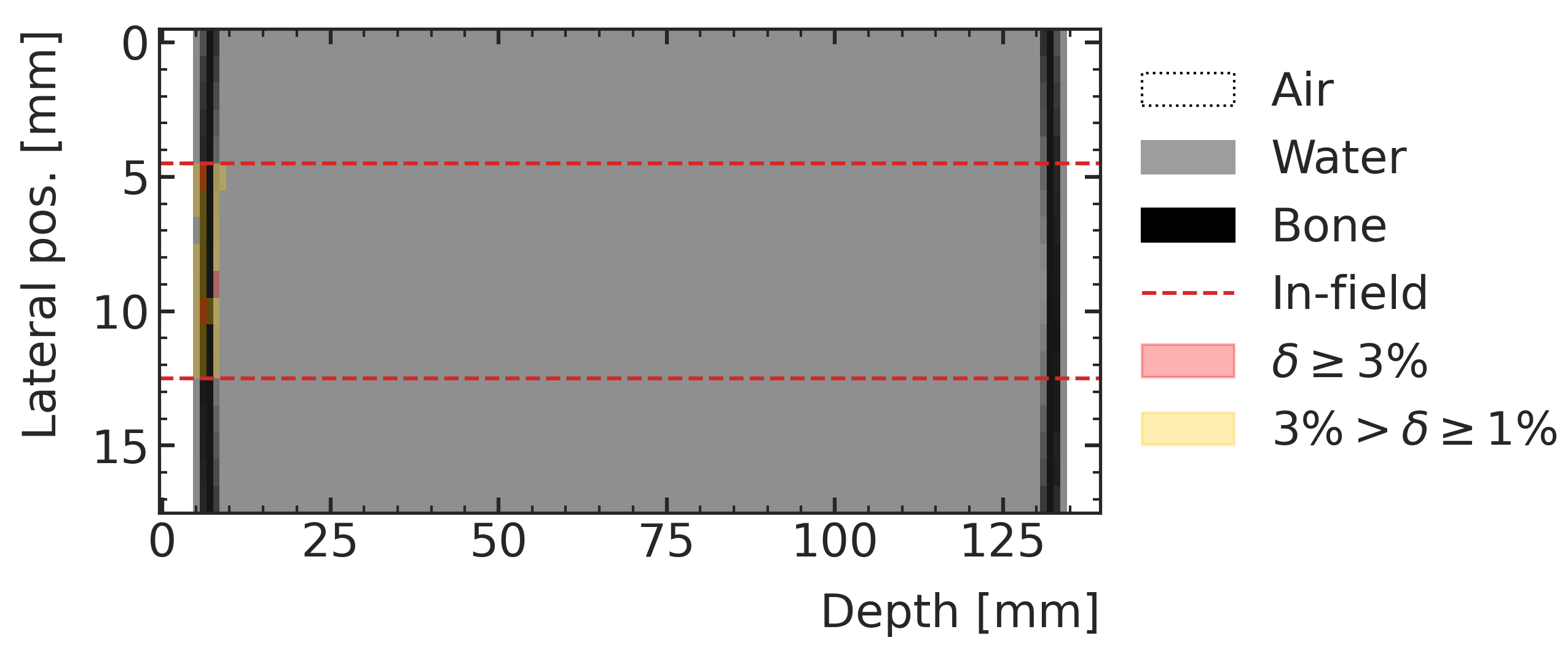}
	\caption{}
	\label{fig:dy273400:a}
\end{subfigure}
\begin{subfigure}[t]{.49\textwidth}
	\includegraphics[width=\linewidth]{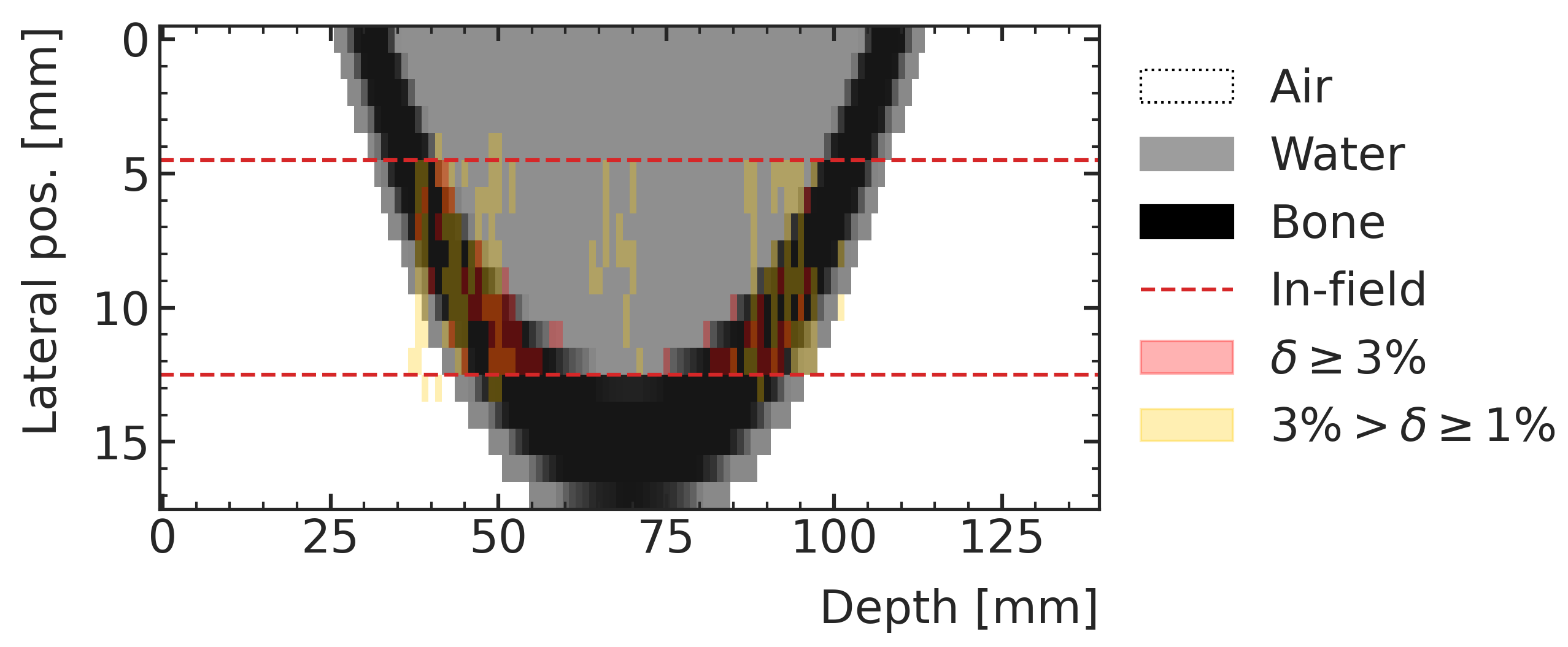}
	\caption{}
	\label{fig:dy273400:b}
\end{subfigure}
\caption{2D slices of the phantom with a translations of (a) $t = 2\,$mm as part of the test data and (b) $t = -65\,$mm as the worst performing case with training data.
	Voxels with deviations in their dose predictions of larger than 3\% (1\%) are shown in red (yellow). Whenever the deviation is smaller than 1\%, the gray-scale color of the material density matrix is shown. The in-field part of the beam is located between the red dashed lines.
	Voxels with more than 3\%$D_\mathrm{max}$ dose prediction deviation (red) and voxels with less than 3\%$D_\mathrm{max}$ but more than 1\%$D_\mathrm{max}$ dose deviation (yellow). The grey scale colours show the densities of the voxels, the red dashed lines indicate the beam edges.\bigskip}
\label{fig:dy273400}
\end{figure}

\begin{figure}[!tb]
\centering
\begin{subfigure}[t]{.32\textwidth}
	\includegraphics[width=\linewidth]{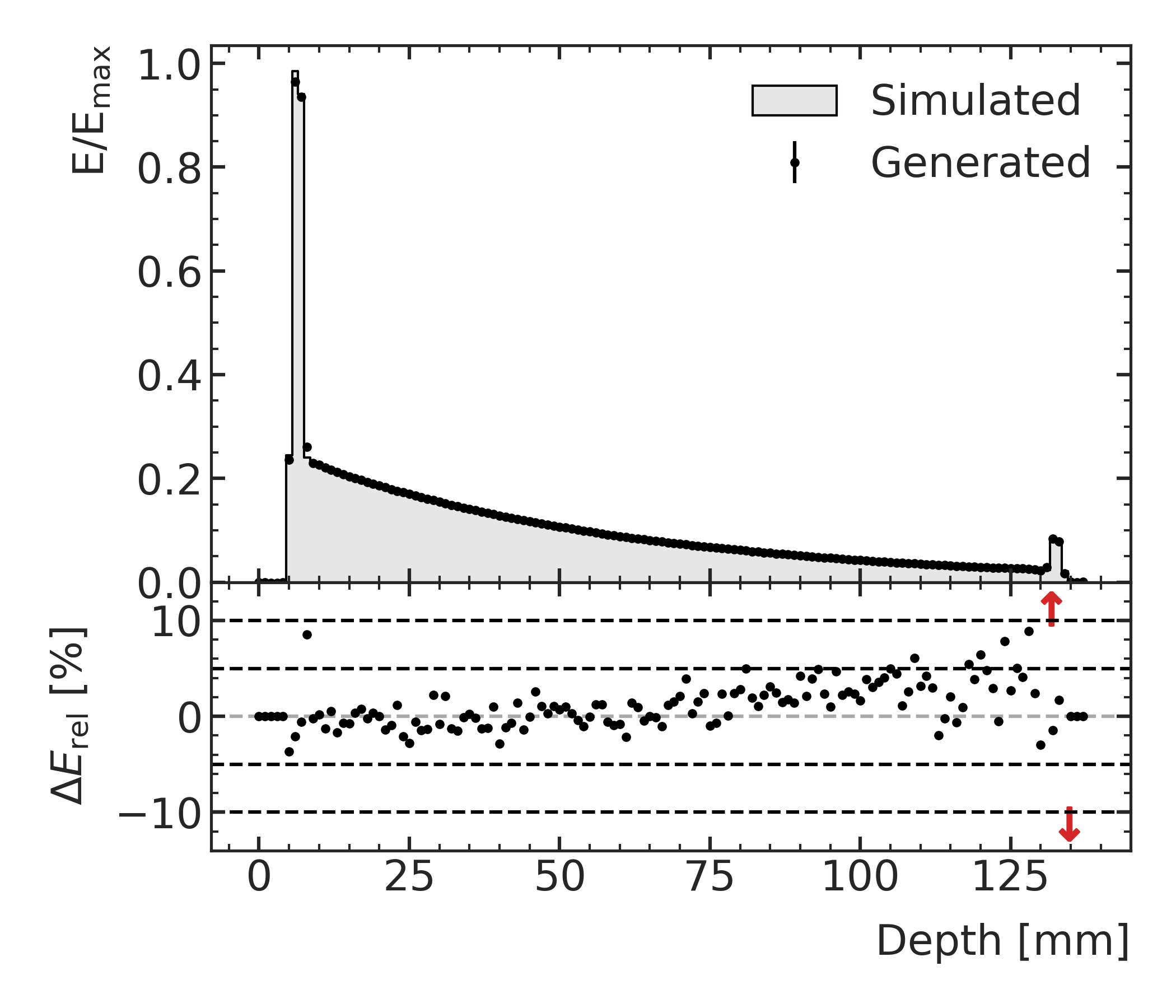}
	\caption{}
	\label{fig:simple_head:depth_dose_curves:a}
\end{subfigure}
\begin{subfigure}[t]{.32\textwidth}
	\includegraphics[width=\linewidth]{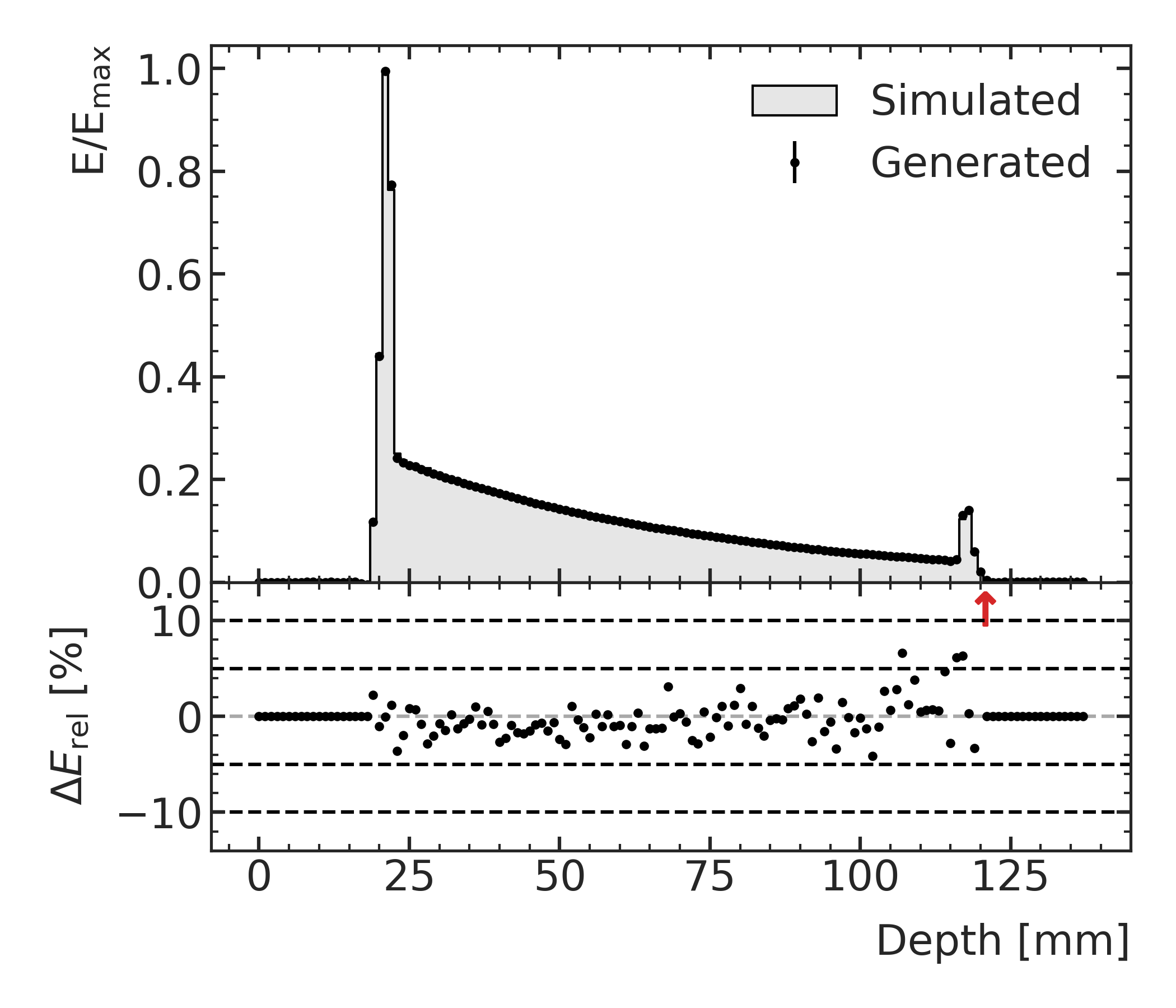}
	\caption{}
	\label{fig:simple_head:depth_dose_curves:b}
\end{subfigure}
\begin{subfigure}[t]{.32\textwidth}
	\includegraphics[width=\linewidth]{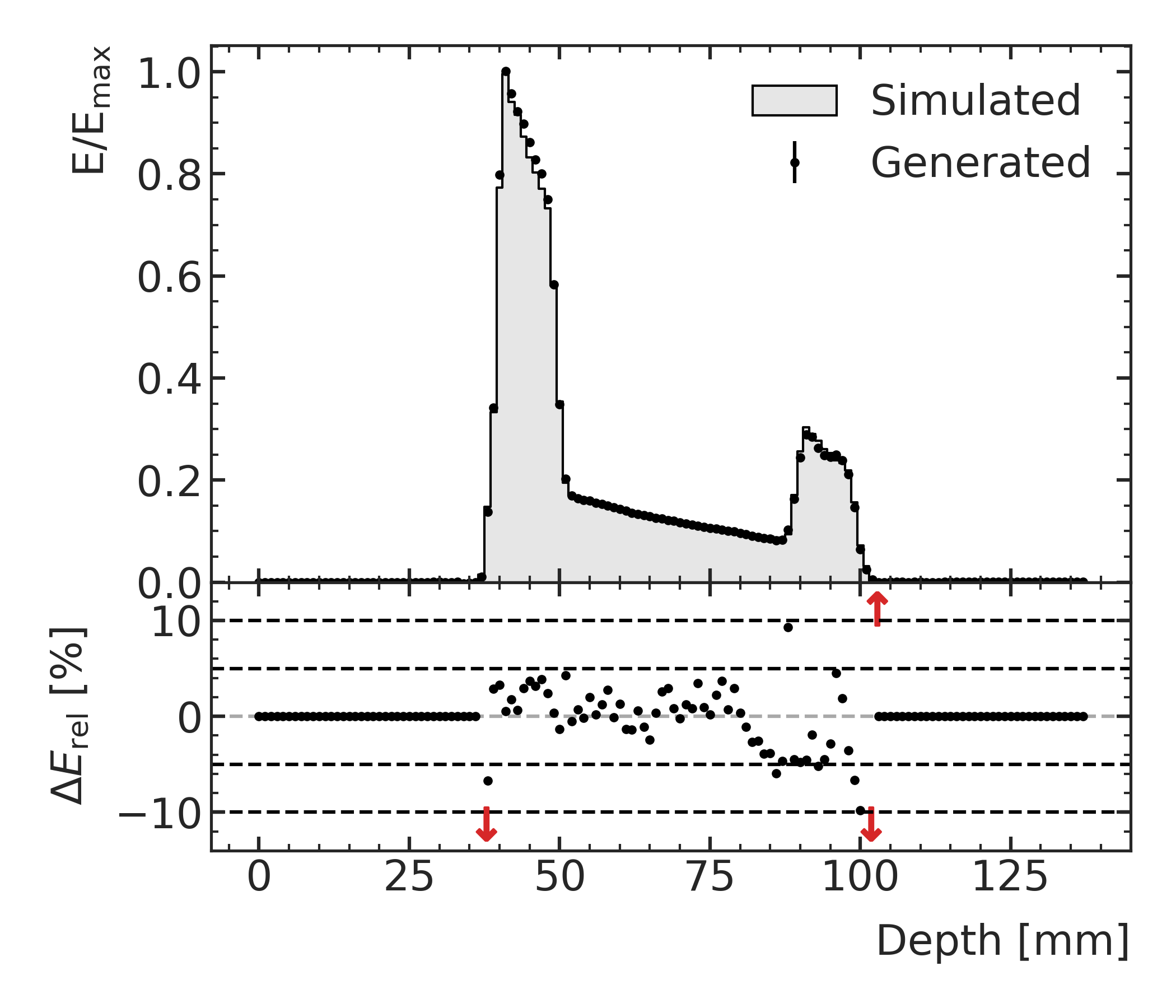}
	\caption{}
	\label{fig:simple_head:depth_dose_curves:c}
\end{subfigure}
\begin{subfigure}[t]{.32\textwidth}
	\includegraphics[width=\linewidth]{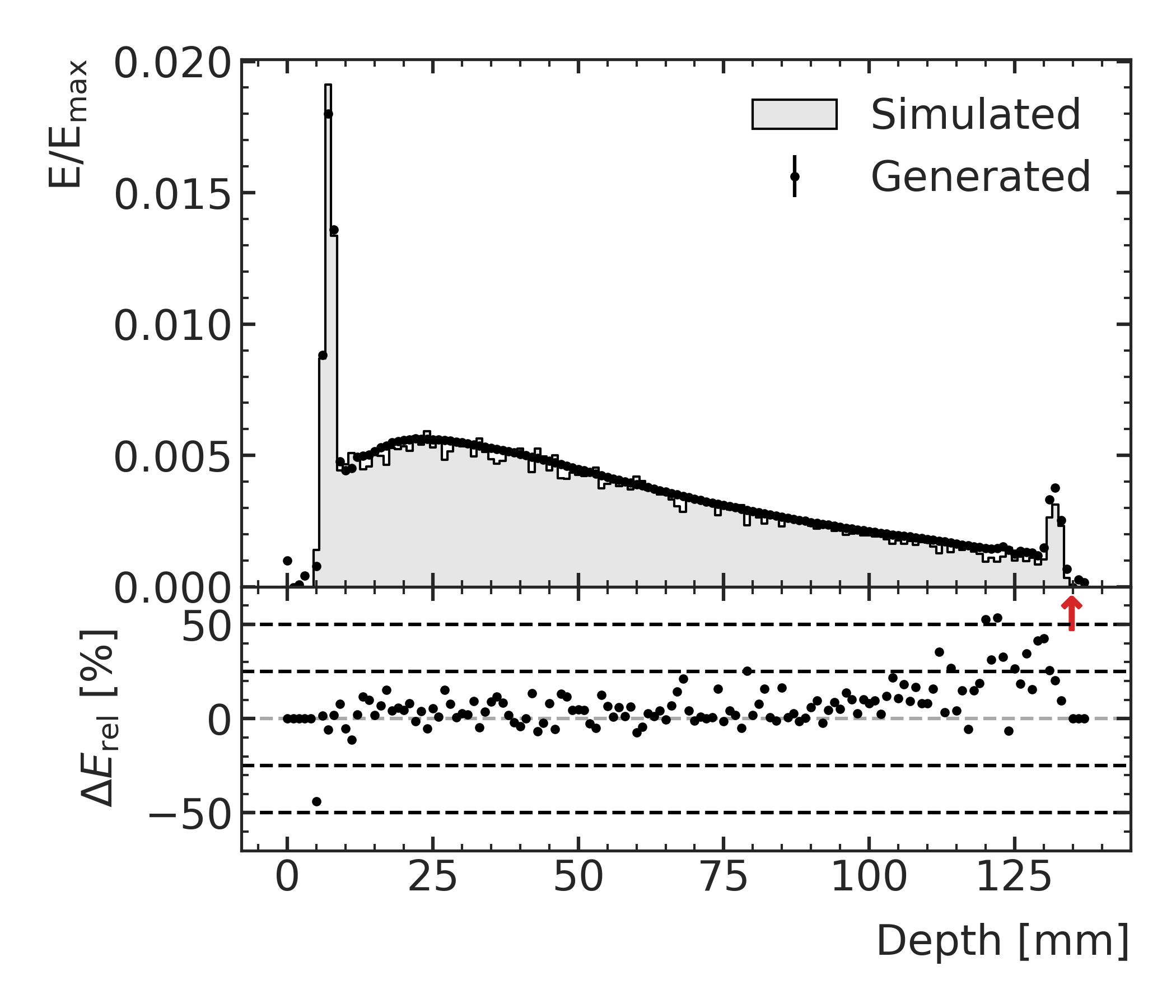}
	\caption{}
	\label{fig:simple_head:depth_dose_curves:d}
\end{subfigure}
\begin{subfigure}[t]{.32\textwidth}
	\includegraphics[width=\linewidth]{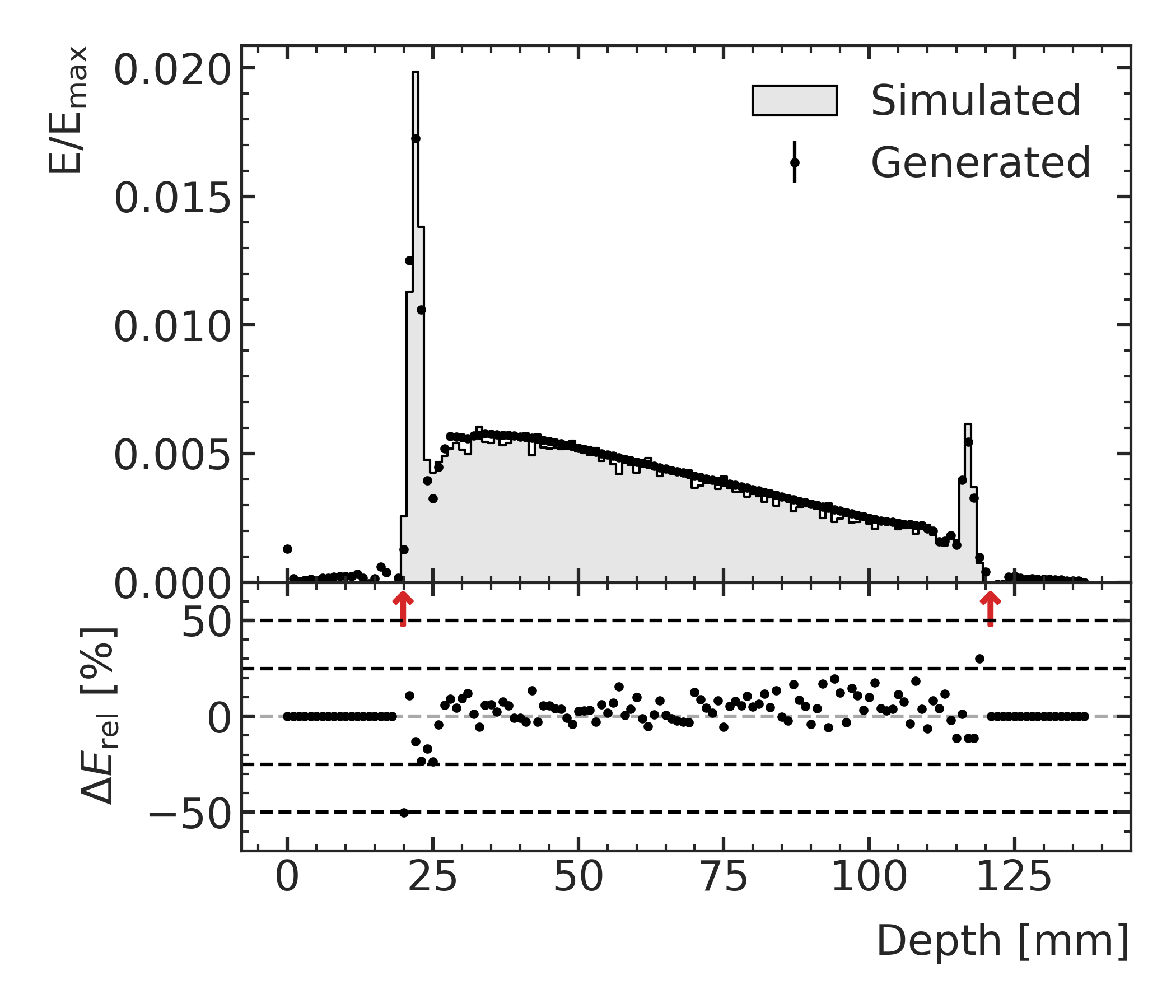}
	\caption{}
	\label{fig:simple_head:depth_dose_curves:e}
\end{subfigure}
\begin{subfigure}[t]{.32\textwidth}
	\includegraphics[width=\linewidth]{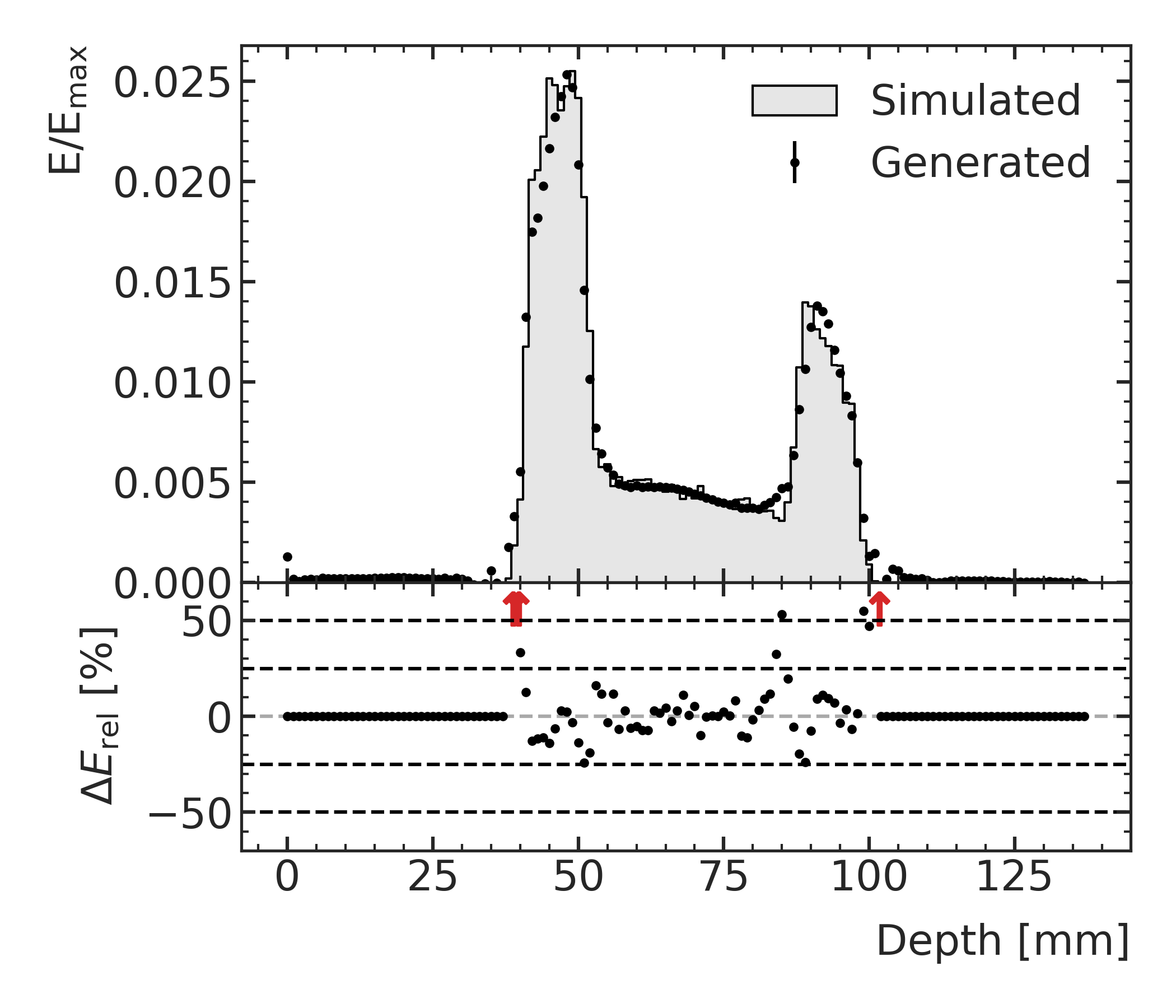}
	\caption{}
	\label{fig:simple_head:depth_dose_curves:f}
\end{subfigure}
\caption{Comparisons of normalized simulated and generated energy depositions inside the phantom along the beam using the test data for the in-field (a-c) and out-of-field (d-f) region of the beam with phantom translations of $t = [2, 48, -65]$\,mm (a+d, b+e, c+f). The lower part of the plots shows the relative energy deviation $\Delta E_\mathrm{rel}$ in percent. Outliers are indicated with a red arrow.\bigskip}
\label{fig:simple_head:depth_dose_curves}
\end{figure}

During the training of the model using the data of the simple paediatric head phantom, the highest 1\% passing rate on the validation data is achieved at 6,900 training epochs as shown in Figure \ref{fig:simple_head:performance_train_val_test:a}.\\
\noindent
As the dose in the bone of the skull is much higher compared to the dose in the water, which is mimicking the brain tissue, any dose deviations in the brain tissue would have negligible impact on the delta index, which is normalized to the total maximal simulated dose.
%
Therefore, we calculate an additional delta index $\delta_\mathrm{brain}$ that is normalized to the maximal simulated dose inside the brain and the corresponding passing rates are only taking the voxels of the brain tissue into account. The global passing rates of the best model averaged over the translation of the head are summarized in Table~\ref{tab:simpleHead_results} for the training, validation and test data.
\begin{table*}[!htb]
	\begin{center}
		\captionv{10}{}{Delta index passing rates with standard deviation for training, validation and test data for the simple pediatric head phantom. 
			\label{tab:simpleHead_results}
			\vspace*{2ex}
		}
		\begin{widetable}{\columnwidth}{r
				S[table-format=2.1(2)]
				S[table-format=2.1(2)]
				S[table-format=2.1(2)]
			}
			\toprule
			{}& \multicolumn{3}{c}{Passing rate [\%]}\\
			{} &    {Training} &  {Validation} &          {Test}\\
			\midrule
			$\delta < 1\%$         &      93\pm7 &  96.0\pm1.7 &    96.3\pm0.9 \\
			$\delta < 3\%$         &  99.3\pm1.5 &  99.4\pm0.5 &  99.6\pm0.1 \\
			$\delta_\mathrm{brain} < 1\%$ &     87\pm10 &      90\pm6 &    90.4\pm2.4\\
			$\delta_\mathrm{brain} < 3\%$ &  98.9\pm2.1 &  99.2\pm0.4 &  99.5\pm0.2 \\
			\bottomrule
		\end{widetable}
\end{center}
\end{table*}
The generated doses of more than 96\% (99\%) of the voxels deviate by less than 1\% (3\%) of the maximal dose from the simulated dose using the test data, if the whole phantom is considered. Although the maximal dose inside the brain is much smaller, the 3\% passing rate of the brain is very similar, while the 1\% passing rate is reduced to 90\%.
In Figure \ref{fig:simple_head:performance_train_val_test:b} the passing rates are shown as a function of the translation of the head for the training, validation and test data.
The 3\% passing rate is very stable around 99\% for the whole phantom as well as for the brain except for the extreme cases of the maximum translation of $t = \pm 65$\,mm. These cases are in particular challenging to predict, because part of the beam traverses only bone and the angle and thickness of the bone change most drastically, which results in steep local gradients of the dose deposition.
The 1\% passing rate shows a stronger dependence on the head position with a decrease to higher absolute translation distances.
The model generalizes well to the unseen translation distances of the test and validation data with minimal differences in performance to the training data.
%
Figure \ref{fig:dy273400} shows exemplary the relative dose deviations as a 2D slices of the head for a central case of the test data with $t = 2\,$mm (a) and the worst performing case of the training data with $t = -65\,$mm (b).  
For the central case (Fig. \ref{fig:dy273400:a}) only few voxels mostly at the first bone and behind show deviations of larger than 1\%. For the worst case (Fig. \ref{fig:dy273400:b}), where the beam is passing through the edge of the head, many voxels inside the bone and close to it show deviations of more than 1\% and several voxels in particular at thicker bone structures deviate by more than 3\% from the simulation.
Figure \ref{fig:simple_head:depth_dose_curves} shows the comparison of the normalized simulated and generated energy depositions inside the phantom along the beam for the in-field (a-c) and out-of-field (d-f) region of the beam with head translations of $t = [2, 48, -65]$\,mm (a+d, b+e, c+f).
The translations of $t = [2, 48]$\,mm correspond to test data, while the translation of $t = -65$\,mm represents training data but shows the worst performing case.
The relative deviations between simulated and generated energy depositions of the test data (a,b,d,e) are mostly within 5\% (25\%) for the in-field (out-of-field) region and the model shows a good ability to interpolate to unknown phantom geometries.
The deviations of the worst performing case (c,f) are only at the transition regions of the bones larger than 5\% (25\%) in the in-field (out-of-field) region and the energies are generally even for this case well predicted by the GAN.

\section{Discussion\label{section:Discussion}}
We developed, trained and evaluated a GAN model for dose predictions of a photon beam inside three variable phantom geometries  with increasing complexity. 
\\
On the first phantom, a water cube with a bone slab of fixed thickness inserted at various rotation angles, our model achieved on the test data a global 1\% and 3\% passing rate of ($96.9\pm0.7$)\% and ($99.6\pm0.2$)\%, respectively, and demonstrated stable performance over the full range of the varied rotation angles. In the in-field region of the beam the relative voxel-by-voxel deviations of the predicted energy depositions were below 5\%, except for voxels at the interface of water and bone. Out-of-field, the relative deviations were mostly within 25\%, however, the absolute deposited energies were also approximately two orders of magnitude smaller and had hence less impact on the loss function during the training. The model was able to interpolate the learnt energy depositions of the training data to energy depositions inside bone slabs with unlearnt rotation angles of the validation and test data with little loss in performance.
\\
In the second phantom, the bone thickness of the slab was varied in addition to the rotation angle. This increase of complexity reduced the global 1\% and 3\% passing rate to ($87.2\pm1.4$)\% and ($98.5\pm0.4$)\%, respectively, on the test data, but both rates were relatively stable over the variation of bone slab rotation angle and thickness with small decrease for bone thicknesses that were only part of the test data.
The deposited energies were accurately predicted on a voxel-by-voxel basis mostly within 5\% (25\%) of relative deviation to the simulated energy deposition in the in-field (out-of-field) region of the beam.
We found that the model was able to predict the deposited energy with comparable accuracy for geometries (bone slab thickness and rotation angle) that were not part of the training data. This demonstrates that the model is able to generalize the learnt energy deposition prediction to unknown but similar phantom geometries. 
The model showed a small trend to systematically overestimates the energy depositions behind the bone slab insert, in particular for thicker bones, which indicates that the model did not fully learn the attenuation effect of the bone slab on the beam.
\\
The last phantom, a simple pediatric head, was shifted orthogonal to the beam in order to emulate dynamically changing phantom geometries in multiple dimensions.
The obtained 1\% and 3\% passing rates on the test data of ($96.3\pm0.9$)\% and ($99.6\pm0.1$)\%, respectively, were at a high level and stable over the full range of phantom translation except for the extreme case of {$t=-65$\,mm}, for which the beam passes through the edge of the head. If only the voxels of the brain tissue were considered, the 1\% passing rate was reduced by roughy 6\% and showed stronger dependency on the translation, while the 3\% passing rate was comparable to the one of the whole phantom.
The relative deviations of the generated energies from the simulations were mostly below 5\% (25\%) for most voxels of the in-field (out-of-field) region of the beam. Larger deviations were found at the interfaces of the bones.
No significant difference were found in the passing rates or in the energy deposition deviations between those obtained from training data and from validation or test data. This establishes that the presented model is able to generalize very well to dynamically changing geometries, which are unknown to the model.\\
Table~\ref{tab:execution_times} shows a comparison of the execution times to obtain the deposited energies inside a phantom either using full Geant4 Monte Carlo simulation or the GAN model on a CPU ({Intel Xeon E5-2630 v4 @ 2.2 GHz}) and a GPU (\textit{Nvidia GeForce GTX 1080i}). 
\begin{table}[t]
	\begin{center}
		\captionv{10}{}{Execution times of the GAN and the Geant4 MC simulations compared. CPU refers to an \textit{Intel Xeon E5-2630 v4 @ 2.2 GHz}, GPU to a \textit{Nvidia GeForce GTX 1080i}.
			\label{tab:execution_times}
			\vspace*{2ex}
		}
		\begin{widetable}{\columnwidth}{r
            S[table-format=2.1(2)]
            S[table-format=2.1(2)]
            S[table-format=2.1(2)]
        }
			
			\toprule
			{Model} & {Time per prediction [s]} & {Rel. speed}\\ 
			\midrule
			{Geant4 (1 CPU)}  & {$9.5 \cdot 10^5$} & {1}\\
			{GAN (1 CPU)} & {0.6} & ${\sim 1.6\cdot10^6}$ \\
			{GAN (1 GPU)} & {0.1} & ${\sim 9.5\cdot10^6}$ \\
			
			\bottomrule
		\end{widetable}
	\end{center}
\end{table}
While the Geant4 MC simulation takes roughly 264 computing hours, the GAN is able to make the predictions in only 0.6 seconds on the same CPU, which corresponds to a speed-up factor of roughly $1.6\cdot 10^6$. Using a GPU instead for the GAN prediction, the deposited energies are obtained in 0.1 seconds, which is almost ten million times faster than the Geant4 MC simulation and also significantly faster than the 30 minutes as reported by a previously developed method~\cite{Donzelli2018}.
Compared to the fastest previously published dose computation method\cite{Poole2017}, the presented model shows significantly better spatial resolution (the fastest previous model does not perform voxel by voxel predictions at all) and more accurate predictions at material interfaces and generally in the case of heterogeneities in the phantom at similar execution speeds.
Compared to generalist approximations such as GPUMCD \cite{Hissoiny2011} and the hybrid approaches for MRT\cite{Donzelli2018}, the presented model does not rely on approximations like averaged attenuation coefficients introduced before generating simulation data. As a consequence, any physical process can be taken into account for the model prediction as long as it can be simulated accurately, e.g. by Monte Carlo methods for radiation physics.
\\
As the training data is generated using full Monte Carlo simulations, it can be adapted closely to realistic treatment scenarios. Fast dose prediction approximations which are currently used in treatment planning programmes have been found to require greater attention and corrections when it comes to commissioning of the system for clinical application (discussed e. g. in the studies by Montenari et al.~\cite{Montanari2014} or Jia et al.~\cite{Jia2012}). The constructed MC simulation application used in the presented study was validated against experimental measurements in a previous publication\cite{Dipuglia2019} and it was successfully used for dose predictions in an in vivo experiment with rat irradiations\cite{Engels2020}. These are important research steps, which pose a solid foundation for an eventual, future commissioning of the proposed TPS engine. To fully commission the proposed TPS engine, the pathway would be to commissioning the Monte Carlo simulations used for the ML training, against dosimetric experimental measurements.
\\
The high accuracy of the predictions, the ability to interpolate and generalize to unknown phantom geometries and the fast execution times encourage to further develop the model and adapt it for the use in MRT. The model demonstrated flexibility and scalability due to the conditioning on the beam properties and on the geometry of the phantom. 
At this stage, the presented model is trained only at one fixed resolution. To use different voxel and prediction volume sizes will be subject of the next stage of the project, where we adapt the developed ML algorithm to microbeams.

%
\section{Conclusion\label{section:Conclusion}}
We present a proof-of-concept study to use a 3D-UNet GAN to predict the dose or energy depositions of a photon beam inside a variety of phantoms. The GAN is conditioned on the beam and phantom properties and is trained using a Geant4 Monte Carlo simulation of the full radiation transport. The presented model is successfully trained on three different phantoms with changing geometry without further need for optimization. The model predicts with adequate accuracy the energy depositions and deviates by less than 1\% of the maximal deposited energy from the simulation in more than 96\% of the in-field voxels for the most realistic phantom - a simple paediatric head. 
The model demonstrates good ability to generalize the predictions to unknown but similar phantom geometries.
\\
Our approach differs from previously published studies on dose estimations with machine learning methods by relying purely on Monte Carlo simulations for radiation physics as training data. This allows the development of a more general approach that can be deployed to novel treatment methods, for which there are no large number of available patient CT scans or existing treatment plans already, such as  mini\cite{Deman2012} or microbeam radiation therapy.\\
For an application in treatment planning systems, dose simulations with satisfactory statistical accuracy using MC techniques are usually computationally too demanding, however, approximations of the physics processes translate to large dose calculation inaccuracies in heterogeneous phantoms, especially at material interfaces.
In the adopted dosimetric calculation scenario, the presented GAN model is able to achieve almost MC simulation precision in only a fraction of a second for a variety of radiation scenarios, which makes it an ideal candidate for future treatment planning systems.
Encouraged by these results, future studies will evaluate the presented machine learning approach using simulations with more realistic patient geometries, deriving for example from the ICRP110 phantom\cite{HG2009}, digital phantoms which allow customization via parameterization\cite{Arce2021, Large2020}, and finally patient data for the application in treatment planning.
 

\section{Acknowledgements}
The authors gratefully acknowledge the computing time provided on the Linux HPC cluster at TU Dortmund University (LiDO3), partially funded in the course of the Large-Scale Equipment Initiative by the German Research Foundation (DFG) as project 271512359.
\section{Bibliography}
\bibliographystyle{ama}     
\bibliography{references}   






\end{document}